\documentclass[acmsmall,screen,nonacm]{acmart}
\setcopyright{none}
\makeatletter
\@ifundefined{hyxmp@parse@acmart}{}{\let\hyxmp@parse@acmart\relax}
\makeatother

\usepackage{graphicx}
\usepackage{amsmath,amsthm,mathtools,booktabs,tabularx,array,enumitem,microtype,placeins,longtable}
\newcolumntype{Y}{>{\raggedright\arraybackslash}X}
\newtheorem{definition}{Definition}

\hypersetup{
  hidelinks,
  keeppdfinfo,
  pdftitle={Accepted but Wrong: Silent Failures of LLM Code Repairs after API Migrations},
  pdfauthor={Xiaobo Chen, Genliang Zhu, Ziyuan Wang, Chu Wang},
  pdfsubject={Software engineering (cs.SE): release gates and test oracles for LLM-based program repair after runtime and API migrations},
  pdfkeywords={LLM program repair, API migration, runtime upgrade, release notes, edge-case test inputs, test oracle adequacy, release gate, LLM test generation, accepted-but-wrong repairs, silent behavioral breaking changes, patch overfitting, differential testing, preregistered empirical study, Python 3.12, Node.js 24, SQLite, GPT-5.6, Qwen3, gpt-oss},
  pdfcreator={LaTeX}
}
\setlist[itemize]{leftmargin=1.35em}
\setlist[enumerate]{leftmargin=1.55em}

\title[Accepted but Wrong]{Accepted but Wrong: Silent Failures of LLM Code Repairs after API Migrations}

\author{Xiaobo Chen}
\affiliation{%
  \institution{University of Massachusetts Amherst}
  \city{Amherst}
  \state{Massachusetts}
  \country{USA}}

\author{Genliang Zhu}
\affiliation{%
  \institution{Accentrust}
  \city{Vancouver}
  \country{Canada}}
\affiliation{%
  \institution{Georgia Institute of Technology}
  \department{School of Computer Science}
  \city{Atlanta}
  \state{Georgia}
  \country{USA}}

\author{Ziyuan Wang}
\affiliation{%
  \institution{Duke University}
  \city{Durham}
  \state{North Carolina}
  \country{USA}}

\author{Chu Wang}
\affiliation{%
  \institution{Accentrust}
  \city{Vancouver}
  \country{Canada}}
\affiliation{%
  \institution{University of Illinois Urbana-Champaign}
  \department{Department of Computer Science}
  \city{Urbana}
  \state{Illinois}
  \country{USA}}

\begin{document}
\newcommand{\PriorBFourJoint}{96/100}
\newcommand{\PriorBOneJoint}{88/100}
\newcommand{\PriorBOneRJoint}{96/100}
\newcommand{\PriorBZeroJoint}{24/100}
\newcommand{\PriorBZeroSilent}{48}
\newcommand{\PriorProcedureMatches}{100/100}
\newcommand{\PriorTestLineages}{20}
\newcommand{\SAAdqConsumerCompleteFA}{0.36}
\newcommand{\SAAdqConsumerCompleteFATasks}{18}
\newcommand{\SAAdqConsumerCompleteFR}{0.00}
\newcommand{\SAAdqConsumerPartialFA}{0.50}
\newcommand{\SAAdqConsumerPartialFATasks}{20}
\newcommand{\SAAdqConsumerPartialFR}{0.00}
\newcommand{\SAAdqDifferentialFA}{0.36}
\newcommand{\SAAdqDifferentialFATasks}{18}
\newcommand{\SAAdqDifferentialFR}{0.05}
\newcommand{\SAAdqLlmContractFA}{0.34}
\newcommand{\SAAdqLlmContractFATasks}{17}
\newcommand{\SAAdqLlmContractFR}{0.05}
\newcommand{\SAAdqNoneFA}{1.00}
\newcommand{\SAAdqNoneFATasks}{24}
\newcommand{\SAAdqNoneFR}{0.00}
\newcommand{\SAAdqRuntimeFA}{0.80}
\newcommand{\SAAdqRuntimeFATasks}{24}
\newcommand{\SAAdqRuntimeFR}{0.00}
\newcommand{\SAAdqTasks}{24}
\newcommand{\SAAllBoundary}{4}
\newcommand{\SAAllControl}{5}
\newcommand{\SAAllExplicit}{12}
\newcommand{\SAAllSilent}{12}
\newcommand{\SACPython}{22}
\newcommand{\SACodexBoundaryAbstained}{7}
\newcommand{\SACodexBoundaryAttempts}{12}
\newcommand{\SACodexBoundaryCC}{0/12}
\newcommand{\SACodexBoundaryConsumerComplete}{0.25}
\newcommand{\SACodexBoundaryConsumerPartial}{0.42}
\newcommand{\SACodexBoundaryDifferential}{0.25}
\newcommand{\SACodexBoundaryFANone}{5/12}
\newcommand{\SACodexBoundaryLlmContract}{0.25}
\newcommand{\SACodexBoundaryNone}{0.42}
\newcommand{\SACodexBoundaryRuntime}{0.42}
\newcommand{\SACodexCCDiffOff}{48/54}
\newcommand{\SACodexCCDiffOn}{48/54}
\newcommand{\SACodexCCFull}{48/54}
\newcommand{\SACodexCCPatch}{48/54}
\newcommand{\SACodexCalls}{122}
\newcommand{\SACodexConsumerCompleteAbst}{9}
\newcommand{\SACodexConsumerCompleteBlocked}{0}
\newcommand{\SACodexConsumerCompleteCC}{96}
\newcommand{\SACodexConsumerCompleteFA}{3}
\newcommand{\SACodexConsumerCompleteFR}{0}
\newcommand{\SACodexConsumerCompleteSFA}{0}
\newcommand{\SACodexConsumerPartialAbst}{7}
\newcommand{\SACodexConsumerPartialBlocked}{0}
\newcommand{\SACodexConsumerPartialCC}{96}
\newcommand{\SACodexConsumerPartialFA}{5}
\newcommand{\SACodexConsumerPartialFR}{0}
\newcommand{\SACodexConsumerPartialSFA}{2}
\newcommand{\SACodexControlCC}{12/12}
\newcommand{\SACodexControlFANone}{0/12}
\newcommand{\SACodexDifferentialAbst}{9}
\newcommand{\SACodexDifferentialBlocked}{4}
\newcommand{\SACodexDifferentialCC}{92}
\newcommand{\SACodexDifferentialFA}{3}
\newcommand{\SACodexDifferentialFR}{4}
\newcommand{\SACodexDifferentialSFA}{0}
\newcommand{\SACodexExplicitCC}{40/40}
\newcommand{\SACodexExplicitFANone}{0/40}
\newcommand{\SACodexFATasksCC}{\texttt{py-distutils-looseversion}}
\newcommand{\SACodexFATasksRuntime}{\texttt{py-distutils-looseversion}, \texttt{py-shuffle-random-param}}
\newcommand{\SACodexFRTasksDiff}{\texttt{py-mkdtemp-absolute}}
\newcommand{\SACodexFRTasksLLM}{\texttt{node-fs-truncate-fd}}
\newcommand{\SACodexHFourDiff}{0.000}
\newcommand{\SACodexHFourDisc}{0}
\newcommand{\SACodexHFourHigh}{0.000}
\newcommand{\SACodexHFourHolm}{--}
\newcommand{\SACodexHFourLow}{0.000}
\newcommand{\SACodexHFourP}{--}
\newcommand{\SACodexHOneDiff}{$-$0.019}
\newcommand{\SACodexHOneDisc}{1}
\newcommand{\SACodexHOneHigh}{0.000}
\newcommand{\SACodexHOneHolm}{1.000}
\newcommand{\SACodexHOneLow}{$-$0.058}
\newcommand{\SACodexHOneP}{1.000}
\newcommand{\SACodexHThreeDiff}{0.000}
\newcommand{\SACodexHThreeDisc}{0}
\newcommand{\SACodexHThreeHigh}{0.000}
\newcommand{\SACodexHThreeHolm}{--}
\newcommand{\SACodexHThreeLow}{0.000}
\newcommand{\SACodexHThreeP}{--}
\newcommand{\SACodexHTwoDiff}{0.000}
\newcommand{\SACodexHTwoDisc}{0}
\newcommand{\SACodexHTwoHigh}{0.000}
\newcommand{\SACodexHTwoHolm}{--}
\newcommand{\SACodexHTwoLow}{0.000}
\newcommand{\SACodexHTwoP}{--}
\newcommand{\SACodexInputTokens}{841{,}267}
\newcommand{\SACodexLlmContractAbst}{9}
\newcommand{\SACodexLlmContractBlocked}{4}
\newcommand{\SACodexLlmContractCC}{92}
\newcommand{\SACodexLlmContractFA}{3}
\newcommand{\SACodexLlmContractFR}{4}
\newcommand{\SACodexLlmContractSFA}{0}
\newcommand{\SACodexMeanEditsCC}{1.01}
\newcommand{\SACodexMeanSeconds}{14.6}
\newcommand{\SACodexModel}{codex-gpt-5.6-luna-high}
\newcommand{\SACodexN}{108}
\newcommand{\SACodexNoneAbst}{7}
\newcommand{\SACodexNoneBlocked}{0}
\newcommand{\SACodexNoneCC}{96}
\newcommand{\SACodexNoneFA}{5}
\newcommand{\SACodexNoneFR}{0}
\newcommand{\SACodexNoneSFA}{2}
\newcommand{\SACodexOutputTokens}{56{,}416}
\newcommand{\SACodexOverEditCC}{0.02}
\newcommand{\SACodexRevisedCC}{2}
\newcommand{\SACodexRuntimeAbst}{7}
\newcommand{\SACodexRuntimeBlocked}{0}
\newcommand{\SACodexRuntimeCC}{96}
\newcommand{\SACodexRuntimeFA}{5}
\newcommand{\SACodexRuntimeFR}{0}
\newcommand{\SACodexRuntimeSFA}{2}
\newcommand{\SACodexSilentCC}{44/44}
\newcommand{\SACodexSilentFANone}{0/44}
\newcommand{\SACodexVoid}{0}
\newcommand{\SADevClusters}{6}
\newcommand{\SADevTasks}{6}
\newcommand{\SADomainMax}{32}
\newcommand{\SADomainTotal}{236}
\newcommand{\SALLMAssertions}{88}
\newcommand{\SALLMCalls}{33}
\newcommand{\SALLMDropped}{0}
\newcommand{\SANode}{11}
\newcommand{\SAPilotCC}{20/24}
\newcommand{\SAPilotCalls}{27}
\newcommand{\SAPilotHalfWidth}{0.04}
\newcommand{\SAQwenBoundaryAbstained}{2}
\newcommand{\SAQwenBoundaryAttempts}{12}
\newcommand{\SAQwenBoundaryCC}{0/12}
\newcommand{\SAQwenBoundaryConsumerComplete}{0.00}
\newcommand{\SAQwenBoundaryConsumerPartial}{0.33}
\newcommand{\SAQwenBoundaryDifferential}{0.00}
\newcommand{\SAQwenBoundaryFANone}{10/12}
\newcommand{\SAQwenBoundaryLlmContract}{0.00}
\newcommand{\SAQwenBoundaryNone}{0.83}
\newcommand{\SAQwenBoundaryRuntime}{0.33}
\newcommand{\SAQwenCCDiffOff}{34/54}
\newcommand{\SAQwenCCDiffOn}{34/54}
\newcommand{\SAQwenCCFull}{34/54}
\newcommand{\SAQwenCCPatch}{34/54}
\newcommand{\SAQwenCalls}{230}
\newcommand{\SAQwenConsumerCompleteAbst}{2}
\newcommand{\SAQwenConsumerCompleteBlocked}{15}
\newcommand{\SAQwenConsumerCompleteCC}{68}
\newcommand{\SAQwenConsumerCompleteFA}{20}
\newcommand{\SAQwenConsumerCompleteFR}{0}
\newcommand{\SAQwenConsumerCompleteSFA}{16}
\newcommand{\SAQwenConsumerPartialAbst}{2}
\newcommand{\SAQwenConsumerPartialBlocked}{10}
\newcommand{\SAQwenConsumerPartialCC}{67}
\newcommand{\SAQwenConsumerPartialFA}{26}
\newcommand{\SAQwenConsumerPartialFR}{0}
\newcommand{\SAQwenConsumerPartialSFA}{22}
\newcommand{\SAQwenControlCC}{11/12}
\newcommand{\SAQwenControlFANone}{1/12}
\newcommand{\SAQwenDifferentialAbst}{5}
\newcommand{\SAQwenDifferentialBlocked}{16}
\newcommand{\SAQwenDifferentialCC}{64}
\newcommand{\SAQwenDifferentialFA}{20}
\newcommand{\SAQwenDifferentialFR}{1}
\newcommand{\SAQwenDifferentialSFA}{16}
\newcommand{\SAQwenExplicitCC}{31/40}
\newcommand{\SAQwenExplicitFANone}{9/40}
\newcommand{\SAQwenFATasksCC}{\texttt{node-buffer-max-length}, \texttt{node-control-path-posix}, \texttt{py-enum-contains-value}, \texttt{py-int-max-str-digits}, \texttt{py-intflag-invert}, \texttt{py-random-sample-set}, \texttt{py-randrange-float}, \texttt{py-strenum-format}, \texttt{py-sum-neumaier}}
\newcommand{\SAQwenFATasksRuntime}{\texttt{node-buffer-max-length}, \texttt{node-control-path-posix}, \texttt{py-enum-contains-value}, \texttt{py-int-max-str-digits}, \texttt{py-intenum-str}, \texttt{py-intflag-invert}, \texttt{py-mkdtemp-absolute}, \texttt{py-random-sample-set}, \texttt{py-randrange-float}, \texttt{py-shuffle-random-param}, \texttt{py-strenum-format}, \texttt{py-sum-neumaier}}
\newcommand{\SAQwenFRTasksDiff}{\texttt{py-mkdtemp-absolute}}
\newcommand{\SAQwenFRTasksLLM}{\texttt{node-fs-truncate-fd}}
\newcommand{\SAQwenHFourDiff}{$-$0.019}
\newcommand{\SAQwenHFourDisc}{2}
\newcommand{\SAQwenHFourHigh}{0.058}
\newcommand{\SAQwenHFourHolm}{1.000}
\newcommand{\SAQwenHFourLow}{$-$0.115}
\newcommand{\SAQwenHFourP}{1.000}
\newcommand{\SAQwenHOneDiff}{$-$0.082}
\newcommand{\SAQwenHOneDisc}{4}
\newcommand{\SAQwenHOneHigh}{$-$0.010}
\newcommand{\SAQwenHOneHolm}{0.375}
\newcommand{\SAQwenHOneLow}{$-$0.178}
\newcommand{\SAQwenHOneP}{0.125}
\newcommand{\SAQwenHThreeDiff}{0.014}
\newcommand{\SAQwenHThreeDisc}{2}
\newcommand{\SAQwenHThreeHigh}{0.038}
\newcommand{\SAQwenHThreeHolm}{0.999}
\newcommand{\SAQwenHThreeLow}{0.000}
\newcommand{\SAQwenHThreeP}{0.499}
\newcommand{\SAQwenHTwoDiff}{0.000}
\newcommand{\SAQwenHTwoDisc}{0}
\newcommand{\SAQwenHTwoHigh}{0.000}
\newcommand{\SAQwenHTwoHolm}{--}
\newcommand{\SAQwenHTwoLow}{0.000}
\newcommand{\SAQwenHTwoP}{--}
\newcommand{\SAQwenLlmContractAbst}{3}
\newcommand{\SAQwenLlmContractBlocked}{19}
\newcommand{\SAQwenLlmContractCC}{63}
\newcommand{\SAQwenLlmContractFA}{20}
\newcommand{\SAQwenLlmContractFR}{4}
\newcommand{\SAQwenLlmContractSFA}{16}
\newcommand{\SAQwenMeanEditsCC}{0.97}
\newcommand{\SAQwenModel}{qwen3-8b-4bit}
\newcommand{\SAQwenN}{108}
\newcommand{\SAQwenNoneAbst}{2}
\newcommand{\SAQwenNoneBlocked}{0}
\newcommand{\SAQwenNoneCC}{65}
\newcommand{\SAQwenNoneFA}{37}
\newcommand{\SAQwenNoneFR}{0}
\newcommand{\SAQwenNoneSFA}{27}
\newcommand{\SAQwenOverEditCC}{0.12}
\newcommand{\SAQwenQualDecode}{38.7}
\newcommand{\SAQwenQualExact}{12/12}
\newcommand{\SAQwenQualPrefill}{970}
\newcommand{\SAQwenRevisedCC}{21}
\newcommand{\SAQwenRuntimeAbst}{2}
\newcommand{\SAQwenRuntimeBlocked}{6}
\newcommand{\SAQwenRuntimeCC}{66}
\newcommand{\SAQwenRuntimeFA}{31}
\newcommand{\SAQwenRuntimeFR}{0}
\newcommand{\SAQwenRuntimeSFA}{27}
\newcommand{\SAQwenSilentCC}{26/44}
\newcommand{\SAQwenSilentFANone}{17/44}
\newcommand{\SATasks}{33}
\newcommand{\SATestAffected}{24}
\newcommand{\SATestBoundary}{3}
\newcommand{\SATestClusters}{26}
\newcommand{\SATestControl}{3}
\newcommand{\SATestExplicit}{10}
\newcommand{\SATestSilent}{11}
\newcommand{\SATestTasks}{27}
\newcommand{\SBBInfFullBeh}{48/48}
\newcommand{\SBBInfFullNone}{48/48}
\newcommand{\SBBInfPatchBeh}{47/48}
\newcommand{\SBBInfPatchNone}{47/48}
\newcommand{\SBBTwoFullBeh}{45/48}
\newcommand{\SBBTwoFullNone}{45/48}
\newcommand{\SBBTwoPatchBeh}{46/48}
\newcommand{\SBBTwoPatchNone}{46/48}
\newcommand{\SBCalls}{192}
\newcommand{\ZBaseIndist}{0.36}
\newcommand{\ZChangedCount}{18}
\newcommand{\ZCodexAffN}{96}
\newcommand{\ZCodexBndConsumerCompleteBlk}{0}
\newcommand{\ZCodexBndConsumerCompleteRel}{3/12}
\newcommand{\ZCodexBndConsumerPartialBlk}{0}
\newcommand{\ZCodexBndConsumerPartialRel}{5/12}
\newcommand{\ZCodexBndDifferentialBlk}{0}
\newcommand{\ZCodexBndDifferentialRel}{3/12}
\newcommand{\ZCodexBndLlmContractBlk}{0}
\newcommand{\ZCodexBndLlmContractRel}{3/12}
\newcommand{\ZCodexBndNoneBlk}{0}
\newcommand{\ZCodexBndNoneRel}{5/12}
\newcommand{\ZCodexBndRuntimeBlk}{0}
\newcommand{\ZCodexBndRuntimeRel}{5/12}
\newcommand{\ZCodexConsumerCompleteCCAff}{84}
\newcommand{\ZCodexConsumerCompleteCCR}{84}
\newcommand{\ZCodexConsumerCompleteFAAff}{3}
\newcommand{\ZCodexConsumerCompleteFAR}{3}
\newcommand{\ZCodexConsumerPartialCCAff}{84}
\newcommand{\ZCodexConsumerPartialCCR}{86}
\newcommand{\ZCodexConsumerPartialFAAff}{5}
\newcommand{\ZCodexConsumerPartialFAR}{3}
\newcommand{\ZCodexDifferentialCCAff}{80}
\newcommand{\ZCodexDifferentialCCR}{80}
\newcommand{\ZCodexDifferentialFAAff}{3}
\newcommand{\ZCodexDifferentialFAR}{3}
\newcommand{\ZCodexLlmContractCCAff}{80}
\newcommand{\ZCodexLlmContractCCR}{80}
\newcommand{\ZCodexLlmContractFAAff}{3}
\newcommand{\ZCodexLlmContractFAR}{3}
\newcommand{\ZCodexNoneCCAff}{84}
\newcommand{\ZCodexNoneCCR}{86}
\newcommand{\ZCodexNoneFAAff}{5}
\newcommand{\ZCodexNoneFAR}{3}
\newcommand{\ZCodexOutTokPerCorrect}{455}
\newcommand{\ZCodexRevAbst}{6}
\newcommand{\ZCodexRevCtoW}{1}
\newcommand{\ZCodexRevWtoC}{0}
\newcommand{\ZCodexRevisions}{14}
\newcommand{\ZCodexRuntimeCCAff}{84}
\newcommand{\ZCodexRuntimeCCR}{86}
\newcommand{\ZCodexRuntimeFAAff}{5}
\newcommand{\ZCodexRuntimeFAR}{3}
\newcommand{\ZCodexSecPerCorrect}{16.4}
\newcommand{\ZCodexWrongIndist}{3}
\newcommand{\ZCodexWrongStepOne}{5}
\newcommand{\ZDegenerateCount}{5}
\newcommand{\ZIndistCount}{3}
\newcommand{\ZIndistTasks}{\texttt{node-buffer-max-length}, \texttt{py-intflag-invert}, \texttt{py-sum-neumaier}}
\newcommand{\ZQwenAffN}{96}
\newcommand{\ZQwenBndConsumerCompleteBlk}{10}
\newcommand{\ZQwenBndConsumerCompleteRel}{0/12}
\newcommand{\ZQwenBndConsumerPartialBlk}{6}
\newcommand{\ZQwenBndConsumerPartialRel}{4/12}
\newcommand{\ZQwenBndDifferentialBlk}{7}
\newcommand{\ZQwenBndDifferentialRel}{0/12}
\newcommand{\ZQwenBndLlmContractBlk}{9}
\newcommand{\ZQwenBndLlmContractRel}{0/12}
\newcommand{\ZQwenBndNoneBlk}{0}
\newcommand{\ZQwenBndNoneRel}{10/12}
\newcommand{\ZQwenBndRuntimeBlk}{6}
\newcommand{\ZQwenBndRuntimeRel}{4/12}
\newcommand{\ZQwenConsumerCompleteCCAff}{57}
\newcommand{\ZQwenConsumerCompleteCCR}{74}
\newcommand{\ZQwenConsumerCompleteFAAff}{19}
\newcommand{\ZQwenConsumerCompleteFAR}{5}
\newcommand{\ZQwenConsumerPartialCCAff}{56}
\newcommand{\ZQwenConsumerPartialCCR}{76}
\newcommand{\ZQwenConsumerPartialFAAff}{25}
\newcommand{\ZQwenConsumerPartialFAR}{6}
\newcommand{\ZQwenDifferentialCCAff}{53}
\newcommand{\ZQwenDifferentialCCR}{67}
\newcommand{\ZQwenDifferentialFAAff}{19}
\newcommand{\ZQwenDifferentialFAR}{5}
\newcommand{\ZQwenLlmContractCCAff}{52}
\newcommand{\ZQwenLlmContractCCR}{66}
\newcommand{\ZQwenLlmContractFAAff}{19}
\newcommand{\ZQwenLlmContractFAR}{5}
\newcommand{\ZQwenNoneCCAff}{54}
\newcommand{\ZQwenNoneCCR}{73}
\newcommand{\ZQwenNoneFAAff}{36}
\newcommand{\ZQwenNoneFAR}{17}
\newcommand{\ZQwenOutTokPerCorrect}{46}
\newcommand{\ZQwenRevAbst}{4}
\newcommand{\ZQwenRevCtoW}{3}
\newcommand{\ZQwenRevWtoC}{11}
\newcommand{\ZQwenRevisions}{95}
\newcommand{\ZQwenRuntimeCCAff}{55}
\newcommand{\ZQwenRuntimeCCR}{74}
\newcommand{\ZQwenRuntimeFAAff}{30}
\newcommand{\ZQwenRuntimeFAR}{11}
\newcommand{\ZQwenSecPerCorrect}{2.0}
\newcommand{\ZQwenWrongIndist}{19}
\newcommand{\ZQwenWrongStepOne}{36}
\newcommand{\ZRConsumerCompleteFA}{0.10}
\newcommand{\ZRConsumerCompleteFR}{0.00}
\newcommand{\ZRConsumerPartialFA}{0.23}
\newcommand{\ZRConsumerPartialFR}{0.00}
\newcommand{\ZRDifferentialFA}{0.05}
\newcommand{\ZRDifferentialFR}{0.26}
\newcommand{\ZRNoneFA}{1.00}
\newcommand{\ZRNoneFR}{0.00}
\newcommand{\ZRRuntimeFA}{0.41}
\newcommand{\ZRRuntimeFR}{0.00}
\newcommand{\ZRTasks}{18}
\newcommand{\ZWConsumerComplete}{0.38}
\newcommand{\ZWConsumerPartial}{0.54}
\newcommand{\ZWDifferential}{0.38}
\newcommand{\ZWLlmContract}{0.38}
\newcommand{\ZWNone}{1.00}
\newcommand{\ZWRuntime}{0.77}

\newcommand{\VAffClusters}{42}
\newcommand{\VAffected}{48}
\newcommand{\VAllBoundary}{7}
\newcommand{\VAllControl}{7}
\newcommand{\VAllExplicit}{24}
\newcommand{\VAllSilent}{17}
\newcommand{\VCFConsumerFA}{0.054}
\newcommand{\VCFDiffFR}{0.048}
\newcommand{\VCPython}{31}
\newcommand{\VCodexBndBoundary}{0/16}
\newcommand{\VCodexBndLlm}{0/16}
\newcommand{\VCodexBndNoneBoundary}{2/16}
\newcommand{\VCodexBndNoneLlm}{2/16}
\newcommand{\VCodexBndNoneTypical}{7/16}
\newcommand{\VCodexBndTypical}{4/16}
\newcommand{\VCodexBoundaryConsumerCompleteAbst}{16}
\newcommand{\VCodexBoundaryConsumerCompleteBlk}{0}
\newcommand{\VCodexBoundaryConsumerCompleteCC}{148}
\newcommand{\VCodexBoundaryConsumerCompleteFA}{4}
\newcommand{\VCodexBoundaryConsumerCompleteFR}{0}
\newcommand{\VCodexBoundaryDifferentialAbst}{17}
\newcommand{\VCodexBoundaryDifferentialBlk}{3}
\newcommand{\VCodexBoundaryDifferentialCC}{144}
\newcommand{\VCodexBoundaryDifferentialFA}{4}
\newcommand{\VCodexBoundaryDifferentialFR}{3}
\newcommand{\VCodexBoundaryNoneAbst}{14}
\newcommand{\VCodexBoundaryNoneBlk}{0}
\newcommand{\VCodexBoundaryNoneCC}{145}
\newcommand{\VCodexBoundaryNoneFA}{8}
\newcommand{\VCodexBoundaryNoneFR}{0}
\newcommand{\VCodexBoundaryRuntimeAbst}{14}
\newcommand{\VCodexBoundaryRuntimeBlk}{0}
\newcommand{\VCodexBoundaryRuntimeCC}{146}
\newcommand{\VCodexBoundaryRuntimeFA}{8}
\newcommand{\VCodexBoundaryRuntimeFR}{0}
\newcommand{\VCodexCalls}{782}
\newcommand{\VCodexHOneA}{0.029}
\newcommand{\VCodexHOneB}{0.066}
\newcommand{\VCodexHOneDecision}{not positive}
\newcommand{\VCodexHOneDiff}{$-$0.037}
\newcommand{\VCodexHOneDisc}{2}
\newcommand{\VCodexHOneHigh}{0.000}
\newcommand{\VCodexHOneHolm}{1.000}
\newcommand{\VCodexHOneLow}{$-$0.103}
\newcommand{\VCodexHThreeA}{0.853}
\newcommand{\VCodexHThreeB}{0.838}
\newcommand{\VCodexHThreeDiff}{0.015}
\newcommand{\VCodexHThreeDisc}{2}
\newcommand{\VCodexHThreeHigh}{0.037}
\newcommand{\VCodexHThreeHolm}{1.000}
\newcommand{\VCodexHThreeLow}{0.000}
\newcommand{\VCodexHTwoA}{0.007}
\newcommand{\VCodexHTwoB}{0.066}
\newcommand{\VCodexHTwoDecision}{not positive}
\newcommand{\VCodexHTwoDiff}{$-$0.059}
\newcommand{\VCodexHTwoDisc}{4}
\newcommand{\VCodexHTwoHigh}{0.007}
\newcommand{\VCodexHTwoHolm}{1.000}
\newcommand{\VCodexHTwoLow}{$-$0.147}
\newcommand{\VCodexLlmConsumerCompleteAbst}{23}
\newcommand{\VCodexLlmConsumerCompleteBlk}{2}
\newcommand{\VCodexLlmConsumerCompleteCC}{142}
\newcommand{\VCodexLlmConsumerCompleteFA}{1}
\newcommand{\VCodexLlmConsumerCompleteFR}{2}
\newcommand{\VCodexLlmDifferentialAbst}{23}
\newcommand{\VCodexLlmDifferentialBlk}{7}
\newcommand{\VCodexLlmDifferentialCC}{138}
\newcommand{\VCodexLlmDifferentialFA}{0}
\newcommand{\VCodexLlmDifferentialFR}{7}
\newcommand{\VCodexLlmNoneAbst}{16}
\newcommand{\VCodexLlmNoneBlk}{0}
\newcommand{\VCodexLlmNoneCC}{145}
\newcommand{\VCodexLlmNoneFA}{7}
\newcommand{\VCodexLlmNoneFR}{0}
\newcommand{\VCodexLlmRuntimeAbst}{16}
\newcommand{\VCodexLlmRuntimeBlk}{0}
\newcommand{\VCodexLlmRuntimeCC}{146}
\newcommand{\VCodexLlmRuntimeFA}{6}
\newcommand{\VCodexLlmRuntimeFR}{0}
\newcommand{\VCodexMeanSeconds}{5.4}
\newcommand{\VCodexN}{168}
\newcommand{\VCodexProbeCPy}{25/31}
\newcommand{\VCodexProbeNode}{9/11}
\newcommand{\VCodexProbeSQFiveFour}{3/3}
\newcommand{\VCodexProbeSQFiveThree}{3/3}
\newcommand{\VCodexTypicalConsumerCompleteAbst}{13}
\newcommand{\VCodexTypicalConsumerCompleteBlk}{0}
\newcommand{\VCodexTypicalConsumerCompleteCC}{146}
\newcommand{\VCodexTypicalConsumerCompleteFA}{9}
\newcommand{\VCodexTypicalConsumerCompleteFR}{0}
\newcommand{\VCodexTypicalDifferentialAbst}{16}
\newcommand{\VCodexTypicalDifferentialBlk}{1}
\newcommand{\VCodexTypicalDifferentialCC}{142}
\newcommand{\VCodexTypicalDifferentialFA}{9}
\newcommand{\VCodexTypicalDifferentialFR}{1}
\newcommand{\VCodexTypicalNoneAbst}{10}
\newcommand{\VCodexTypicalNoneBlk}{0}
\newcommand{\VCodexTypicalNoneCC}{145}
\newcommand{\VCodexTypicalNoneFA}{12}
\newcommand{\VCodexTypicalNoneFR}{0}
\newcommand{\VCodexTypicalRuntimeAbst}{10}
\newcommand{\VCodexTypicalRuntimeBlk}{0}
\newcommand{\VCodexTypicalRuntimeCC}{146}
\newcommand{\VCodexTypicalRuntimeFA}{12}
\newcommand{\VCodexTypicalRuntimeFR}{0}
\newcommand{\VCodexVoid}{0}
\newcommand{\VCodexWithDocCPy}{26/31}
\newcommand{\VCodexWithDocNode}{9/11}
\newcommand{\VCodexWithDocSQFiveFour}{3/3}
\newcommand{\VCodexWithDocSQFiveThree}{2/3}
\newcommand{\VControls}{7}
\newcommand{\VGptOssAdmitted}{admitted}
\newcommand{\VGptOssBndBoundary}{0/16}
\newcommand{\VGptOssBndLlm}{1/16}
\newcommand{\VGptOssBndNoneBoundary}{9/16}
\newcommand{\VGptOssBndNoneLlm}{11/16}
\newcommand{\VGptOssBndNoneTypical}{9/16}
\newcommand{\VGptOssBndTypical}{5/16}
\newcommand{\VGptOssBoundaryConsumerCompleteAbst}{23}
\newcommand{\VGptOssBoundaryConsumerCompleteBlk}{5}
\newcommand{\VGptOssBoundaryConsumerCompleteCC}{131}
\newcommand{\VGptOssBoundaryConsumerCompleteFA}{4}
\newcommand{\VGptOssBoundaryConsumerCompleteFR}{0}
\newcommand{\VGptOssBoundaryDifferentialAbst}{22}
\newcommand{\VGptOssBoundaryDifferentialBlk}{10}
\newcommand{\VGptOssBoundaryDifferentialCC}{127}
\newcommand{\VGptOssBoundaryDifferentialFA}{4}
\newcommand{\VGptOssBoundaryDifferentialFR}{3}
\newcommand{\VGptOssBoundaryNoneAbst}{14}
\newcommand{\VGptOssBoundaryNoneBlk}{0}
\newcommand{\VGptOssBoundaryNoneCC}{124}
\newcommand{\VGptOssBoundaryNoneFA}{24}
\newcommand{\VGptOssBoundaryNoneFR}{0}
\newcommand{\VGptOssBoundaryRuntimeAbst}{18}
\newcommand{\VGptOssBoundaryRuntimeBlk}{0}
\newcommand{\VGptOssBoundaryRuntimeCC}{125}
\newcommand{\VGptOssBoundaryRuntimeFA}{20}
\newcommand{\VGptOssBoundaryRuntimeFR}{0}
\newcommand{\VGptOssCalls}{921}
\newcommand{\VGptOssHOneA}{0.029}
\newcommand{\VGptOssHOneB}{0.169}
\newcommand{\VGptOssHOneDecision}{positive}
\newcommand{\VGptOssHOneDiff}{$-$0.140}
\newcommand{\VGptOssHOneDisc}{8}
\newcommand{\VGptOssHOneHigh}{$-$0.059}
\newcommand{\VGptOssHOneHolm}{0.023}
\newcommand{\VGptOssHOneLow}{$-$0.235}
\newcommand{\VGptOssHThreeA}{0.768}
\newcommand{\VGptOssHThreeB}{0.717}
\newcommand{\VGptOssHThreeDiff}{0.051}
\newcommand{\VGptOssHThreeDisc}{6}
\newcommand{\VGptOssHThreeHigh}{0.110}
\newcommand{\VGptOssHThreeHolm}{0.125}
\newcommand{\VGptOssHThreeLow}{0.000}
\newcommand{\VGptOssHTwoA}{0.037}
\newcommand{\VGptOssHTwoB}{0.169}
\newcommand{\VGptOssHTwoDecision}{positive}
\newcommand{\VGptOssHTwoDiff}{$-$0.132}
\newcommand{\VGptOssHTwoDisc}{8}
\newcommand{\VGptOssHTwoHigh}{$-$0.051}
\newcommand{\VGptOssHTwoHolm}{0.023}
\newcommand{\VGptOssHTwoLow}{$-$0.228}
\newcommand{\VGptOssLlmConsumerCompleteAbst}{22}
\newcommand{\VGptOssLlmConsumerCompleteBlk}{12}
\newcommand{\VGptOssLlmConsumerCompleteCC}{126}
\newcommand{\VGptOssLlmConsumerCompleteFA}{5}
\newcommand{\VGptOssLlmConsumerCompleteFR}{2}
\newcommand{\VGptOssLlmDifferentialAbst}{20}
\newcommand{\VGptOssLlmDifferentialBlk}{18}
\newcommand{\VGptOssLlmDifferentialCC}{123}
\newcommand{\VGptOssLlmDifferentialFA}{4}
\newcommand{\VGptOssLlmDifferentialFR}{6}
\newcommand{\VGptOssLlmNoneAbst}{7}
\newcommand{\VGptOssLlmNoneBlk}{0}
\newcommand{\VGptOssLlmNoneCC}{129}
\newcommand{\VGptOssLlmNoneFA}{28}
\newcommand{\VGptOssLlmNoneFR}{0}
\newcommand{\VGptOssLlmRuntimeAbst}{10}
\newcommand{\VGptOssLlmRuntimeBlk}{0}
\newcommand{\VGptOssLlmRuntimeCC}{130}
\newcommand{\VGptOssLlmRuntimeFA}{25}
\newcommand{\VGptOssLlmRuntimeFR}{0}
\newcommand{\VGptOssN}{168}
\newcommand{\VGptOssProbeCPy}{17/31}
\newcommand{\VGptOssProbeNode}{5/11}
\newcommand{\VGptOssProbeSQFiveFour}{2/3}
\newcommand{\VGptOssProbeSQFiveThree}{0/3}
\newcommand{\VGptOssQualDecode}{59.2}
\newcommand{\VGptOssQualExact}{12/12}
\newcommand{\VGptOssQualPrefill}{1337}
\newcommand{\VGptOssTypicalConsumerCompleteAbst}{12}
\newcommand{\VGptOssTypicalConsumerCompleteBlk}{4}
\newcommand{\VGptOssTypicalConsumerCompleteCC}{126}
\newcommand{\VGptOssTypicalConsumerCompleteFA}{23}
\newcommand{\VGptOssTypicalConsumerCompleteFR}{0}
\newcommand{\VGptOssTypicalDifferentialAbst}{12}
\newcommand{\VGptOssTypicalDifferentialBlk}{7}
\newcommand{\VGptOssTypicalDifferentialCC}{122}
\newcommand{\VGptOssTypicalDifferentialFA}{24}
\newcommand{\VGptOssTypicalDifferentialFR}{4}
\newcommand{\VGptOssTypicalNoneAbst}{10}
\newcommand{\VGptOssTypicalNoneBlk}{0}
\newcommand{\VGptOssTypicalNoneCC}{123}
\newcommand{\VGptOssTypicalNoneFA}{30}
\newcommand{\VGptOssTypicalNoneFR}{0}
\newcommand{\VGptOssTypicalRuntimeAbst}{10}
\newcommand{\VGptOssTypicalRuntimeBlk}{1}
\newcommand{\VGptOssTypicalRuntimeCC}{124}
\newcommand{\VGptOssTypicalRuntimeFA}{30}
\newcommand{\VGptOssTypicalRuntimeFR}{0}
\newcommand{\VGptOssWithDocCPy}{18/31}
\newcommand{\VGptOssWithDocNode}{8/11}
\newcommand{\VGptOssWithDocSQFiveFour}{3/3}
\newcommand{\VGptOssWithDocSQFiveThree}{1/3}
\newcommand{\VHMOneA}{0.065}
\newcommand{\VHMOneB}{0.377}
\newcommand{\VHMOneDecision}{positive}
\newcommand{\VHMOneDiff}{$-$0.312}
\newcommand{\VHMOneDisc}{25}
\newcommand{\VHMOneFRInc}{0.000}
\newcommand{\VHMOneHigh}{$-$0.220}
\newcommand{\VHMOneHolm}{<0.001}
\newcommand{\VHMOneLow}{$-$0.409}
\newcommand{\VHMTwoA}{0.034}
\newcommand{\VHMTwoB}{0.377}
\newcommand{\VHMTwoDecision}{not positive}
\newcommand{\VHMTwoDiff}{$-$0.343}
\newcommand{\VHMTwoDisc}{25}
\newcommand{\VHMTwoFRInc}{0.061}
\newcommand{\VHMTwoHigh}{$-$0.237}
\newcommand{\VHMTwoHolm}{<0.001}
\newcommand{\VHMTwoLow}{$-$0.454}
\newcommand{\VLLMMean}{2.8}
\newcommand{\VLLMOverlap}{9}
\newcommand{\VLLMTasks}{54}
\newcommand{\VLinCpythonBoundary}{0.07}
\newcommand{\VLinCpythonLlm}{0.06}
\newcommand{\VLinCpythonTypical}{0.37}
\newcommand{\VLinNodeBoundary}{0.00}
\newcommand{\VLinNodeLlm}{0.00}
\newcommand{\VLinNodeTypical}{0.46}
\newcommand{\VLinSqliteBoundary}{0.15}
\newcommand{\VLinSqliteLlm}{0.00}
\newcommand{\VLinSqliteTypical}{0.33}
\newcommand{\VMFConsumerCompleteBoundaryFA}{0.07}
\newcommand{\VMFConsumerCompleteBoundaryFR}{0.015}
\newcommand{\VMFConsumerCompleteLlmFA}{0.04}
\newcommand{\VMFConsumerCompleteLlmFR}{0.076}
\newcommand{\VMFConsumerCompleteTypicalFA}{0.39}
\newcommand{\VMFConsumerCompleteTypicalFR}{0.015}
\newcommand{\VMFDifferentialBoundaryFA}{0.07}
\newcommand{\VMFDifferentialBoundaryFR}{0.045}
\newcommand{\VMFDifferentialLlmFA}{0.04}
\newcommand{\VMFDifferentialLlmFR}{0.106}
\newcommand{\VMFDifferentialTypicalFA}{0.39}
\newcommand{\VMFDifferentialTypicalFR}{0.045}
\newcommand{\VMFRuntimeBoundaryFA}{0.74}
\newcommand{\VMFRuntimeBoundaryFR}{0.000}
\newcommand{\VMFRuntimeLlmFA}{0.74}
\newcommand{\VMFRuntimeLlmFR}{0.000}
\newcommand{\VMFRuntimeTypicalFA}{0.76}
\newcommand{\VMFRuntimeTypicalFR}{0.000}
\newcommand{\VNewTasks}{22}
\newcommand{\VNode}{11}
\newcommand{\VPHClusEdgeFA}{0.065}
\newcommand{\VPHClusEdgeFR}{0.008}
\newcommand{\VPHClusLLMFA}{0.034}
\newcommand{\VPHClusLLMFR}{0.075}
\newcommand{\VPHClusTypicalFA}{0.377}
\newcommand{\VPHClusTypicalFR}{0.008}
\newcommand{\VPHEdgeExact}{0}
\newcommand{\VPHEdgeLeafAll}{7}
\newcommand{\VPHEdgeLeafShare}{0.28}
\newcommand{\VPHEdgeN}{37}
\newcommand{\VPHEdgeRatioHigh}{5}
\newcommand{\VPHEdgeRatioMean}{0.71}
\newcommand{\VPHEdgeRatioMedian}{0.73}
\newcommand{\VPHHMTwoDecision}{not positive}
\newcommand{\VPHHMTwoDiff}{$-$0.343}
\newcommand{\VPHHMTwoFR}{0.076}
\newcommand{\VPHHMTwoFRInc}{0.061}
\newcommand{\VPHHMTwoHolm}{<0.001}
\newcommand{\VPHOverlapTasks}{5}
\newcommand{\VPHOverlapTest}{5}
\newcommand{\VPHRandCPy}{6.5/31}
\newcommand{\VPHRandNode}{3.0/11}
\newcommand{\VPHRandSQFiveFour}{0.8/3}
\newcommand{\VPHRandSQFiveThree}{0.6/3}
\newcommand{\VPHTaskEdgeFA}{0.066}
\newcommand{\VPHTaskEdgeFR}{0.015}
\newcommand{\VPHTaskLLMFA}{0.039}
\newcommand{\VPHTaskLLMFR}{0.076}
\newcommand{\VPHTaskTypicalFA}{0.386}
\newcommand{\VPHTaskTypicalFR}{0.015}
\newcommand{\VPHTruncCodex}{0}
\newcommand{\VPHTruncGptOss}{9}
\newcommand{\VPHTruncQwen}{2}
\newcommand{\VPHVerdictsChanged}{0}
\newcommand{\VPostCutoff}{3}
\newcommand{\VQwenAdmitted}{admitted}
\newcommand{\VQwenBndBoundary}{4/16}
\newcommand{\VQwenBndLlm}{4/16}
\newcommand{\VQwenBndNoneBoundary}{14/16}
\newcommand{\VQwenBndNoneLlm}{13/16}
\newcommand{\VQwenBndNoneTypical}{14/16}
\newcommand{\VQwenBndTypical}{4/16}
\newcommand{\VQwenBoundaryConsumerCompleteAbst}{8}
\newcommand{\VQwenBoundaryConsumerCompleteBlk}{34}
\newcommand{\VQwenBoundaryConsumerCompleteCC}{111}
\newcommand{\VQwenBoundaryConsumerCompleteFA}{12}
\newcommand{\VQwenBoundaryConsumerCompleteFR}{0}
\newcommand{\VQwenBoundaryDifferentialAbst}{11}
\newcommand{\VQwenBoundaryDifferentialBlk}{29}
\newcommand{\VQwenBoundaryDifferentialCC}{113}
\newcommand{\VQwenBoundaryDifferentialFA}{12}
\newcommand{\VQwenBoundaryDifferentialFR}{3}
\newcommand{\VQwenBoundaryNoneAbst}{2}
\newcommand{\VQwenBoundaryNoneBlk}{0}
\newcommand{\VQwenBoundaryNoneCC}{103}
\newcommand{\VQwenBoundaryNoneFA}{60}
\newcommand{\VQwenBoundaryNoneFR}{0}
\newcommand{\VQwenBoundaryRuntimeAbst}{12}
\newcommand{\VQwenBoundaryRuntimeBlk}{6}
\newcommand{\VQwenBoundaryRuntimeCC}{105}
\newcommand{\VQwenBoundaryRuntimeFA}{42}
\newcommand{\VQwenBoundaryRuntimeFR}{0}
\newcommand{\VQwenCalls}{1209}
\newcommand{\VQwenHOneA}{0.085}
\newcommand{\VQwenHOneB}{0.228}
\newcommand{\VQwenHOneDecision}{positive}
\newcommand{\VQwenHOneDiff}{$-$0.143}
\newcommand{\VQwenHOneDisc}{10}
\newcommand{\VQwenHOneHigh}{$-$0.059}
\newcommand{\VQwenHOneHolm}{0.006}
\newcommand{\VQwenHOneLow}{$-$0.246}
\newcommand{\VQwenHThreeA}{0.618}
\newcommand{\VQwenHThreeB}{0.603}
\newcommand{\VQwenHThreeDiff}{0.015}
\newcommand{\VQwenHThreeDisc}{8}
\newcommand{\VQwenHThreeHigh}{0.081}
\newcommand{\VQwenHThreeHolm}{0.875}
\newcommand{\VQwenHThreeLow}{$-$0.066}
\newcommand{\VQwenHTwoA}{0.081}
\newcommand{\VQwenHTwoB}{0.228}
\newcommand{\VQwenHTwoDecision}{not positive}
\newcommand{\VQwenHTwoDiff}{$-$0.147}
\newcommand{\VQwenHTwoDisc}{11}
\newcommand{\VQwenHTwoHigh}{$-$0.055}
\newcommand{\VQwenHTwoHolm}{0.006}
\newcommand{\VQwenHTwoLow}{$-$0.261}
\newcommand{\VQwenLlmConsumerCompleteAbst}{6}
\newcommand{\VQwenLlmConsumerCompleteBlk}{50}
\newcommand{\VQwenLlmConsumerCompleteCC}{96}
\newcommand{\VQwenLlmConsumerCompleteFA}{12}
\newcommand{\VQwenLlmConsumerCompleteFR}{10}
\newcommand{\VQwenLlmDifferentialAbst}{7}
\newcommand{\VQwenLlmDifferentialBlk}{48}
\newcommand{\VQwenLlmDifferentialCC}{97}
\newcommand{\VQwenLlmDifferentialFA}{12}
\newcommand{\VQwenLlmDifferentialFR}{14}
\newcommand{\VQwenLlmNoneAbst}{2}
\newcommand{\VQwenLlmNoneBlk}{0}
\newcommand{\VQwenLlmNoneCC}{98}
\newcommand{\VQwenLlmNoneFA}{63}
\newcommand{\VQwenLlmNoneFR}{0}
\newcommand{\VQwenLlmRuntimeAbst}{7}
\newcommand{\VQwenLlmRuntimeBlk}{6}
\newcommand{\VQwenLlmRuntimeCC}{100}
\newcommand{\VQwenLlmRuntimeFA}{51}
\newcommand{\VQwenLlmRuntimeFR}{0}
\newcommand{\VQwenN}{168}
\newcommand{\VQwenProbeCPy}{15/31}
\newcommand{\VQwenProbeNode}{6/11}
\newcommand{\VQwenProbeSQFiveFour}{1/3}
\newcommand{\VQwenProbeSQFiveThree}{1/3}
\newcommand{\VQwenQualDecode}{46.3}
\newcommand{\VQwenQualExact}{12/12}
\newcommand{\VQwenQualPrefill}{1185}
\newcommand{\VQwenTypicalConsumerCompleteAbst}{2}
\newcommand{\VQwenTypicalConsumerCompleteBlk}{20}
\newcommand{\VQwenTypicalConsumerCompleteCC}{110}
\newcommand{\VQwenTypicalConsumerCompleteFA}{33}
\newcommand{\VQwenTypicalConsumerCompleteFR}{0}
\newcommand{\VQwenTypicalDifferentialAbst}{5}
\newcommand{\VQwenTypicalDifferentialBlk}{21}
\newcommand{\VQwenTypicalDifferentialCC}{106}
\newcommand{\VQwenTypicalDifferentialFA}{33}
\newcommand{\VQwenTypicalDifferentialFR}{1}
\newcommand{\VQwenTypicalNoneAbst}{2}
\newcommand{\VQwenTypicalNoneBlk}{0}
\newcommand{\VQwenTypicalNoneCC}{98}
\newcommand{\VQwenTypicalNoneFA}{64}
\newcommand{\VQwenTypicalNoneFR}{0}
\newcommand{\VQwenTypicalRuntimeAbst}{2}
\newcommand{\VQwenTypicalRuntimeBlk}{12}
\newcommand{\VQwenTypicalRuntimeCC}{99}
\newcommand{\VQwenTypicalRuntimeFA}{52}
\newcommand{\VQwenTypicalRuntimeFR}{0}
\newcommand{\VQwenWithDocCPy}{15/31}
\newcommand{\VQwenWithDocNode}{8/11}
\newcommand{\VQwenWithDocSQFiveFour}{0/3}
\newcommand{\VQwenWithDocSQFiveThree}{0/3}
\newcommand{\VROnlyBoundaryShift}{\texttt{node-createcipher} (4), \texttt{py-binascii-hqx} (5), \texttt{py-distutils-looseversion} (2), \texttt{py-shuffle-random-param} (5)}
\newcommand{\VROnlyBoundaryShiftCount}{4}
\newcommand{\VSQFiveFourBoundary}{0.00}
\newcommand{\VSQFiveFourLlm}{0.00}
\newcommand{\VSQFiveFourTypical}{0.11}
\newcommand{\VSQFiveThreeBoundary}{0.20}
\newcommand{\VSQFiveThreeLlm}{0.00}
\newcommand{\VSQFiveThreeTypical}{0.58}
\newcommand{\VSQLite}{6}
\newcommand{\VStaleBoundary}{37}
\newcommand{\VStaleLLM}{37}
\newcommand{\VStaleOf}{37}
\newcommand{\VStaleTypical}{26}
\newcommand{\VTasks}{55}
\newcommand{\VTestAff}{37}
\newcommand{\VTestClusters}{34}

\newcommand{\FAffClusters}{60}
\newcommand{\FAffected}{66}
\newcommand{\FAllBoundary}{7}
\newcommand{\FAllControl}{7}
\newcommand{\FAllExplicit}{34}
\newcommand{\FAllSilent}{25}
\newcommand{\FAuditAgree}{0.91}
\newcommand{\FAuditCalls}{146}
\newcommand{\FAuditFlagged}{34}
\newcommand{\FAuditFlaggedTest}{24}
\newcommand{\FAuditHMClusters}{25}
\newcommand{\FAuditHMDiff}{$-$0.370}
\newcommand{\FAuditHMP}{<0.001}
\newcommand{\FAuditKappa}{0.73}
\newcommand{\FAuditMFourADiff}{$-$0.270}
\newcommand{\FAuditMFourAP}{<0.001}
\newcommand{\FAuditNoFive}{7}
\newcommand{\FAuditNoFour}{20}
\newcommand{\FAuditNoOne}{0}
\newcommand{\FAuditNoThree}{10}
\newcommand{\FAuditNoTwo}{3}
\newcommand{\FAuditSoft}{20}
\newcommand{\FAuditUnparse}{0}
\newcommand{\FAuditYesFive}{0.82}
\newcommand{\FAuditYesFour}{0.51}
\newcommand{\FAuditYesOne}{0.98}
\newcommand{\FAuditYesThree}{0.81}
\newcommand{\FAuditYesTwo}{0.92}
\newcommand{\FCPython}{45}
\newcommand{\FConfirmatoryCount}{2}
\newcommand{\FControls}{7}
\newcommand{\FDevAff}{14}
\newcommand{\FEdgeRepMet}{met}
\newcommand{\FEdgeRepPositive}{3}
\newcommand{\FGemmaAdmitted}{not admitted}
\newcommand{\FGemmaQualDecode}{20.7}
\newcommand{\FGemmaQualExact}{0/12}
\newcommand{\FGemmaQualPrefill}{572}
\newcommand{\FGptOssBothAbst}{51}
\newcommand{\FGptOssBothCC}{351}
\newcommand{\FGptOssBothCallsPerCell}{1.24}
\newcommand{\FGptOssBothFA}{8}
\newcommand{\FGptOssBothFR}{0}
\newcommand{\FGptOssBothNoGateFA}{89}
\newcommand{\FGptOssCalls}{2965}
\newcommand{\FGptOssCapability}{0.677}
\newcommand{\FGptOssDevFA}{0.114}
\newcommand{\FGptOssEAbst}{46}
\newcommand{\FGptOssECC}{353}
\newcommand{\FGptOssECallsPerCell}{1.25}
\newcommand{\FGptOssEFA}{8}
\newcommand{\FGptOssEFR}{0}
\newcommand{\FGptOssENoGateFA}{99}
\newcommand{\FGptOssHFourA}{0.700}
\newcommand{\FGptOssHFourB}{0.677}
\newcommand{\FGptOssHFourDiff}{0.023}
\newcommand{\FGptOssHFourDisc}{17}
\newcommand{\FGptOssHFourHigh}{0.069}
\newcommand{\FGptOssHFourHolm}{0.385}
\newcommand{\FGptOssHFourLow}{$-$0.018}
\newcommand{\FGptOssHOneA}{0.020}
\newcommand{\FGptOssHOneB}{0.181}
\newcommand{\FGptOssHOneDecision}{positive}
\newcommand{\FGptOssHOneDiff}{$-$0.161}
\newcommand{\FGptOssHOneDisc}{15}
\newcommand{\FGptOssHOneHigh}{$-$0.084}
\newcommand{\FGptOssHOneHolm}{<0.001}
\newcommand{\FGptOssHOneLow}{$-$0.247}
\newcommand{\FGptOssHThreeA}{0.020}
\newcommand{\FGptOssHThreeB}{0.020}
\newcommand{\FGptOssHThreeDecision}{not positive}
\newcommand{\FGptOssHThreeDiff}{0.000}
\newcommand{\FGptOssHThreeDisc}{0}
\newcommand{\FGptOssHThreeHigh}{0.000}
\newcommand{\FGptOssHThreeHolm}{--}
\newcommand{\FGptOssHThreeLow}{0.000}
\newcommand{\FGptOssHTwoA}{0.117}
\newcommand{\FGptOssHTwoB}{0.181}
\newcommand{\FGptOssHTwoDecision}{not positive}
\newcommand{\FGptOssHTwoDiff}{$-$0.064}
\newcommand{\FGptOssHTwoDisc}{12}
\newcommand{\FGptOssHTwoHigh}{$-$0.010}
\newcommand{\FGptOssHTwoHolm}{0.088}
\newcommand{\FGptOssHTwoLow}{$-$0.128}
\newcommand{\FGptOssMFBothFA}{0.046}
\newcommand{\FGptOssMFBothFR}{0.010}
\newcommand{\FGptOssMFRawFA}{0.149}
\newcommand{\FGptOssMFRawFR}{0.031}
\newcommand{\FGptOssMFSFA}{0.235}
\newcommand{\FGptOssMFSFR}{0.010}
\newcommand{\FGptOssN}{456}
\newcommand{\FGptOssProbeCPy}{29/45}
\newcommand{\FGptOssProbeNode}{7/13}
\newcommand{\FGptOssProbeSQFiveFour}{3/4}
\newcommand{\FGptOssProbeSQFiveThree}{2/4}
\newcommand{\FGptOssQualDecode}{61.4}
\newcommand{\FGptOssQualExact}{12/12}
\newcommand{\FGptOssQualPrefill}{1269}
\newcommand{\FGptOssSAbst}{46}
\newcommand{\FGptOssSCC}{327}
\newcommand{\FGptOssSCallsPerCell}{1.20}
\newcommand{\FGptOssSFA}{47}
\newcommand{\FGptOssSFR}{1}
\newcommand{\FGptOssSNoGateFA}{112}
\newcommand{\FGptOssSelfBinding}{93}
\newcommand{\FGptOssSelfKept}{128}
\newcommand{\FGptOssSelfOverlap}{1}
\newcommand{\FGptOssSelfParse}{7}
\newcommand{\FGptOssSelfSim}{0.69}
\newcommand{\FGptOssSelfTasks}{57}
\newcommand{\FGptOssStatus}{confirmatory}
\newcommand{\FGptOssStepGOneFA}{0.339}
\newcommand{\FGptOssStepGOneFR}{0.000}
\newcommand{\FGptOssStepGThreeFA}{0.266}
\newcommand{\FGptOssStepGThreeFR}{0.000}
\newcommand{\FGptOssStepGTwoFA}{0.266}
\newcommand{\FGptOssStepGTwoFR}{0.000}
\newcommand{\FGptOssStepSelfRawFA}{0.339}
\newcommand{\FGptOssStepSelfRawFR}{0.030}
\newcommand{\FGptOssStepSelfSFA}{0.376}
\newcommand{\FGptOssStepSelfSFR}{0.000}
\newcommand{\FGptOssStepWOneFA}{0.156}
\newcommand{\FGptOssStepWOneFR}{0.000}
\newcommand{\FGptOssStepWThreeFA}{0.083}
\newcommand{\FGptOssStepWThreeFR}{0.000}
\newcommand{\FGptOssStepWTwoFA}{0.083}
\newcommand{\FGptOssStepWTwoFR}{0.000}
\newcommand{\FGptOssTAbst}{36}
\newcommand{\FGptOssTCC}{320}
\newcommand{\FGptOssTCallsPerCell}{1.15}
\newcommand{\FGptOssTFA}{71}
\newcommand{\FGptOssTFR}{0}
\newcommand{\FGptOssTNoGateFA}{111}
\newcommand{\FGptOssTrunc}{7}
\newcommand{\FGptOssVoided}{0}
\newcommand{\FGptOssWithDocCPy}{28/45}
\newcommand{\FGptOssWithDocNode}{8/13}
\newcommand{\FGptOssWithDocSQFiveFour}{3/4}
\newcommand{\FGptOssWithDocSQFiveThree}{1/4}
\newcommand{\FHMAllDiff}{$-$0.342}
\newcommand{\FHMAllDisc}{36}
\newcommand{\FLunaBothAbst}{36}
\newcommand{\FLunaBothCC}{409}
\newcommand{\FLunaBothCallsPerCell}{1.09}
\newcommand{\FLunaBothFA}{0}
\newcommand{\FLunaBothFR}{0}
\newcommand{\FLunaBothNoGateFA}{38}
\newcommand{\FLunaCalls}{2626}
\newcommand{\FLunaCapability}{0.852}
\newcommand{\FLunaDevFA}{0.000}
\newcommand{\FLunaEAbst}{36}
\newcommand{\FLunaECC}{408}
\newcommand{\FLunaECallsPerCell}{1.07}
\newcommand{\FLunaEFA}{7}
\newcommand{\FLunaEFR}{0}
\newcommand{\FLunaENoGateFA}{37}
\newcommand{\FLunaHFourA}{0.865}
\newcommand{\FLunaHFourB}{0.852}
\newcommand{\FLunaHFourDiff}{0.013}
\newcommand{\FLunaHFourDisc}{5}
\newcommand{\FLunaHFourHigh}{0.051}
\newcommand{\FLunaHFourHolm}{1.000}
\newcommand{\FLunaHFourLow}{$-$0.013}
\newcommand{\FLunaHOneA}{0.018}
\newcommand{\FLunaHOneB}{0.056}
\newcommand{\FLunaHOneDecision}{not positive}
\newcommand{\FLunaHOneDiff}{$-$0.038}
\newcommand{\FLunaHOneDisc}{5}
\newcommand{\FLunaHOneHigh}{$-$0.003}
\newcommand{\FLunaHOneHolm}{0.499}
\newcommand{\FLunaHOneLow}{$-$0.089}
\newcommand{\FLunaHThreeA}{0.000}
\newcommand{\FLunaHThreeB}{0.018}
\newcommand{\FLunaHThreeDecision}{not positive}
\newcommand{\FLunaHThreeDiff}{$-$0.018}
\newcommand{\FLunaHThreeDisc}{1}
\newcommand{\FLunaHThreeHigh}{0.000}
\newcommand{\FLunaHThreeHolm}{1.000}
\newcommand{\FLunaHThreeLow}{$-$0.054}
\newcommand{\FLunaHTwoA}{0.023}
\newcommand{\FLunaHTwoB}{0.056}
\newcommand{\FLunaHTwoDecision}{not positive}
\newcommand{\FLunaHTwoDiff}{$-$0.033}
\newcommand{\FLunaHTwoDisc}{4}
\newcommand{\FLunaHTwoHigh}{0.003}
\newcommand{\FLunaHTwoHolm}{0.749}
\newcommand{\FLunaHTwoLow}{$-$0.082}
\newcommand{\FLunaMFBothFA}{0.029}
\newcommand{\FLunaMFBothFR}{0.010}
\newcommand{\FLunaMFRawFA}{0.054}
\newcommand{\FLunaMFRawFR}{0.031}
\newcommand{\FLunaMFSFA}{0.153}
\newcommand{\FLunaMFSFR}{0.010}
\newcommand{\FLunaN}{456}
\newcommand{\FLunaProbeCPy}{37/45}
\newcommand{\FLunaProbeNode}{10/13}
\newcommand{\FLunaProbeSQFiveFour}{3/4}
\newcommand{\FLunaProbeSQFiveThree}{3/4}
\newcommand{\FLunaSAbst}{36}
\newcommand{\FLunaSCC}{403}
\newcommand{\FLunaSCallsPerCell}{1.07}
\newcommand{\FLunaSFA}{9}
\newcommand{\FLunaSFR}{0}
\newcommand{\FLunaSNoGateFA}{38}
\newcommand{\FLunaSelfBinding}{153}
\newcommand{\FLunaSelfKept}{187}
\newcommand{\FLunaSelfOverlap}{5}
\newcommand{\FLunaSelfParse}{0}
\newcommand{\FLunaSelfSim}{0.69}
\newcommand{\FLunaSelfTasks}{71}
\newcommand{\FLunaStatus}{descriptive}
\newcommand{\FLunaStepGOneFA}{0.311}
\newcommand{\FLunaStepGOneFR}{0.000}
\newcommand{\FLunaStepGThreeFA}{0.178}
\newcommand{\FLunaStepGThreeFR}{0.000}
\newcommand{\FLunaStepGTwoFA}{0.178}
\newcommand{\FLunaStepGTwoFR}{0.000}
\newcommand{\FLunaStepSelfRawFA}{0.044}
\newcommand{\FLunaStepSelfRawFR}{0.023}
\newcommand{\FLunaStepSelfSFA}{0.178}
\newcommand{\FLunaStepSelfSFR}{0.000}
\newcommand{\FLunaStepWOneFA}{0.178}
\newcommand{\FLunaStepWOneFR}{0.000}
\newcommand{\FLunaStepWThreeFA}{0.044}
\newcommand{\FLunaStepWThreeFR}{0.000}
\newcommand{\FLunaStepWTwoFA}{0.044}
\newcommand{\FLunaStepWTwoFR}{0.000}
\newcommand{\FLunaTAbst}{30}
\newcommand{\FLunaTCC}{398}
\newcommand{\FLunaTCallsPerCell}{1.05}
\newcommand{\FLunaTFA}{22}
\newcommand{\FLunaTFR}{0}
\newcommand{\FLunaTNoGateFA}{45}
\newcommand{\FLunaTrunc}{0}
\newcommand{\FLunaVoided}{0}
\newcommand{\FLunaWithDocCPy}{37/45}
\newcommand{\FLunaWithDocNode}{10/13}
\newcommand{\FLunaWithDocSQFiveFour}{3/4}
\newcommand{\FLunaWithDocSQFiveThree}{2/4}
\newcommand{\FMFEdgeAdded}{1.00}
\newcommand{\FMFEdgeCFA}{0.064}
\newcommand{\FMFEdgeFA}{0.065}
\newcommand{\FMFEdgeFR}{0.010}
\newcommand{\FMFEdgeStale}{52}
\newcommand{\FMFGOneAdded}{0.73}
\newcommand{\FMFGOneCFA}{0.208}
\newcommand{\FMFGOneFA}{0.206}
\newcommand{\FMFGOneFR}{0.010}
\newcommand{\FMFGOneStale}{42}
\newcommand{\FMFGThreeAdded}{1.85}
\newcommand{\FMFGThreeCFA}{0.161}
\newcommand{\FMFGThreeFA}{0.157}
\newcommand{\FMFGThreeFR}{0.010}
\newcommand{\FMFGThreeStale}{42}
\newcommand{\FMFGTwoAdded}{1.40}
\newcommand{\FMFGTwoCFA}{0.166}
\newcommand{\FMFGTwoFA}{0.161}
\newcommand{\FMFGTwoFR}{0.010}
\newcommand{\FMFGTwoStale}{42}
\newcommand{\FMFTypAdded}{0.00}
\newcommand{\FMFTypCFA}{0.406}
\newcommand{\FMFTypFA}{0.410}
\newcommand{\FMFTypFR}{0.010}
\newcommand{\FMFTypStale}{34}
\newcommand{\FMFWOneAdded}{0.96}
\newcommand{\FMFWOneCFA}{0.127}
\newcommand{\FMFWOneFA}{0.130}
\newcommand{\FMFWOneFR}{0.010}
\newcommand{\FMFWOneStale}{49}
\newcommand{\FMFWThreeAdded}{2.46}
\newcommand{\FMFWThreeCFA}{0.070}
\newcommand{\FMFWThreeFA}{0.070}
\newcommand{\FMFWThreeFR}{0.010}
\newcommand{\FMFWThreeStale}{50}
\newcommand{\FMFWTwoAdded}{1.85}
\newcommand{\FMFWTwoCFA}{0.075}
\newcommand{\FMFWTwoFA}{0.075}
\newcommand{\FMFWTwoFR}{0.010}
\newcommand{\FMFWTwoStale}{50}
\newcommand{\FMFourAA}{0.161}
\newcommand{\FMFourAB}{0.406}
\newcommand{\FMFourAClusters}{49}
\newcommand{\FMFourADecision}{not positive}
\newcommand{\FMFourADiff}{$-$0.245}
\newcommand{\FMFourADisc}{23}
\newcommand{\FMFourAHigh}{$-$0.158}
\newcommand{\FMFourAHolm}{<0.001}
\newcommand{\FMFourALow}{$-$0.339}
\newcommand{\FMFourBA}{0.006}
\newcommand{\FMFourBB}{0.006}
\newcommand{\FMFourBClusters}{45}
\newcommand{\FMFourBDecision}{not positive}
\newcommand{\FMFourBDiff}{0.000}
\newcommand{\FMFourBDisc}{0}
\newcommand{\FMFourBHigh}{0.000}
\newcommand{\FMFourBHolm}{--}
\newcommand{\FMFourBLow}{0.000}
\newcommand{\FMFourCA}{0.061}
\newcommand{\FMFourCB}{0.471}
\newcommand{\FMFourCClusters}{15}
\newcommand{\FMFourCDecision}{positive}
\newcommand{\FMFourCDiff}{$-$0.410}
\newcommand{\FMFourCDisc}{11}
\newcommand{\FMFourCHigh}{$-$0.238}
\newcommand{\FMFourCHolm}{<0.001}
\newcommand{\FMFourCLow}{$-$0.589}
\newcommand{\FNewClusters}{18}
\newcommand{\FNewFAEdge}{0.061}
\newcommand{\FNewFATyp}{0.471}
\newcommand{\FNewFREdge}{0.000}
\newcommand{\FNewFRTyp}{0.000}
\newcommand{\FNewTasks}{18}
\newcommand{\FNewTestClusters}{15}
\newcommand{\FNode}{13}
\newcommand{\FPostCutoff}{4}
\newcommand{\FPostNoPChanged}{0}
\newcommand{\FPostNoPDChanged}{0}
\newcommand{\FPowerGptOss}{0.94}
\newcommand{\FPowerLuna}{0.06}
\newcommand{\FPowerQwenEight}{0.99}
\newcommand{\FQwenEightBothAbst}{13}
\newcommand{\FQwenEightBothCC}{268}
\newcommand{\FQwenEightBothCallsPerCell}{1.42}
\newcommand{\FQwenEightBothFA}{20}
\newcommand{\FQwenEightBothFR}{0}
\newcommand{\FQwenEightBothNoGateFA}{181}
\newcommand{\FQwenEightCalls}{3266}
\newcommand{\FQwenEightCapability}{0.546}
\newcommand{\FQwenEightDevFA}{0.170}
\newcommand{\FQwenEightEAbst}{11}
\newcommand{\FQwenEightECC}{270}
\newcommand{\FQwenEightECallsPerCell}{1.41}
\newcommand{\FQwenEightEFA}{24}
\newcommand{\FQwenEightEFR}{0}
\newcommand{\FQwenEightENoGateFA}{181}
\newcommand{\FQwenEightHFourA}{0.556}
\newcommand{\FQwenEightHFourB}{0.546}
\newcommand{\FQwenEightHFourDiff}{0.010}
\newcommand{\FQwenEightHFourDisc}{4}
\newcommand{\FQwenEightHFourHigh}{0.031}
\newcommand{\FQwenEightHFourHolm}{0.998}
\newcommand{\FQwenEightHFourLow}{$-$0.005}
\newcommand{\FQwenEightHOneA}{0.057}
\newcommand{\FQwenEightHOneB}{0.233}
\newcommand{\FQwenEightHOneDecision}{positive}
\newcommand{\FQwenEightHOneDiff}{$-$0.176}
\newcommand{\FQwenEightHOneDisc}{17}
\newcommand{\FQwenEightHOneHigh}{$-$0.094}
\newcommand{\FQwenEightHOneHolm}{<0.001}
\newcommand{\FQwenEightHOneLow}{$-$0.267}
\newcommand{\FQwenEightHThreeA}{0.047}
\newcommand{\FQwenEightHThreeB}{0.057}
\newcommand{\FQwenEightHThreeDecision}{not positive}
\newcommand{\FQwenEightHThreeDiff}{$-$0.010}
\newcommand{\FQwenEightHThreeDisc}{1}
\newcommand{\FQwenEightHThreeHigh}{0.000}
\newcommand{\FQwenEightHThreeHolm}{1.000}
\newcommand{\FQwenEightHThreeLow}{$-$0.031}
\newcommand{\FQwenEightHTwoA}{0.188}
\newcommand{\FQwenEightHTwoB}{0.233}
\newcommand{\FQwenEightHTwoDecision}{not positive}
\newcommand{\FQwenEightHTwoDiff}{$-$0.046}
\newcommand{\FQwenEightHTwoDisc}{3}
\newcommand{\FQwenEightHTwoHigh}{0.000}
\newcommand{\FQwenEightHTwoHolm}{0.749}
\newcommand{\FQwenEightHTwoLow}{$-$0.102}
\newcommand{\FQwenEightMFBothFA}{0.055}
\newcommand{\FQwenEightMFBothFR}{0.010}
\newcommand{\FQwenEightMFRawFA}{0.240}
\newcommand{\FQwenEightMFRawFR}{0.031}
\newcommand{\FQwenEightMFSFA}{0.355}
\newcommand{\FQwenEightMFSFR}{0.010}
\newcommand{\FQwenEightN}{456}
\newcommand{\FQwenEightProbeCPy}{19/45}
\newcommand{\FQwenEightProbeNode}{6/13}
\newcommand{\FQwenEightProbeSQFiveFour}{1/4}
\newcommand{\FQwenEightProbeSQFiveThree}{1/4}
\newcommand{\FQwenEightQualDecode}{38.9}
\newcommand{\FQwenEightQualExact}{12/12}
\newcommand{\FQwenEightQualPrefill}{1119}
\newcommand{\FQwenEightSAbst}{5}
\newcommand{\FQwenEightSCC}{271}
\newcommand{\FQwenEightSCallsPerCell}{1.28}
\newcommand{\FQwenEightSFA}{80}
\newcommand{\FQwenEightSFR}{0}
\newcommand{\FQwenEightSNoGateFA}{183}
\newcommand{\FQwenEightSelfBinding}{27}
\newcommand{\FQwenEightSelfKept}{49}
\newcommand{\FQwenEightSelfOverlap}{2}
\newcommand{\FQwenEightSelfParse}{9}
\newcommand{\FQwenEightSelfSim}{0.71}
\newcommand{\FQwenEightSelfTasks}{30}
\newcommand{\FQwenEightStatus}{confirmatory}
\newcommand{\FQwenEightStepGOneFA}{0.341}
\newcommand{\FQwenEightStepGOneFR}{0.000}
\newcommand{\FQwenEightStepGThreeFA}{0.261}
\newcommand{\FQwenEightStepGThreeFR}{0.000}
\newcommand{\FQwenEightStepGTwoFA}{0.261}
\newcommand{\FQwenEightStepGTwoFR}{0.000}
\newcommand{\FQwenEightStepSelfRawFA}{0.330}
\newcommand{\FQwenEightStepSelfRawFR}{0.038}
\newcommand{\FQwenEightStepSelfSFA}{0.455}
\newcommand{\FQwenEightStepSelfSFR}{0.000}
\newcommand{\FQwenEightStepWOneFA}{0.261}
\newcommand{\FQwenEightStepWOneFR}{0.000}
\newcommand{\FQwenEightStepWThreeFA}{0.170}
\newcommand{\FQwenEightStepWThreeFR}{0.000}
\newcommand{\FQwenEightStepWTwoFA}{0.170}
\newcommand{\FQwenEightStepWTwoFR}{0.000}
\newcommand{\FQwenEightTAbst}{3}
\newcommand{\FQwenEightTCC}{268}
\newcommand{\FQwenEightTCallsPerCell}{1.24}
\newcommand{\FQwenEightTFA}{98}
\newcommand{\FQwenEightTFR}{0}
\newcommand{\FQwenEightTNoGateFA}{182}
\newcommand{\FQwenEightTrunc}{6}
\newcommand{\FQwenEightVoided}{0}
\newcommand{\FQwenEightWithDocCPy}{24/45}
\newcommand{\FQwenEightWithDocNode}{9/13}
\newcommand{\FQwenEightWithDocSQFiveFour}{0/4}
\newcommand{\FQwenEightWithDocSQFiveThree}{0/4}
\newcommand{\FQwenMoeBothAbst}{44}
\newcommand{\FQwenMoeBothCC}{379}
\newcommand{\FQwenMoeBothCallsPerCell}{1.20}
\newcommand{\FQwenMoeBothFA}{8}
\newcommand{\FQwenMoeBothFR}{0}
\newcommand{\FQwenMoeBothNoGateFA}{82}
\newcommand{\FQwenMoeCalls}{2885}
\newcommand{\FQwenMoeCapability}{0.753}
\newcommand{\FQwenMoeDevFA}{0.148}
\newcommand{\FQwenMoeEAbst}{51}
\newcommand{\FQwenMoeECC}{368}
\newcommand{\FQwenMoeECallsPerCell}{1.21}
\newcommand{\FQwenMoeEFA}{7}
\newcommand{\FQwenMoeEFR}{0}
\newcommand{\FQwenMoeENoGateFA}{89}
\newcommand{\FQwenMoeHFourA}{0.776}
\newcommand{\FQwenMoeHFourB}{0.753}
\newcommand{\FQwenMoeHFourDiff}{0.023}
\newcommand{\FQwenMoeHFourDisc}{9}
\newcommand{\FQwenMoeHFourHigh}{0.048}
\newcommand{\FQwenMoeHFourHolm}{0.256}
\newcommand{\FQwenMoeHFourLow}{0.001}
\newcommand{\FQwenMoeHOneA}{0.018}
\newcommand{\FQwenMoeHOneB}{0.120}
\newcommand{\FQwenMoeHOneDecision}{positive}
\newcommand{\FQwenMoeHOneDiff}{$-$0.102}
\newcommand{\FQwenMoeHOneDisc}{11}
\newcommand{\FQwenMoeHOneHigh}{$-$0.041}
\newcommand{\FQwenMoeHOneHolm}{0.004}
\newcommand{\FQwenMoeHOneLow}{$-$0.176}
\newcommand{\FQwenMoeHThreeA}{0.020}
\newcommand{\FQwenMoeHThreeB}{0.018}
\newcommand{\FQwenMoeHThreeDecision}{not positive}
\newcommand{\FQwenMoeHThreeDiff}{0.003}
\newcommand{\FQwenMoeHThreeDisc}{1}
\newcommand{\FQwenMoeHThreeHigh}{0.008}
\newcommand{\FQwenMoeHThreeHolm}{1.000}
\newcommand{\FQwenMoeHThreeLow}{0.000}
\newcommand{\FQwenMoeHTwoA}{0.071}
\newcommand{\FQwenMoeHTwoB}{0.120}
\newcommand{\FQwenMoeHTwoDecision}{not positive}
\newcommand{\FQwenMoeHTwoDiff}{$-$0.048}
\newcommand{\FQwenMoeHTwoDisc}{6}
\newcommand{\FQwenMoeHTwoHigh}{$-$0.003}
\newcommand{\FQwenMoeHTwoHolm}{0.256}
\newcommand{\FQwenMoeHTwoLow}{$-$0.110}
\newcommand{\FQwenMoeMFBothFA}{0.043}
\newcommand{\FQwenMoeMFBothFR}{0.010}
\newcommand{\FQwenMoeMFRawFA}{0.100}
\newcommand{\FQwenMoeMFRawFR}{0.031}
\newcommand{\FQwenMoeMFSFA}{0.215}
\newcommand{\FQwenMoeMFSFR}{0.010}
\newcommand{\FQwenMoeN}{456}
\newcommand{\FQwenMoeProbeCPy}{32/45}
\newcommand{\FQwenMoeProbeNode}{10/13}
\newcommand{\FQwenMoeProbeSQFiveFour}{2/4}
\newcommand{\FQwenMoeProbeSQFiveThree}{2/4}
\newcommand{\FQwenMoeQualDecode}{104.0}
\newcommand{\FQwenMoeQualExact}{12/12}
\newcommand{\FQwenMoeQualPrefill}{1240}
\newcommand{\FQwenMoeSAbst}{35}
\newcommand{\FQwenMoeSCC}{363}
\newcommand{\FQwenMoeSCallsPerCell}{1.16}
\newcommand{\FQwenMoeSFA}{30}
\newcommand{\FQwenMoeSFR}{0}
\newcommand{\FQwenMoeSNoGateFA}{89}
\newcommand{\FQwenMoeSelfBinding}{122}
\newcommand{\FQwenMoeSelfKept}{168}
\newcommand{\FQwenMoeSelfOverlap}{3}
\newcommand{\FQwenMoeSelfParse}{3}
\newcommand{\FQwenMoeSelfSim}{0.68}
\newcommand{\FQwenMoeSelfTasks}{67}
\newcommand{\FQwenMoeStatus}{descriptive}
\newcommand{\FQwenMoeStepGOneFA}{0.380}
\newcommand{\FQwenMoeStepGOneFR}{0.000}
\newcommand{\FQwenMoeStepGThreeFA}{0.337}
\newcommand{\FQwenMoeStepGThreeFR}{0.000}
\newcommand{\FQwenMoeStepGTwoFA}{0.337}
\newcommand{\FQwenMoeStepGTwoFR}{0.000}
\newcommand{\FQwenMoeStepSelfRawFA}{0.326}
\newcommand{\FQwenMoeStepSelfRawFR}{0.026}
\newcommand{\FQwenMoeStepSelfSFA}{0.337}
\newcommand{\FQwenMoeStepSelfSFR}{0.000}
\newcommand{\FQwenMoeStepWOneFA}{0.098}
\newcommand{\FQwenMoeStepWOneFR}{0.000}
\newcommand{\FQwenMoeStepWThreeFA}{0.054}
\newcommand{\FQwenMoeStepWThreeFR}{0.000}
\newcommand{\FQwenMoeStepWTwoFA}{0.054}
\newcommand{\FQwenMoeStepWTwoFR}{0.000}
\newcommand{\FQwenMoeTAbst}{33}
\newcommand{\FQwenMoeTCC}{354}
\newcommand{\FQwenMoeTCallsPerCell}{1.13}
\newcommand{\FQwenMoeTFA}{49}
\newcommand{\FQwenMoeTFR}{0}
\newcommand{\FQwenMoeTNoGateFA}{97}
\newcommand{\FQwenMoeTrunc}{1}
\newcommand{\FQwenMoeVoided}{0}
\newcommand{\FQwenMoeWithDocCPy}{34/45}
\newcommand{\FQwenMoeWithDocNode}{6/13}
\newcommand{\FQwenMoeWithDocSQFiveFour}{2/4}
\newcommand{\FQwenMoeWithDocSQFiveThree}{3/4}
\newcommand{\FRandCPy}{9.9/45}
\newcommand{\FRandNode}{3.8/13}
\newcommand{\FRandSQFiveFour}{1.0/4}
\newcommand{\FRandSQFiveThree}{0.8/4}
\newcommand{\FSQLite}{8}
\newcommand{\FScalePAbs}{0.083}
\newcommand{\FScalePRel}{0.750}
\newcommand{\FScaleRhoAbs}{$-$1.00}
\newcommand{\FScaleRhoRel}{$-$0.40}
\newcommand{\FSelfRepMet}{not met}
\newcommand{\FSelfRepPositive}{0}
\newcommand{\FTasks}{73}
\newcommand{\FTestAff}{52}
\newcommand{\FTestClusters}{49}
\newcommand{\FWriterDropDup}{12}
\newcommand{\FWriterDropNondet}{2}
\newcommand{\FWriterDropOld}{13}
\newcommand{\FWriterDropped}{27}
\newcommand{\FWriterKept}{176}
\newcommand{\FWriterNonBinding}{35}
\newcommand{\FWriterOverlap}{3}
\newcommand{\FWriterSeconds}{19.4}
\newcommand{\FWriterTasks}{73}
\newcommand{\FWriterWithKept}{69}

\newcommand{\FPHClassExplicitEFA}{0.053}
\newcommand{\FPHClassExplicitEP}{<0.001}
\newcommand{\FPHClassExplicitTFA}{0.297}
\newcommand{\FPHClassExplicitTasks}{25}
\newcommand{\FPHClassExplicitWP}{<0.001}
\newcommand{\FPHClassExplicitWThreeFA}{0.055}
\newcommand{\FPHClassNoSafeEFA}{0.050}
\newcommand{\FPHClassNoSafeEP}{0.500}
\newcommand{\FPHClassNoSafeTFA}{0.192}
\newcommand{\FPHClassNoSafeTasks}{4}
\newcommand{\FPHClassNoSafeWP}{0.500}
\newcommand{\FPHClassNoSafeWThreeFA}{0.050}
\newcommand{\FPHClassSilentEFA}{0.080}
\newcommand{\FPHClassSilentEP}{<0.001}
\newcommand{\FPHClassSilentTFA}{0.571}
\newcommand{\FPHClassSilentTasks}{23}
\newcommand{\FPHClassSilentWP}{<0.001}
\newcommand{\FPHClassSilentWThreeFA}{0.091}
\newcommand{\FPHCommonTasks}{21}
\newcommand{\FPHDiagFA}{0.260}
\newcommand{\FPHGptOssHFourStrict}{0.770}
\newcommand{\FPHGptOssHOneStrict}{<0.001}
\newcommand{\FPHGptOssHThreeStrict}{1.000}
\newcommand{\FPHGptOssHTwoStrict}{0.132}
\newcommand{\FPHGptOssMatchedSrawFA}{0.126}
\newcommand{\FPHGptOssMatchedStepN}{96}
\newcommand{\FPHGptOssMatchedStepSraw}{0.292}
\newcommand{\FPHGptOssMatchedStepWThree}{0.021}
\newcommand{\FPHGptOssMatchedTasks}{42}
\newcommand{\FPHGptOssMatchedWThreeFA}{0.075}
\newcommand{\FPHLunaHFourStrict}{1.000}
\newcommand{\FPHLunaHOneStrict}{0.499}
\newcommand{\FPHLunaHThreeStrict}{1.000}
\newcommand{\FPHLunaHTwoStrict}{0.749}
\newcommand{\FPHLunaMatchedSrawFA}{0.054}
\newcommand{\FPHLunaMatchedStepN}{45}
\newcommand{\FPHLunaMatchedStepSraw}{0.044}
\newcommand{\FPHLunaMatchedStepWThree}{0.044}
\newcommand{\FPHLunaMatchedTasks}{52}
\newcommand{\FPHLunaMatchedWThreeFA}{0.070}
\newcommand{\FPHMFourAStrict}{<0.001}
\newcommand{\FPHMFourCStrict}{0.002}
\newcommand{\FPHNoOverlapTasks}{3}
\newcommand{\FPHNoOverlapWOneFA}{0.149}
\newcommand{\FPHNoOverlapWOneFR}{0.010}
\newcommand{\FPHNoOverlapWThreeFA}{0.090}
\newcommand{\FPHNoOverlapWThreeFR}{0.010}
\newcommand{\FPHNoOverlapWTwoFA}{0.094}
\newcommand{\FPHNoOverlapWTwoFR}{0.010}
\newcommand{\FPHOffDiagFA}{0.246}
\newcommand{\FPHQwenEightHFourStrict}{0.998}
\newcommand{\FPHQwenEightHOneStrict}{<0.001}
\newcommand{\FPHQwenEightHThreeStrict}{1.000}
\newcommand{\FPHQwenEightHTwoStrict}{0.749}
\newcommand{\FPHQwenEightMatchedSrawFA}{0.099}
\newcommand{\FPHQwenEightMatchedStepN}{76}
\newcommand{\FPHQwenEightMatchedStepSraw}{0.132}
\newcommand{\FPHQwenEightMatchedStepWThree}{0.026}
\newcommand{\FPHQwenEightMatchedTasks}{23}
\newcommand{\FPHQwenEightMatchedWThreeFA}{0.063}
\newcommand{\FPHQwenMoeHFourStrict}{0.256}
\newcommand{\FPHQwenMoeHOneStrict}{0.004}
\newcommand{\FPHQwenMoeHThreeStrict}{1.000}
\newcommand{\FPHQwenMoeHTwoStrict}{0.256}
\newcommand{\FPHQwenMoeMatchedSrawFA}{0.086}
\newcommand{\FPHQwenMoeMatchedStepN}{85}
\newcommand{\FPHQwenMoeMatchedStepSraw}{0.271}
\newcommand{\FPHQwenMoeMatchedStepWThree}{0.000}
\newcommand{\FPHQwenMoeMatchedTasks}{48}
\newcommand{\FPHQwenMoeMatchedWThreeFA}{0.066}
\newcommand{\FPHWriterSimExact}{3}
\newcommand{\FPHWriterSimHigh}{20}
\newcommand{\FPHWriterSimMean}{0.70}
\newcommand{\FPHWriterSimMedian}{0.72}
\newcommand{\FPHWriterSimN}{128}

\newcommand{\FLAdded}{1086}
\newcommand{\FLFreezeTest}{db2f355dba79}
\newcommand{\FLFreezeZero}{2c95616bd805}
\newcommand{\FLGptOssCalls}{1403}
\newcommand{\FLGptOssECC}{334}
\newcommand{\FLGptOssEFA}{15}
\newcommand{\FLGptOssEFR}{0}
\newcommand{\FLGptOssHOneA}{0.038}
\newcommand{\FLGptOssHOneB}{0.212}
\newcommand{\FLGptOssHOneDecision}{positive}
\newcommand{\FLGptOssHOneDiff}{$-$0.173}
\newcommand{\FLGptOssHOneHolm}{<0.001}
\newcommand{\FLGptOssHThreeA}{0.652}
\newcommand{\FLGptOssHThreeB}{0.677}
\newcommand{\FLGptOssHThreeDiff}{$-$0.026}
\newcommand{\FLGptOssHThreeHolm}{0.335}
\newcommand{\FLGptOssHTwoA}{0.212}
\newcommand{\FLGptOssHTwoB}{0.181}
\newcommand{\FLGptOssHTwoDecision}{not positive}
\newcommand{\FLGptOssHTwoDiff}{0.031}
\newcommand{\FLGptOssHTwoHolm}{0.335}
\newcommand{\FLGptOssN}{456}
\newcommand{\FLGptOssStatus}{confirmatory}
\newcommand{\FLGptOssStepGThree}{0.234}
\newcommand{\FLGptOssStepSraw}{0.379}
\newcommand{\FLGptOssStepWThree}{0.089}
\newcommand{\FLGptOssTCC}{310}
\newcommand{\FLGptOssTFA}{85}
\newcommand{\FLGptOssTFR}{0}
\newcommand{\FLLMOneA}{0.040}
\newcommand{\FLLMOneB}{0.271}
\newcommand{\FLLMOneDecision}{positive}
\newcommand{\FLLMOneDiff}{$-$0.230}
\newcommand{\FLLMOneDisc}{37}
\newcommand{\FLLMOneHolm}{<0.001}
\newcommand{\FLLMThreeA}{0.271}
\newcommand{\FLLMThreeB}{0.406}
\newcommand{\FLLMThreeDecision}{not positive}
\newcommand{\FLLMThreeDiff}{$-$0.135}
\newcommand{\FLLMThreeDisc}{38}
\newcommand{\FLLMThreeHolm}{<0.001}
\newcommand{\FLLMTwoA}{0.050}
\newcommand{\FLLMTwoB}{0.271}
\newcommand{\FLLMTwoDecision}{positive}
\newcommand{\FLLMTwoDiff}{$-$0.221}
\newcommand{\FLLMTwoDisc}{34}
\newcommand{\FLLMTwoHolm}{<0.001}
\newcommand{\FLLunaCalls}{1242}
\newcommand{\FLLunaECC}{401}
\newcommand{\FLLunaEFA}{12}
\newcommand{\FLLunaEFR}{0}
\newcommand{\FLLunaHOneA}{0.031}
\newcommand{\FLLunaHOneB}{0.074}
\newcommand{\FLLunaHOneDecision}{not positive}
\newcommand{\FLLunaHOneDiff}{$-$0.043}
\newcommand{\FLLunaHOneHolm}{0.378}
\newcommand{\FLLunaHThreeA}{0.834}
\newcommand{\FLLunaHThreeB}{0.852}
\newcommand{\FLLunaHThreeDiff}{$-$0.018}
\newcommand{\FLLunaHThreeHolm}{0.378}
\newcommand{\FLLunaHTwoA}{0.074}
\newcommand{\FLLunaHTwoB}{0.056}
\newcommand{\FLLunaHTwoDecision}{not positive}
\newcommand{\FLLunaHTwoDiff}{0.018}
\newcommand{\FLLunaHTwoHolm}{0.378}
\newcommand{\FLLunaN}{456}
\newcommand{\FLLunaStatus}{descriptive}
\newcommand{\FLLunaStepGThree}{0.265}
\newcommand{\FLLunaStepSraw}{0.082}
\newcommand{\FLLunaStepWThree}{0.082}
\newcommand{\FLLunaTCC}{391}
\newcommand{\FLLunaTFA}{29}
\newcommand{\FLLunaTFR}{0}
\newcommand{\FLMFEdgeFA}{0.045}
\newcommand{\FLMFEdgeFR}{0.014}
\newcommand{\FLMFGOneFA}{0.135}
\newcommand{\FLMFGOneFR}{0.036}
\newcommand{\FLMFGThreeFA}{0.106}
\newcommand{\FLMFGThreeFR}{0.043}
\newcommand{\FLMFGTwoFA}{0.107}
\newcommand{\FLMFGTwoFR}{0.043}
\newcommand{\FLMFTypFA}{0.278}
\newcommand{\FLMFTypFR}{0.014}
\newcommand{\FLMFWOneFA}{0.089}
\newcommand{\FLMFWOneFR}{0.036}
\newcommand{\FLMFWThreeFA}{0.053}
\newcommand{\FLMFWThreeFR}{0.043}
\newcommand{\FLMFWTwoFA}{0.054}
\newcommand{\FLMFWTwoFR}{0.043}
\newcommand{\FLPins}{63}
\newcommand{\FLProcedures}{1986}
\newcommand{\FLQwenEightCalls}{1599}
\newcommand{\FLQwenEightECC}{244}
\newcommand{\FLQwenEightEFA}{27}
\newcommand{\FLQwenEightEFR}{0}
\newcommand{\FLQwenEightHOneA}{0.062}
\newcommand{\FLQwenEightHOneB}{0.253}
\newcommand{\FLQwenEightHOneDecision}{positive}
\newcommand{\FLQwenEightHOneDiff}{$-$0.190}
\newcommand{\FLQwenEightHOneHolm}{<0.001}
\newcommand{\FLQwenEightHThreeA}{0.499}
\newcommand{\FLQwenEightHThreeB}{0.546}
\newcommand{\FLQwenEightHThreeDiff}{$-$0.047}
\newcommand{\FLQwenEightHThreeHolm}{0.546}
\newcommand{\FLQwenEightHTwoA}{0.253}
\newcommand{\FLQwenEightHTwoB}{0.233}
\newcommand{\FLQwenEightHTwoDecision}{not positive}
\newcommand{\FLQwenEightHTwoDiff}{0.019}
\newcommand{\FLQwenEightHTwoHolm}{0.634}
\newcommand{\FLQwenEightN}{456}
\newcommand{\FLQwenEightStatus}{confirmatory}
\newcommand{\FLQwenEightStepGThree}{0.241}
\newcommand{\FLQwenEightStepSraw}{0.382}
\newcommand{\FLQwenEightStepWThree}{0.161}
\newcommand{\FLQwenEightTCC}{241}
\newcommand{\FLQwenEightTFA}{108}
\newcommand{\FLQwenEightTFR}{0}
\newcommand{\FLQwenMoeCalls}{1387}
\newcommand{\FLQwenMoeECC}{344}
\newcommand{\FLQwenMoeEFA}{9}
\newcommand{\FLQwenMoeEFR}{0}
\newcommand{\FLQwenMoeHOneA}{0.023}
\newcommand{\FLQwenMoeHOneB}{0.177}
\newcommand{\FLQwenMoeHOneDecision}{positive}
\newcommand{\FLQwenMoeHOneDiff}{$-$0.154}
\newcommand{\FLQwenMoeHOneHolm}{<0.001}
\newcommand{\FLQwenMoeHThreeA}{0.684}
\newcommand{\FLQwenMoeHThreeB}{0.753}
\newcommand{\FLQwenMoeHThreeDiff}{$-$0.069}
\newcommand{\FLQwenMoeHThreeHolm}{0.047}
\newcommand{\FLQwenMoeHTwoA}{0.177}
\newcommand{\FLQwenMoeHTwoB}{0.120}
\newcommand{\FLQwenMoeHTwoDecision}{positive}
\newcommand{\FLQwenMoeHTwoDiff}{0.057}
\newcommand{\FLQwenMoeHTwoHolm}{0.047}
\newcommand{\FLQwenMoeN}{456}
\newcommand{\FLQwenMoeStatus}{descriptive}
\newcommand{\FLQwenMoeStepGThree}{0.288}
\newcommand{\FLQwenMoeStepSraw}{0.297}
\newcommand{\FLQwenMoeStepWThree}{0.034}
\newcommand{\FLQwenMoeTCC}{327}
\newcommand{\FLQwenMoeTFA}{72}
\newcommand{\FLQwenMoeTFR}{0}
\newcommand{\FLRepMet}{met}
\newcommand{\FLRepPositive}{3}
\newcommand{\FLSmallTasks}{25}
\newcommand{\FLTotal}{1491}
\newcommand{\FLVFour}{405}
\newcommand{\FLVFourTypFA}{0.410}

\newcommand{\FQCommonTasks}{21}
\newcommand{\FQDiagCommon}{0.161}
\newcommand{\FQFreezeTestDate}{2026-10-18}
\newcommand{\FQFreezeTestFiles}{923}
\newcommand{\FQFreezeTestTime}{03:48}
\newcommand{\FQFreezeZeroDate}{2026-10-18}
\newcommand{\FQFreezeZeroFiles}{192}
\newcommand{\FQFreezeZeroTime}{03:27}
\newcommand{\FQGptOssLargeEAny}{2}
\newcommand{\FQGptOssLargeTAny}{16}
\newcommand{\FQGptOssQualDecode}{55.9}
\newcommand{\FQGptOssQualExact}{12/12}
\newcommand{\FQGptOssSmallEAny}{1}
\newcommand{\FQGptOssSmallTAny}{16}
\newcommand{\FQGptOssTrunc}{8}
\newcommand{\FQLargeEdgeAny}{17}
\newcommand{\FQLargeEdgeMean}{0.88}
\newcommand{\FQLargeEdgeTotal}{46}
\newcommand{\FQLargeProgMedian}{30}
\newcommand{\FQLargeTypAny}{42}
\newcommand{\FQLargeTypMean}{8.56}
\newcommand{\FQLargeTypTotal}{445}
\newcommand{\FQLargeWThreeAny}{15}
\newcommand{\FQLargeWThreeMean}{1.19}
\newcommand{\FQLargeWThreeTotal}{62}
\newcommand{\FQLargeWrongMedian}{26.5}
\newcommand{\FQLunaLargeEAny}{3}
\newcommand{\FQLunaLargeTAny}{7}
\newcommand{\FQLunaSmallEAny}{1}
\newcommand{\FQLunaSmallTAny}{5}
\newcommand{\FQMenuMedian}{27}
\newcommand{\FQOffCommon}{0.187}
\newcommand{\FQQwenEightLargeEAny}{5}
\newcommand{\FQQwenEightLargeTAny}{20}
\newcommand{\FQQwenEightQualDecode}{40.7}
\newcommand{\FQQwenEightQualExact}{12/12}
\newcommand{\FQQwenEightSmallEAny}{5}
\newcommand{\FQQwenEightSmallTAny}{19}
\newcommand{\FQQwenEightTrunc}{5}
\newcommand{\FQQwenMoeLargeEAny}{2}
\newcommand{\FQQwenMoeLargeTAny}{17}
\newcommand{\FQQwenMoeQualDecode}{74.3}
\newcommand{\FQQwenMoeQualExact}{12/12}
\newcommand{\FQQwenMoeSmallEAny}{2}
\newcommand{\FQQwenMoeSmallTAny}{11}
\newcommand{\FQQwenMoeTrunc}{3}
\newcommand{\FQRepairWrongCommonMax}{70}
\newcommand{\FQRepairWrongCommonMin}{30}
\newcommand{\FQSmallEdgeAny}{11}
\newcommand{\FQSmallEdgeMean}{0.27}
\newcommand{\FQSmallEdgeTotal}{14}
\newcommand{\FQSmallProgMedian}{5}
\newcommand{\FQSmallTypAny}{41}
\newcommand{\FQSmallTypMean}{2.17}
\newcommand{\FQSmallTypTotal}{113}
\newcommand{\FQSmallWThreeAny}{12}
\newcommand{\FQSmallWThreeMean}{0.29}
\newcommand{\FQSmallWThreeTotal}{15}
\newcommand{\FQSmallWrongMedian}{4.0}
\newcommand{\FQTasks}{52}

\newcommand{\FHFragments}{1689}
\newcommand{\FHFreezeTest}{5f4ca8417951}
\newcommand{\FHFreezeTestFiles}{928}
\newcommand{\FHFreezeTestTime}{2026-10-19 02:54}
\newcommand{\FHFreezeZero}{bf681a0b7291}
\newcommand{\FHFreezeZeroFiles}{197}
\newcommand{\FHFreezeZeroTime}{2026-10-19 02:22}
\newcommand{\FHGptOssCalls}{1404}
\newcommand{\FHGptOssDecode}{62.0}
\newcommand{\FHGptOssExact}{12/12}
\newcommand{\FHGptOssHAOneA}{0.043}
\newcommand{\FHGptOssHAOneB}{0.218}
\newcommand{\FHGptOssHAOneDecision}{positive}
\newcommand{\FHGptOssHAOneDiff}{$-$0.175}
\newcommand{\FHGptOssHAOneHolm}{<0.001}
\newcommand{\FHGptOssHATwoA}{0.218}
\newcommand{\FHGptOssHATwoB}{0.212}
\newcommand{\FHGptOssHATwoDecision}{not positive}
\newcommand{\FHGptOssHATwoDiff}{0.006}
\newcommand{\FHGptOssHATwoHolm}{0.757}
\newcommand{\FHGptOssN}{456}
\newcommand{\FHGptOssStatus}{confirmatory}
\newcommand{\FHGptOssStepSraw}{0.416}
\newcommand{\FHGptOssStepWThree}{0.104}
\newcommand{\FHGptOssStepXEight}{0.384}
\newcommand{\FHGptOssStepXMoe}{0.392}
\newcommand{\FHGptOssTrunc}{4}
\newcommand{\FHHDOneA}{0.053}
\newcommand{\FHHDOneB}{0.328}
\newcommand{\FHHDOneDecision}{positive}
\newcommand{\FHHDOneHolm}{<0.001}
\newcommand{\FHHDThreeA}{0.328}
\newcommand{\FHHDThreeB}{0.271}
\newcommand{\FHHDThreeDecision}{positive}
\newcommand{\FHHDThreeHolm}{<0.001}
\newcommand{\FHHDTwoA}{0.056}
\newcommand{\FHHDTwoB}{0.328}
\newcommand{\FHHDTwoDecision}{positive}
\newcommand{\FHHDTwoHolm}{<0.001}
\newcommand{\FHHard}{233}
\newcommand{\FHHardPrograms}{359}
\newcommand{\FHHardTasks}{26}
\newcommand{\FHHardWrong}{323}
\newcommand{\FHLunaCalls}{1238}
\newcommand{\FHLunaHAOneA}{0.031}
\newcommand{\FHLunaHAOneB}{0.107}
\newcommand{\FHLunaHAOneDecision}{not positive}
\newcommand{\FHLunaHAOneDiff}{$-$0.077}
\newcommand{\FHLunaHAOneHolm}{0.063}
\newcommand{\FHLunaHATwoA}{0.107}
\newcommand{\FHLunaHATwoB}{0.074}
\newcommand{\FHLunaHATwoDecision}{not positive}
\newcommand{\FHLunaHATwoDiff}{0.033}
\newcommand{\FHLunaHATwoHolm}{0.094}
\newcommand{\FHLunaN}{456}
\newcommand{\FHLunaStatus}{descriptive}
\newcommand{\FHLunaStepSraw}{0.117}
\newcommand{\FHLunaStepWThree}{0.117}
\newcommand{\FHLunaStepXEight}{0.300}
\newcommand{\FHLunaStepXMoe}{0.433}
\newcommand{\FHLunaTrunc}{0}
\newcommand{\FHMFEdgeFA}{0.058}
\newcommand{\FHMFEdgeFR}{0.010}
\newcommand{\FHMFEdgeHard}{0.168}
\newcommand{\FHMFGThreeFA}{0.135}
\newcommand{\FHMFGThreeFR}{0.042}
\newcommand{\FHMFGThreeHard}{0.492}
\newcommand{\FHMFTypFA}{0.337}
\newcommand{\FHMFTypFR}{0.010}
\newcommand{\FHMFTypHard}{0.990}
\newcommand{\FHMFWThreeFA}{0.058}
\newcommand{\FHMFWThreeFR}{0.042}
\newcommand{\FHMFWThreeHard}{0.153}
\newcommand{\FHProcedures}{2310}
\newcommand{\FHQwenEightCalls}{1593}
\newcommand{\FHQwenEightDecode}{48.4}
\newcommand{\FHQwenEightExact}{12/12}
\newcommand{\FHQwenEightHAOneA}{0.050}
\newcommand{\FHQwenEightHAOneB}{0.250}
\newcommand{\FHQwenEightHAOneDecision}{positive}
\newcommand{\FHQwenEightHAOneDiff}{$-$0.200}
\newcommand{\FHQwenEightHAOneHolm}{<0.001}
\newcommand{\FHQwenEightHATwoA}{0.250}
\newcommand{\FHQwenEightHATwoB}{0.253}
\newcommand{\FHQwenEightHATwoDecision}{not positive}
\newcommand{\FHQwenEightHATwoDiff}{$-$0.003}
\newcommand{\FHQwenEightHATwoHolm}{1.000}
\newcommand{\FHQwenEightN}{456}
\newcommand{\FHQwenEightStatus}{confirmatory}
\newcommand{\FHQwenEightStepSraw}{0.395}
\newcommand{\FHQwenEightStepWThree}{0.173}
\newcommand{\FHQwenEightStepXEight}{0.378}
\newcommand{\FHQwenEightStepXMoe}{0.303}
\newcommand{\FHQwenEightTrunc}{4}
\newcommand{\FHQwenMoeCalls}{1391}
\newcommand{\FHQwenMoeDecode}{94.7}
\newcommand{\FHQwenMoeExact}{12/12}
\newcommand{\FHQwenMoeHAOneA}{0.028}
\newcommand{\FHQwenMoeHAOneB}{0.198}
\newcommand{\FHQwenMoeHAOneDecision}{positive}
\newcommand{\FHQwenMoeHAOneDiff}{$-$0.170}
\newcommand{\FHQwenMoeHAOneHolm}{<0.001}
\newcommand{\FHQwenMoeHATwoA}{0.198}
\newcommand{\FHQwenMoeHATwoB}{0.177}
\newcommand{\FHQwenMoeHATwoDecision}{not positive}
\newcommand{\FHQwenMoeHATwoDiff}{0.020}
\newcommand{\FHQwenMoeHATwoHolm}{0.225}
\newcommand{\FHQwenMoeN}{456}
\newcommand{\FHQwenMoeStatus}{descriptive}
\newcommand{\FHQwenMoeStepSraw}{0.317}
\newcommand{\FHQwenMoeStepWThree}{0.048}
\newcommand{\FHQwenMoeStepXEight}{0.349}
\newcommand{\FHQwenMoeStepXMoe}{0.373}
\newcommand{\FHQwenMoeTrunc}{4}
\newcommand{\FHRepMet}{met}
\newcommand{\FHRepPositive}{3}
\newcommand{\FHTypFRTask}{0.010}
\newcommand{\FHWOneFA}{0.094}
\newcommand{\FHWOneFR}{0.083}
\newcommand{\FHWROneA}{0.094}
\newcommand{\FHWROneB}{0.328}
\newcommand{\FHWROneDecision}{not positive}
\newcommand{\FHWROneHolm}{<0.001}
\newcommand{\FHWRThreeA}{0.138}
\newcommand{\FHWRThreeB}{0.097}
\newcommand{\FHWRThreeHolm}{0.196}
\newcommand{\FHWRTwoA}{0.182}
\newcommand{\FHWRTwoB}{0.328}
\newcommand{\FHWRTwoDecision}{not positive}
\newcommand{\FHWRTwoHolm}{<0.001}
\newcommand{\FHWTwoFA}{0.182}
\newcommand{\FHWTwoFR}{0.049}
\newcommand{\FHWriterEightDecode}{43.8}
\newcommand{\FHWriterEightExact}{12/12}
\newcommand{\FHWriterEightKept}{105}
\newcommand{\FHWriterEightOverlap}{2}
\newcommand{\FHWriterEightParse}{21}
\newcommand{\FHWriterEightTasks}{44}
\newcommand{\FHWriterMoeDecode}{65.1}
\newcommand{\FHWriterMoeExact}{12/12}
\newcommand{\FHWriterMoeKept}{171}
\newcommand{\FHWriterMoeOverlap}{6}
\newcommand{\FHWriterMoeParse}{8}
\newcommand{\FHWriterMoeTasks}{61}

\newcommand{\FFHardIndEdge}{77}
\newcommand{\FFHardIndTyp}{709}
\newcommand{\FFHardIndWThree}{82}
\newcommand{\FFProgOne}{484}
\newcommand{\FFProgThree}{2310}
\newcommand{\FFProgTwo}{1986}

\newcommand{\ISSQualifiedComponents}{10}
\newcommand{\ISSTargetComponents}{20}
\newcommand{\ISSMaxAchievable}{18}
\newcommand{\ISSRegisteredCandidates}{26}
\newcommand{\ISSRoots}{40}
\newcommand{\ISSTestRoots}{24}
\newcommand{\ISSTestComponents}{6}
\newcommand{\ISSTestAffected}{18}
\newcommand{\ISSTestControls}{6}
\newcommand{\ISSOutcomes}{160}
\newcommand{\ISSScienceCalls}{196}
\newcommand{\ISSScienceMinutes}{4.6}
\newcommand{\ISSFullNoDiffAll}{27/40}
\newcommand{\ISSFullNoDiffTest}{17/24}
\newcommand{\ISSFullNoDiffInvalidAll}{7}
\newcommand{\ISSFullNoDiffInvalidTest}{7}
\newcommand{\ISSFullNoDiffTestSubmitted}{17/17}
\newcommand{\ISSFullNoDiffRegress}{3/10}
\newcommand{\ISSFullNoDiffABW}{4}
\newcommand{\ISSFullDiffAll}{28/40}
\newcommand{\ISSFullDiffTest}{17/24}
\newcommand{\ISSFullDiffInvalidAll}{11}
\newcommand{\ISSFullDiffInvalidTest}{7}
\newcommand{\ISSFullDiffTestSubmitted}{17/17}
\newcommand{\ISSFullDiffRegress}{3/10}
\newcommand{\ISSFullDiffABW}{1}
\newcommand{\ISSLocalNoDiffAll}{34/40}
\newcommand{\ISSLocalNoDiffTest}{23/24}
\newcommand{\ISSLocalNoDiffInvalidAll}{0}
\newcommand{\ISSLocalNoDiffInvalidTest}{0}
\newcommand{\ISSLocalNoDiffTestSubmitted}{23/24}
\newcommand{\ISSLocalNoDiffRegress}{0/10}
\newcommand{\ISSLocalNoDiffABW}{3}
\newcommand{\ISSLocalDiffAll}{33/40}
\newcommand{\ISSLocalDiffTest}{21/24}
\newcommand{\ISSLocalDiffInvalidAll}{0}
\newcommand{\ISSLocalDiffInvalidTest}{0}
\newcommand{\ISSLocalDiffTestSubmitted}{21/24}
\newcommand{\ISSLocalDiffRegress}{0/10}
\newcommand{\ISSLocalDiffABW}{4}
\newcommand{\ISSFullInvalidAll}{18}
\newcommand{\ISSFullEditBudgetInvalid}{14}
\newcommand{\ISSFullDomainInvalid}{4}
\newcommand{\ISSLocalInvalidAll}{0}
\newcommand{\ISSPrimaryMean}{0.167}
\newcommand{\ISSPrimaryLow}{0.042}
\newcommand{\ISSPrimaryHigh}{0.333}
\newcommand{\ISSPrimaryDiscordant}{4}
\newcommand{\ISSPrimaryPermP}{0.25}
\newcommand{\ISSPrimaryDiscordantInvalid}{4}
\newcommand{\ISSRepFullMean}{0.000}
\newcommand{\ISSRepLocalMean}{$-$0.083}
\newcommand{\ISSRepHolmMin}{1.00}
\newcommand{\ISSLlamaExact}{11/12}
\newcommand{\ISSLlamaContent}{11/12}
\newcommand{\ISSLlamaMultiline}{0}
\newcommand{\ISSLlamaPrefill}{995}
\newcommand{\ISSLlamaDecode}{46.9}
\newcommand{\ISSMistralExact}{4/12}
\newcommand{\ISSMistralContent}{12/12}
\newcommand{\ISSMistralMultiline}{8}
\newcommand{\ISSMistralPrefill}{946}
\newcommand{\ISSMistralDecode}{41.1}
\newcommand{\ISSQwenPrefill}{1,074}
\newcommand{\ISSQwenDecode}{45.8}

\newcommand{\PHSFrozenFullDiffAll}{28/40}
\newcommand{\PHSFrozenFullDiffTest}{17/24}
\newcommand{\PHSFrozenFullDiffInvalidTest}{7}
\newcommand{\PHSFrozenFullNoDiffAll}{27/40}
\newcommand{\PHSFrozenFullNoDiffTest}{17/24}
\newcommand{\PHSFrozenFullNoDiffInvalidTest}{7}
\newcommand{\PHSFrozenMean}{0.167}
\newcommand{\PHSFrozenLow}{0.042}
\newcommand{\PHSFrozenHigh}{0.333}
\newcommand{\PHSFrozenDiscordant}{4}
\newcommand{\PHSFrozenPermP}{0.25}
\newcommand{\PHSFrozenUnobserved}{0}
\newcommand{\PHSFrozenUnobservedTest}{0}
\newcommand{\PHSBudgetFullDiffAll}{33--34/40}
\newcommand{\PHSBudgetFullDiffTest}{21/24}
\newcommand{\PHSBudgetFullDiffInvalidTest}{3}
\newcommand{\PHSBudgetFullNoDiffAll}{33/40}
\newcommand{\PHSBudgetFullNoDiffTest}{23/24}
\newcommand{\PHSBudgetFullNoDiffInvalidTest}{1}
\newcommand{\PHSBudgetMean}{0.000}
\newcommand{\PHSBudgetLow}{$-$0.167}
\newcommand{\PHSBudgetHigh}{0.167}
\newcommand{\PHSBudgetDiscordant}{4}
\newcommand{\PHSBudgetPermP}{1.00}
\newcommand{\PHSBudgetUnobserved}{1}
\newcommand{\PHSBudgetUnobservedTest}{0}
\newcommand{\PHSNoLimitFullDiffAll}{33--36/40}
\newcommand{\PHSNoLimitFullDiffTest}{21/24}
\newcommand{\PHSNoLimitFullDiffInvalidTest}{3}
\newcommand{\PHSNoLimitFullNoDiffAll}{33/40}
\newcommand{\PHSNoLimitFullNoDiffTest}{23/24}
\newcommand{\PHSNoLimitFullNoDiffInvalidTest}{1}
\newcommand{\PHSNoLimitMean}{0.000}
\newcommand{\PHSNoLimitLow}{$-$0.167}
\newcommand{\PHSNoLimitHigh}{0.167}
\newcommand{\PHSNoLimitDiscordant}{4}
\newcommand{\PHSNoLimitPermP}{1.00}
\newcommand{\PHSNoLimitUnobserved}{3}
\newcommand{\PHSNoLimitUnobservedTest}{0}
\newcommand{\PHSParseOnlyFullDiffAll}{36--39/40}
\newcommand{\PHSParseOnlyFullDiffTest}{24/24}
\newcommand{\PHSParseOnlyFullDiffInvalidTest}{0}
\newcommand{\PHSParseOnlyFullNoDiffAll}{34/40}
\newcommand{\PHSParseOnlyFullNoDiffTest}{24/24}
\newcommand{\PHSParseOnlyFullNoDiffInvalidTest}{0}
\newcommand{\PHSParseOnlyMean}{$-$0.125}
\newcommand{\PHSParseOnlyLow}{$-$0.208}
\newcommand{\PHSParseOnlyHigh}{$-$0.042}
\newcommand{\PHSParseOnlyDiscordant}{3}
\newcommand{\PHSParseOnlyPermP}{0.25}
\newcommand{\PHSParseOnlyUnobserved}{3}
\newcommand{\PHSParseOnlyUnobservedTest}{0}
\newcommand{\PHSBudgetChangedTest}{10}
\newcommand{\PHSBudgetChangedTestSuccess}{10}
\newcommand{\PHSParseChangedTestDomain}{4}

\newcommand{\PVHardWrong}{297}
\newcommand{\PVHardPassE}{31}
\newcommand{\PVHardPassEPooled}{0.104}
\newcommand{\PVHardPassW}{24}
\newcommand{\PVHardPassWPooled}{0.081}
\newcommand{\PVHardPassBoth}{10}
\newcommand{\PVHardPassBothPooled}{0.034}
\newcommand{\PVHardTaskMeanE}{0.168}
\newcommand{\PVHardTasks}{23}
\newcommand{\PVHardPassETasks}{7}
\newcommand{\PVHardRelE}{3}
\newcommand{\PVHardRelT}{12}
\newcommand{\PVOwnLunaWrong}{45}
\newcommand{\PVOwnLunaPassOwn}{2}
\newcommand{\PVOwnLunaPassOwnE}{0}
\newcommand{\PVOwnLunaCaught}{2}
\newcommand{\PVOwnLunaCaughtRate}{1.00}
\newcommand{\PVOwnLunaClusters}{1}
\newcommand{\PVOwnQwenEightWrong}{176}
\newcommand{\PVOwnQwenEightPassOwn}{58}
\newcommand{\PVOwnQwenEightPassOwnE}{20}
\newcommand{\PVOwnQwenEightCaught}{38}
\newcommand{\PVOwnQwenEightCaughtRate}{0.66}
\newcommand{\PVOwnQwenEightClusters}{12}
\newcommand{\PVOwnGptOssWrong}{109}
\newcommand{\PVOwnGptOssPassOwn}{37}
\newcommand{\PVOwnGptOssPassOwnE}{8}
\newcommand{\PVOwnGptOssCaught}{29}
\newcommand{\PVOwnGptOssCaughtRate}{0.78}
\newcommand{\PVOwnGptOssClusters}{8}
\newcommand{\PVOwnQwenThreeSixWrong}{92}
\newcommand{\PVOwnQwenThreeSixPassOwn}{30}
\newcommand{\PVOwnQwenThreeSixPassOwnE}{4}
\newcommand{\PVOwnQwenThreeSixCaught}{26}
\newcommand{\PVOwnQwenThreeSixCaughtRate}{0.87}
\newcommand{\PVOwnQwenThreeSixClusters}{6}

\newcommand{\VSTasks}{52}
\newcommand{\VSClusters}{49}
\newcommand{\VSExcluded}{0}
\newcommand{\VSK}{4}
\newcommand{\VSTemp}{0.7}
\newcommand{\VSMaxTok}{4,096}
\newcommand{\VSQualCand}{484}
\newcommand{\VSQualDiff}{0}
\newcommand{\VSQualHostDiff}{0}
\newcommand{\VSQualExec}{3,241}
\newcommand{\VSQualRawDiff}{43}
\newcommand{\VSQualRawTasks}{4}
\newcommand{\VSGptOssResp}{208}
\newcommand{\VSGptOssParsed}{146}
\newcommand{\VSGptOssParseRate}{0.70}
\newcommand{\VSGptOssClean}{137}
\newcommand{\VSGptOssWrong}{43}
\newcommand{\VSGptOssCorrect}{94}
\newcommand{\VSGptOssWrongClusters}{22}
\newcommand{\VSGptOssFAT}{0.517}
\newcommand{\VSGptOssFAE}{0.114}
\newcommand{\VSGptOssDelta}{$-$0.403}
\newcommand{\VSGptOssDeltaLow}{$-$0.592}
\newcommand{\VSGptOssDeltaHigh}{$-$0.223}
\newcommand{\VSGptOssHOneP}{$<$0.001}
\newcommand{\VSGptOssHOneHolm}{0.001}
\newcommand{\VSGptOssICR}{0.89}
\newcommand{\VSGptOssICRn}{9}
\newcommand{\VSGptOssICRLow}{0.57}
\newcommand{\VSGptOssHTwoP}{$<$0.001}
\newcommand{\VSGptOssHTwoHolm}{0.001}
\newcommand{\VSGptOssICRCaught}{8}
\newcommand{\VSGptOssCorrRelT}{1.00}
\newcommand{\VSGptOssCorrRelE}{1.00}
\newcommand{\VSGptOssWrongRelT}{0.120}
\newcommand{\VSGptOssWrongRelE}{0.029}
\newcommand{\VSGptOssVacuous}{5}
\newcommand{\VSGptOssHardcoded}{0}
\newcommand{\VSGptOssPredFAT}{met}
\newcommand{\VSGptOssPredFAE}{not met}
\newcommand{\VSGptOssFalsFAE}{no}
\newcommand{\VSGptOssFalsCI}{no}
\newcommand{\VSGptOssFalsified}{not falsified}
\newcommand{\VSGptOssPredICR}{met}
\newcommand{\VSGptOssHOneDec}{supported}
\newcommand{\VSGptOssHTwoDec}{supported}
\newcommand{\VSGptOssDescOnly}{no}
\newcommand{\VSQwenThreeSixResp}{208}
\newcommand{\VSQwenThreeSixParsed}{187}
\newcommand{\VSQwenThreeSixParseRate}{0.90}
\newcommand{\VSQwenThreeSixClean}{174}
\newcommand{\VSQwenThreeSixWrong}{65}
\newcommand{\VSQwenThreeSixCorrect}{109}
\newcommand{\VSQwenThreeSixWrongClusters}{25}
\newcommand{\VSQwenThreeSixFAT}{0.653}
\newcommand{\VSQwenThreeSixFAE}{0.113}
\newcommand{\VSQwenThreeSixDelta}{$-$0.540}
\newcommand{\VSQwenThreeSixDeltaLow}{$-$0.707}
\newcommand{\VSQwenThreeSixDeltaHigh}{$-$0.367}
\newcommand{\VSQwenThreeSixHOneP}{$<$0.001}
\newcommand{\VSQwenThreeSixHOneHolm}{$<$0.001}
\newcommand{\VSQwenThreeSixICR}{0.50}
\newcommand{\VSQwenThreeSixICRn}{10}
\newcommand{\VSQwenThreeSixICRLow}{0.11}
\newcommand{\VSQwenThreeSixHTwoP}{0.144}
\newcommand{\VSQwenThreeSixHTwoHolm}{0.144}
\newcommand{\VSQwenThreeSixICRCaught}{5}
\newcommand{\VSQwenThreeSixCorrRelT}{1.00}
\newcommand{\VSQwenThreeSixCorrRelE}{1.00}
\newcommand{\VSQwenThreeSixWrongRelT}{0.226}
\newcommand{\VSQwenThreeSixWrongRelE}{0.043}
\newcommand{\VSQwenThreeSixVacuous}{3}
\newcommand{\VSQwenThreeSixHardcoded}{2}
\newcommand{\VSQwenThreeSixPredFAT}{met}
\newcommand{\VSQwenThreeSixPredFAE}{not met}
\newcommand{\VSQwenThreeSixFalsFAE}{no}
\newcommand{\VSQwenThreeSixFalsCI}{no}
\newcommand{\VSQwenThreeSixFalsified}{not falsified}
\newcommand{\VSQwenThreeSixPredICR}{met}
\newcommand{\VSQwenThreeSixHOneDec}{supported}
\newcommand{\VSQwenThreeSixHTwoDec}{not supported}
\newcommand{\VSQwenThreeSixDescOnly}{no}
\newcommand{\VSLunaResp}{208}
\newcommand{\VSLunaParsed}{203}
\newcommand{\VSLunaParseRate}{0.98}
\newcommand{\VSLunaClean}{199}
\newcommand{\VSLunaWrong}{23}
\newcommand{\VSLunaCorrect}{176}
\newcommand{\VSLunaWrongClusters}{9}
\newcommand{\VSLunaFAT}{0.713}
\newcommand{\VSLunaFAE}{0.481}
\newcommand{\VSLunaDelta}{$-$0.231}
\newcommand{\VSLunaDeltaLow}{$-$0.491}
\newcommand{\VSLunaDeltaHigh}{$-$0.037}
\newcommand{\VSLunaHOneP}{0.250}
\newcommand{\VSLunaHOneHolm}{0.500}
\newcommand{\VSLunaICR}{0.27}
\newcommand{\VSLunaICRn}{15}
\newcommand{\VSLunaICRLow}{0.00}
\newcommand{\VSLunaHTwoP}{0.499}
\newcommand{\VSLunaHTwoHolm}{0.500}
\newcommand{\VSLunaICRCaught}{4}
\newcommand{\VSLunaCorrRelT}{1.00}
\newcommand{\VSLunaCorrRelE}{1.00}
\newcommand{\VSLunaWrongRelT}{0.082}
\newcommand{\VSLunaWrongRelE}{0.058}
\newcommand{\VSLunaVacuous}{7}
\newcommand{\VSLunaHardcoded}{0}
\newcommand{\VSLunaPredFAT}{met}
\newcommand{\VSLunaPredFAE}{not met}
\newcommand{\VSLunaFalsFAE}{yes}
\newcommand{\VSLunaFalsCI}{no}
\newcommand{\VSLunaFalsified}{falsified}
\newcommand{\VSLunaPredICR}{not met}
\newcommand{\VSLunaHOneDec}{descriptive}
\newcommand{\VSLunaHTwoDec}{descriptive}
\newcommand{\VSLunaDescOnly}{yes}
\newcommand{\VSOneGptOssWrong}{43}
\newcommand{\VSOneGptOssFAT}{0.517}
\newcommand{\VSOneGptOssFAE}{0.114}
\newcommand{\VSOneGptOssDelta}{$-$0.403}
\newcommand{\VSOneGptOssHOneHolm}{0.001}
\newcommand{\VSOneGptOssICR}{0.89}
\newcommand{\VSOneGptOssICRn}{9}
\newcommand{\VSOneGptOssICRLow}{0.57}
\newcommand{\VSOneQwenThreeSixWrong}{65}
\newcommand{\VSOneQwenThreeSixFAT}{0.653}
\newcommand{\VSOneQwenThreeSixFAE}{0.113}
\newcommand{\VSOneQwenThreeSixDelta}{$-$0.540}
\newcommand{\VSOneQwenThreeSixHOneHolm}{$<$0.001}
\newcommand{\VSOneQwenThreeSixICR}{0.50}
\newcommand{\VSOneQwenThreeSixICRn}{10}
\newcommand{\VSOneQwenThreeSixICRLow}{0.11}
\newcommand{\VSOneLunaWrong}{23}
\newcommand{\VSOneLunaFAT}{0.713}
\newcommand{\VSOneLunaFAE}{0.481}
\newcommand{\VSOneLunaDelta}{$-$0.231}
\newcommand{\VSOneLunaHOneHolm}{0.500}
\newcommand{\VSOneLunaICR}{0.27}
\newcommand{\VSOneLunaICRn}{15}
\newcommand{\VSOneLunaICRLow}{0.00}
\newcommand{\VSReGptOssExec}{2,137}
\newcommand{\VSReGptOssTOFirst}{0}
\newcommand{\VSReGptOssTOConf}{0}
\newcommand{\VSReGptOssTOCand}{0}
\newcommand{\VSReGptOssDiff}{0}
\newcommand{\VSReGptOssReplaced}{0}
\newcommand{\VSReGptOssKept}{0}
\newcommand{\VSReGptOssChgT}{0}
\newcommand{\VSReGptOssChgE}{0}
\newcommand{\VSReGptOssChgWThree}{0}
\newcommand{\VSReGptOssChgHidden}{0}
\newcommand{\VSReGptOssChgO}{0}
\newcommand{\VSReQwenThreeSixExec}{3,345}
\newcommand{\VSReQwenThreeSixTOFirst}{7}
\newcommand{\VSReQwenThreeSixTOConf}{7}
\newcommand{\VSReQwenThreeSixTOCand}{1}
\newcommand{\VSReQwenThreeSixDiff}{0}
\newcommand{\VSReQwenThreeSixReplaced}{0}
\newcommand{\VSReQwenThreeSixKept}{0}
\newcommand{\VSReQwenThreeSixChgT}{0}
\newcommand{\VSReQwenThreeSixChgE}{0}
\newcommand{\VSReQwenThreeSixChgWThree}{0}
\newcommand{\VSReQwenThreeSixChgHidden}{0}
\newcommand{\VSReQwenThreeSixChgO}{0}
\newcommand{\VSReLunaExec}{2,956}
\newcommand{\VSReLunaTOFirst}{0}
\newcommand{\VSReLunaTOConf}{0}
\newcommand{\VSReLunaTOCand}{0}
\newcommand{\VSReLunaDiff}{0}
\newcommand{\VSReLunaReplaced}{0}
\newcommand{\VSReLunaKept}{0}
\newcommand{\VSReLunaChgT}{0}
\newcommand{\VSReLunaChgE}{0}
\newcommand{\VSReLunaChgWThree}{0}
\newcommand{\VSReLunaChgHidden}{0}
\newcommand{\VSReLunaChgO}{0}
\newcommand{\VSReExec}{8,438}
\newcommand{\VSReTOFirst}{7}
\newcommand{\VSReTOConf}{7}
\newcommand{\VSReTOCand}{1}
\newcommand{\VSReDiff}{0}
\newcommand{\VSReReplaced}{0}
\newcommand{\VSReKept}{0}
\newcommand{\VSPrimary}{v7-001}
\newcommand{\VSSecGptOssUncFAT}{0.460}
\newcommand{\VSSecGptOssUncFAE}{0.097}
\newcommand{\VSSecGptOssUncN}{52}
\newcommand{\VSSecGptOssUncClusters}{24}
\newcommand{\VSSecGptOssORej}{61}
\newcommand{\VSSecGptOssORejN}{94}
\newcommand{\VSSecGptOssORejShare}{0.65}
\newcommand{\VSSecGptOssSqlWrong}{8}
\newcommand{\VSSecGptOssSqlCorrect}{8}
\newcommand{\VSSecGptOssSqlPassT}{5}
\newcommand{\VSSecGptOssSqlPassE}{4}
\newcommand{\VSSecGptOssSqlCorrRelE}{8}
\newcommand{\VSSecQwenThreeSixUncFAT}{0.527}
\newcommand{\VSSecQwenThreeSixUncFAE}{0.098}
\newcommand{\VSSecQwenThreeSixUncN}{78}
\newcommand{\VSSecQwenThreeSixUncClusters}{28}
\newcommand{\VSSecQwenThreeSixORej}{53}
\newcommand{\VSSecQwenThreeSixORejN}{109}
\newcommand{\VSSecQwenThreeSixORejShare}{0.49}
\newcommand{\VSSecQwenThreeSixSqlWrong}{13}
\newcommand{\VSSecQwenThreeSixSqlCorrect}{8}
\newcommand{\VSSecQwenThreeSixSqlPassT}{7}
\newcommand{\VSSecQwenThreeSixSqlPassE}{4}
\newcommand{\VSSecQwenThreeSixSqlCorrRelE}{8}
\newcommand{\VSSecLunaUncFAT}{0.538}
\newcommand{\VSSecLunaUncFAE}{0.348}
\newcommand{\VSSecLunaUncN}{27}
\newcommand{\VSSecLunaUncClusters}{11}
\newcommand{\VSSecLunaORej}{23}
\newcommand{\VSSecLunaORejN}{176}
\newcommand{\VSSecLunaORejShare}{0.13}
\newcommand{\VSSecLunaSqlWrong}{3}
\newcommand{\VSSecLunaSqlCorrect}{21}
\newcommand{\VSSecLunaSqlPassT}{0}
\newcommand{\VSSecLunaSqlPassE}{0}
\newcommand{\VSSecLunaSqlCorrRelE}{21}
\newcommand{\VSSecSqlTasks}{6}

\newcommand{\HRItems}{40}
\newcommand{\HRSampleFlagged}{20}
\newcommand{\HRSampleUnflagged}{20}
\newcommand{\HRReviewers}{three}
\newcommand{\HRAuditRFourNo}{11}
\newcommand{\HRLLMRFourNo}{0}
\newcommand{\HRAuthorRFourNo}{1}
\newcommand{\HRFOneRFourNo}{2}
\newcommand{\HRFTwoRFourNo}{1}
\newcommand{\HRFTwoUnclear}{1}
\newcommand{\HRUpheldAudit}{none}
\newcommand{\HRMajorityNo}{one}
\newcommand{\HRMajorityTask}{py-random-seed-tuple}
\newcommand{\HRMajorityItem}{35}
\newcommand{\HRAnyHumanNo}{3}
\newcommand{\HRAgreeAuthorFOne}{37}
\newcommand{\HRAgreeAuthorFTwo}{38}
\newcommand{\HRAgreeFOneFTwo}{39}
\newcommand{\HRAgreeHumanAuditMin}{27}
\newcommand{\HRAgreeHumanAuditMax}{28}
\newcommand{\HRAgreeHumanLLMMin}{38}
\newcommand{\HRAgreeHumanLLMMax}{39}
\newcommand{\HRFleissKappa}{0.22}
\newcommand{\HRFleissKappaUnclear}{0.37}
\newcommand{\HRLLMFlagged}{2}
\newcommand{\HRAuditFlagged}{20}
\newcommand{\HRLLMTaskAgree}{0.50}
\newcommand{\HRDevFAAll}{0.148}
\newcommand{\HRDevFAExcl}{0.125}
\newcommand{\HRExclChecked}{128}
\newcommand{\HRExclChanged}{none}
\newcommand{\HRExclAnyChanged}{none}
\newcommand{\HRExclLunaCILow}{$-$0.417}
\newcommand{\HRExclLunaCIHigh}{0.000}
\newcommand{\HRExclLunaCIAllHigh}{$-$0.037}
\newcommand{\HRExclLunaHAOneHolmAll}{0.063}
\newcommand{\HRExclLunaHAOneHolm}{0.125}
\newcommand{\HRExclMaxDiff}{0.029}
\newcommand{\HRSolRFourNo}{13}
\newcommand{\HRSolUpheld}{7}
\newcommand{\HRSolAgreeMajority}{28}
\newcommand{\HRSolAgreeLLM}{27}
\newcommand{\HRSolAgreeAudit}{30}
\newcommand{\HRFriendsDoneTime}{2026-11-13 03:39}

\newcommand{\RVHardProgTest}{325}
\newcommand{\RVHardWrongTest}{297}
\newcommand{\RVHardTasksTest}{24}
\newcommand{\RVHardProgDev}{34}
\newcommand{\RVHardWrongDev}{26}
\newcommand{\RVHardTasksDev}{2}
\newcommand{\RVMaxCCDrop}{1}
\newcommand{\RVMaxCCDropN}{456}
\newcommand{\RVFFFreezeHash}{4bf89c190f41}
\newcommand{\RVFFFreezeFiles}{10}
\newcommand{\RVFFFreezeTime}{2026-11-11 20:37}
\newcommand{\RVFFLunaDone}{2026-11-11 21:29}
\newcommand{\RVFFLocalDone}{2026-11-12 12:37}
\newcommand{\RVFFAnalysisTime}{2026-11-12 16:08}
\newcommand{\RVFFAnalysisHash}{5e5351026916}
\newcommand{\RVFFReevalTime}{2026-11-12 23:16}
\newcommand{\RVFFReevalExecHash}{b37c03a400b1}
\newcommand{\RVReviewSampleTime}{2026-11-11 19:23}
\newcommand{\RVLLMReviewTime}{2026-11-12 21:49}
\newcommand{\RVHumanReviewTime}{2026-11-13 02:48}

\begin{abstract}
After an API migration, an LLM-repaired procedure can run cleanly yet return wrong results, and a typical-input release check ships it.  We report a preregistered study of \FTasks{} tasks on \FAffClusters{} documented changes in CPython, Node.js and SQLite.  In the main stages, repairs are choices among documented code fragments, run on the real runtimes against withheld contracts.  Over the complete repair space, the all-assertions gate on typical inputs released \FMFTypFA{} of wrong repairs.  One edge input written from the release note cut this to \FMFEdgeFA{}, and to \FNewFAEdge{} on \FNewTestClusters{} unseen clusters; three inputs from a blind low-cost commercial writer (GPT-5.6-luna) reached \FMFWThreeFA{}.  Two more frozen stages enlarged the menu to \FLTotal{} fragments and added \FHHard{} hard distractors that match legacy output on typical inputs.  The edge input (\FHMFEdgeFA{}) and the GPT-5.6-luna writer (\FHMFWThreeFA{}) passed preregistered tests (two local-model writers did not), yet \FHMFEdgeHard{} and \FHMFWThreeHard{} of hard wrong programs still passed (task means; post hoc: pooled \PVHardPassEPooled{} and \PVHardPassWPooled{}; models rarely chose them).  In all three open-weight families and menus, the edge input cut wrong releases per attempt without materially lowering correct completion; self-tests helped less.  A blind AI audit flagged \FAuditFlagged{} of \FTasks{} contracts (\FAuditNoFour{} as too strict).  Three post-hoc blind human reviewers of strictness (the first author and two non-author software practitioners) upheld none of its \HRAuditRFourNo{} such flags in \HRItems{} sampled contracts; a majority judged one contract (\HRMajorityTask{}) stricter than its requirement.  Dropping disputed checks changed \FPostNoPDChanged{} verdicts; class-label disputes are open.  In a preregistered free-form addendum, models wrote whole programs, run only in a sandbox; for clean-running wrong programs, the edge input cut releases from \VSGptOssFAT{} to \VSGptOssFAE{} for gpt-oss-20b (Holm $p$ \VSGptOssHOneHolm{}) and from \VSQwenThreeSixFAT{} to \VSQwenThreeSixFAE{} for qwen3.6, but only from \VSLunaFAT{} to \VSLunaFAE{} for GPT-5.6-luna (descriptive).
\end{abstract}

\keywords{LLM program repair, API migration, runtime upgrade, release notes, edge-case test inputs, test oracle adequacy, release gate, LLM test generation, accepted-but-wrong repairs, silent behavioral breaking changes, patch overfitting, differential testing, preregistered empirical study, Python 3.12, Node.js 24, SQLite}
\maketitle
\hypersetup{pdfsubject={Software engineering (cs.SE): release gates and test oracles for LLM-based program repair after runtime and API migrations},pdfkeywords={LLM program repair, API migration, runtime upgrade, release notes, edge-case test inputs, test oracle adequacy, release gate, LLM test generation, accepted-but-wrong repairs, silent behavioral breaking changes, patch overfitting, differential testing, preregistered empirical study, Python 3.12, Node.js 24, SQLite, GPT-5.6, Qwen3, gpt-oss}}

\section{Introduction}

When the runtime or API under a business procedure evolves, an approved migration must repair the procedure so that it still meets the same obligation.  Some breakages are loud: a removed function raises an error.  Others are quiet: every call succeeds, yet the result changes.  For example:
\begin{itemize}
  \item an enumeration renders as \texttt{3} instead of \texttt{Priority.HIGH};
  \item \texttt{sum()} returns a more accurate float that no longer matches stored ledger totals;
  \item \texttt{Buffer.poolSize} grows eightfold and silently changes the chunking of every upload.
\end{itemize}
A valid call can still have an invalid effect~\cite{lee2026validtoolcall}.  Agents notice plausible wrong values far less often than explicit errors~\cite{kraishan2026loudquiet}.  In program repair, weak tests let incorrect patches through: the overfitting problem~\cite{smith2015cure,ye2021rgt,aleithan2024swebenchplus}.

Our earlier study~(Section~\ref{sec:prior}) found that a pre-release check fails mainly through its \emph{inputs}.  Checks on the obligation released few wrong repairs once one edge input written from the release note was visible.  Reviewers raised four objections:
\begin{itemize}
  \item the edge inputs were author-written and close to the hidden inputs;
  \item the repair space is a finite menu;
  \item the human contract sign-off never happened;
  \item the test partition had only 34 clusters.
\end{itemize}
This paper answers the first and fourth with new, preregistered experiments.  It answers the third only partly: with an AI audit, and with a post-hoc blind review of one audit criterion (R4, assertion strictness) by three human reviewers, the first author and two non-author software practitioners (Section~\ref{sec:threats}).  Neither is a full human contract sign-off.  It answers the second only partly: a second stage enlarges the menu to a median of \FQMenuMedian{} programs per task, but \FLSmallTasks{} tasks still have fewer than 15 options.  Repairs remain choices among author-written fragments, and that stage shows that the share of wrong programs a gate releases depends on how the menu is built (the preregistered large-menu contrast LM3, Section~\ref{sec:design-lm}).

\textbf{Setting.}  We built \FTasks{} tasks with \FAffected{} affected tasks in \FAffClusters{} documented-change clusters, \FNewClusters{} of them new, on three installed open-source lineages:
\begin{itemize}
  \item CPython 3.9$\to$3.12;
  \item Node.js 20$\to$24;
  \item the SQLite shell 3.51$\to$3.54.
\end{itemize}
Every change is grounded in pinned release-note text.  Every candidate is executed on the real runtimes.  Hidden contracts are never shown to any model.  We compare the visible-input sets that a release gate can use:
\begin{itemize}
  \item typical inputs only;
  \item plus the author's release-note edge input;
  \item plus up to three inputs from a \emph{blind writer}, an isolated LLM that sees only the release note and the visible contract;
  \item plus \emph{self-tests} that the repairing model writes for itself before it repairs.
\end{itemize}
Four model families repair the procedures under equal information: GPT-5.6-luna (a low-cost commercial model, the cost-optimized tier of OpenAI's GPT-5.6 family, called via the Codex CLI)\footnote{OpenAI model page, \url{https://developers.openai.com/api/docs/models/gpt-5.6-luna} (accessed 2026-11-12): ``optimized for cost-sensitive workloads''.}, Qwen3-8B, gpt-oss-20b and qwen3.6-35b-a3b.  A fifth, gemma-4-12b, failed the unchanged admission check and is reported as not admitted.

\textbf{Contributions.}
\begin{enumerate}
  \item \textbf{Exhaustive gate adequacy, replicated on unseen changes.}  Over the complete repair space of \FTestAff{} affected test tasks:
    \begin{itemize}
      \item the all-assertions gate on typical inputs released \FMFTypFA{} of wrong repairs, and one release-note edge input lowered this to \FMFEdgeFA{};
      \item on the \FNewTestClusters{} new test clusters, built after the earlier study, the drop was \FNewFATyp{}$\to$\FNewFAEdge{} (Holm $p$ \FPHMFourCStrict{});
      \item false rejection stayed at \FMFEdgeFR{}.
    \end{itemize}
  \item \textbf{A blind test writer comes close to the author (descriptive), and a preregistered guard fails.}
    \begin{itemize}
      \item Three blind-writer inputs per task reached \FMFWThreeFA{} (\FPHNoOverlapWThreeFA{} without the \FWriterOverlap{} inputs that equal a hidden input), with no increase in false rejection.  This result is descriptive: the unguarded curve was not a preregistered hypothesis, it was known before any repair call, and there is one writer.
      \item The preregistered coverage-aware gate kept only inputs the writer declared legacy-binding.  It reached only \FMFGThreeFA{}, so its preregistered test is not positive.
      \item Writers mislabel change-revealing inputs as non-binding: the label the guard relies on is the label models get wrong.
    \end{itemize}
  \item \textbf{Edge inputs replicate across model families; self-tests do not reach significance.}
    \begin{itemize}
      \item The edge input cut wrong releases per attempt in all three open-weight families (Table~\ref{tab:family}), with no loss of correct completion on this menu.
      \item The preregistered replication criterion is \FEdgeRepMet{}.
      \item GPT-5.6-luna released few wrong repairs in any arm.
      \item Self-tests lowered wrong releases in every family, but no self-test contrast survived Holm correction.  A post-hoc writer $\times$ repairer matrix, restricted to the \FQCommonTasks{} tasks where every writer has tests, is consistent with writer capability mattering more than independence.  Tests from the strongest model we tested (GPT-5.6-luna) gated best, and same-model cells were not worse than cross-model cells.  This is descriptive, with four families and one prompt.
    \end{itemize}
  \item \textbf{A larger repair menu.}  A second frozen stage enlarged the menu from \FLVFour{} to \FLTotal{} documented fragments.  Over that space the edge input (\FLMFEdgeFA{}) and the blind writer (\FLMFWThreeFA{}) both passed their preregistered tests.  The edge-input effect on model repairs replicated in the same three open-weight families.
  \item \textbf{Hard distractors and more writers.}  A third frozen stage added \FHHard{} wrong-looking-right distractors that match the legacy output on typical inputs.  The edge input (\FHMFEdgeFA{}) and the GPT-5.6-luna writer (\FHMFWThreeFA{}) still passed their preregistered tests, and the edge-input effect replicated in all three open-weight families.  Two weaker blind writers did not pass.
  \item \textbf{Independent AI contract audit.}
    \begin{itemize}
      \item Two blind GPT-5.6-luna audits per contract agreed on \FAuditAgree{} of items ($\kappa=\FAuditKappa$).
      \item They flagged \FAuditFlagged{} of \FTasks{} contracts on at least one criterion, \FAuditNoFour{} of them on R4 (an assertion stricter than the requirement), mostly the file-preservation check.  The human review below upheld none of the R4 objections in its sample.
      \item Removing that check, and legacy equality as well, changes \FPostNoPDChanged{} hidden verdicts; the class-label (R2) objections remain open.
      \item We report this as an AI audit, not as a human sign-off.  Three blind human reviewers (the first author and two non-author software practitioners) reviewed the strictness criterion R4 on \HRItems{} sampled contracts: none upheld any of the audit's \HRAuditRFourNo{} R4 rejections, and a majority judged \HRMajorityNo{} contract (\texttt{\HRMajorityTask}) stricter than its requirement.  The review was post hoc and not preregistered; the first author co-wrote the tasks and their contracts and knew the study's results, and the reviewers saw the contracts without the audit's verdicts.  R1--R3 and R5 were reviewed only by a blind LLM (Section~\ref{sec:threats}).
    \end{itemize}
  \item \textbf{Free-form repairs (preregistered addendum).}  Models wrote whole programs, which ran only in a fresh, network-isolated container per execution (Section~\ref{sec:freeform}).
    \begin{itemize}
      \item For gpt-oss-20b (confirmatory), the edge input cut false acceptance among wrong programs that ran cleanly on the typical input from \VSGptOssFAT{} to \VSGptOssFAE{} (Holm $p$ \VSGptOssHOneHolm{}), and it still rejected \VSGptOssICRCaught{} of the \VSGptOssICRn{} wrong programs that passed their own tests.  Both hypotheses are supported, but the predicted false-acceptance rate with the edge input, FA(E) $\le0.10$, was missed.
      \item qwen3.6 replicated the first result (\VSQwenThreeSixFAT{}$\to$\VSQwenThreeSixFAE{}) but not the second.
      \item GPT-5.6-luna's few wrong programs were much harder to catch (\VSLunaFAT{}$\to$\VSLunaFAE{}, descriptive).
    \end{itemize}
\end{enumerate}

\textbf{Scope.}  In the main stages, repairs are choices among documented fragments, and no model-written code is executed.  The free-form addendum executes model-written programs only in a container sandbox; it samples at temperature \VSTemp{}, whereas the main stages decode greedily.

\section{Related Work}

\subsection{Repairing clients after API and dependency evolution}

Behavioral breaking changes occur even without signature changes~\cite{xavier2017breaking,jayasuriya2024behavioral}, and library releases often break interfaces in ways that semantic versioning does not permit~\cite{raemaekers2017semver}.  Migration recommenders recover old-to-new API relations~\cite{wu2010aura,dagenais2009semdiff}, and web-service clients have been repaired from migration patterns~\cite{schmiedmayer2023migration}.

The closest recent work repairs client code after dependency updates:
\begin{itemize}
  \item Fruntke and Krinke fix dependency breaking changes automatically~\cite{fruntke2025fixing}.
  \item Byam combines compilation errors, error localization and API differences with an LLM to fix breaking dependency updates~\cite{reyes2026byam}.
  \item DepRepair observes that the decisive evidence often lies upstream (release notes, API diffs) and that no failing test localizes the break~\cite{yang2026deprepair}.
\end{itemize}
The closest is BBCFixer~\cite{chen2026whobrokeme}.  It targets \emph{behavioral} dependency breaks (BBCBench: 100 Python and JavaScript cases).  It runs the failing test under the old and new library versions, ranks the calls whose return values differ to locate the root API, and filters the library diff accordingly.  Gilesi records API-interaction snapshots during client tests and compares them across library versions to flag behavioral breaking changes~\cite{monce2025gilesi}.  In effect, it applies differential testing at the API boundary.

These systems accept a repair when the client compiles or its own tests pass.  BBCFixer is the exception: it uses old-versus-new execution as \emph{repair evidence}, not as a release criterion.  BBCBench is the closest benchmark: it contains 100 real behavioral dependency breaks in Python and JavaScript projects, each with the failing client test.  We did not evaluate our release gates on it, for three reasons:
\begin{itemize}
  \item its cases require installing project dependencies, which our offline, no-install protocol forbids;
  \item its oracle is the project's own failing test, so the oracle cannot be varied independently of the repair;
  \item its repairs are free-form code, whose space cannot be enumerated, so a gate's false-acceptance rate over all possible repairs is not defined for it.
\end{itemize}
The two designs are complementary.  BBCBench measures whether repair succeeds against the project's test.  Our design asks how often a given release check would let a wrong repair through, which requires oracles that can be swapped.  Over all possible repairs it also requires a finite, enumerable repair space (our fragment stages); over model-written programs it does not, and our free-form addendum measures it there (Section~\ref{sec:freeform}).

Our increment relative to BBCFixer, Gilesi and Byam is fourfold:
\begin{itemize}
  \item we hold the repairer's information fixed and vary only the release gate;
  \item we measure each gate's false-acceptance and false-rejection rates over the complete space of possible repairs and over model-produced repairs;
  \item we treat ``no safe replacement exists'' as a registered outcome;
  \item we compare the sources of the inputs that make a gate adequate: author-written, a blind LLM writer, and the repairing model itself.
\end{itemize}
The release gate is measured against acceptance and preservation contracts that are withheld from the repairer.  Byam's signal (compile errors plus API diffs) is the natural analogue of our runtime/schema gate and diff view.  We therefore do not port its Java-specific pipeline as a baseline; we compare the gate classes it relies on.

Version-aware generation benchmarks are adjacent:
\begin{itemize}
  \item updating model knowledge of API changes~\cite{liu2024codeupdatearena};
  \item generating code for a requested library version~\cite{kuhar2025libevolutioneval,misra2026gitchameleon};
  \item migrating repositories~\cite{liu2026migrationbench,may2026freshbrew}.
\end{itemize}
They evaluate generation or whole-project migration with tests as the oracle.  They do not ask how much a weaker oracle would have let through.

\subsection{Tools that evolve under agents}

\begin{itemize}
  \item Continual documentation adaptation lets agents self-evolve as tools are added, changed or deprecated, by revising tool descriptions~\cite{wu2026contda}.
  \item Drift benchmarks add, rename and deprecate tools and revise their parameter schemas, and measure how task success degrades~\cite{paidi2026tooldriftbench}.
  \item DynamicMCPBench scores effects on live MCP servers~\cite{kaminski2026dynamicmcpbench}.
  \item Kaliyev and Maryanskyy report a gap between in-session verification and held-out conformance for tools that agents synthesize themselves~\cite{kaliyev2026conformancegap}.
  \item ContractRL repairs individual JSON tool calls with contract-masked minimal patches and compares them with full regeneration~\cite{hu2026contractrl}.
\end{itemize}
Bench2Robust augments agent benchmarks with recoverable and unrecoverable failures, including drift and episodes in which no viable path remains~\cite{chen2026bench2robust}.  Kraishan injects typed tool faults and finds that agents flag plausible wrong values far less often than explicit errors~\cite{kraishan2026loudquiet}.  Our setting is different: upstream evolution under a fixed obligation, with the release oracle as the manipulated variable.  We do not train any model.

\subsection{Oracles, contracts and patch overfitting}

Test-based repair produces patches that pass the available tests yet are incorrect.  The problem is known as overfitting~\cite{smith2015cure}, and it motivates patch-correctness assessment beyond the given tests~\cite{ye2021rgt}.  It recurs for LLM agents evaluated on issue-resolution benchmarks such as SWE-bench~\cite{jimenez2024swebench}, whose ``solved'' issues are often accepted by tests too weak to verify the patch~\cite{aleithan2024swebenchplus}.  Our release gates play the role of the given tests.  The migration setting adds two things: a fixed legacy behavior that can serve as an oracle, and inputs whose relevance is named by the change note.

Executable properties and contracts are established ways to state behavior when a full output oracle is unavailable:
\begin{itemize}
  \item property-based and metamorphic testing~\cite{claessen2000quickcheck,segura2016metamorphic};
  \item consumer-driven contract testing~\cite{schwarz2025contracts};
  \item differential regression testing of API versions~\cite{godefroid2020differential}.
\end{itemize}

LLMs now write property-based tests and other tests whose fault-revealing power is assessed, or whose generation is guided, by mutation analysis~\cite{vikram2023pbt,dakhel2024mutap,foster2025ach}.  Structural adequacy is not behavioral adequacy: on ten Java libraries, Paul and Holmes find that a substantial share of documented expected behaviors is never validated, mostly in methods that already have high line coverage and often high mutation kill scores~\cite{paul2026behaviouralgaps}.  A typical-input release gate after a migration is an instance of such a behavioral gap: the documented change is exactly the behavior that no typical input exercises.  BreakGuard uses LLM-generated tests to detect behavioral breaking changes in client code~\cite{raj2026breakguard}.

\textbf{Generated tests as filters, and self-tests.}  CodeT generates tests with the same model and uses agreement with them to rank sampled programs~\cite{chen2023codet}.  Self-debugging lets a model explain and revise its own code from execution feedback~\cite{chen2024selfdebug}.  Both treat model-written tests as a selection signal, and both measure success against held-out tests.  We ask a narrower question that neither isolates.  When a model writes test \emph{inputs} from a release note before repairing, how many of its own wrong repairs does a gate built on those inputs still release?  How does that compare with an independent blind writer and with an author-written edge input?  Our tests are data only.  Expected outputs come from the legacy program on the legacy runtime, so no model-written code or oracle is executed in the fragment stages; the free-form addendum (Section~\ref{sec:freeform}) runs model-written programs in a sandbox and also checks them against the model's own stated outputs.

\textbf{LLM judges as auditors.}  LLM judges agree with human preferences on open-ended answers but show position and self-preference biases~\cite{zheng2023judging}.  We use an LLM to audit contracts against release notes, with two blind replicates and a frozen adjudication rule, and report it as an AI audit.  A post-hoc blind LLM review by a different model and a post-hoc blind review of one criterion by three human reviewers, one of them the first author, check part of that audit (Section~\ref{sec:threats}); neither is a full human contract sign-off.

We use these oracle families as release gates for repairs.  We then report, both over the complete finite repair space and over model-produced repairs, how many wrong repairs each gate lets through and how many correct ones it blocks.  The model-free measurement is the complement of a mutation score computed over the space of possible repairs.

\subsection{Edit formats and over-editing}

Models over-edit when asked for minimal changes, and preservation-aware training or instructions reduce this~\cite{zhu2026edittoomuch}.  Diff and full-rewrite output formats trade off fragility against cost~\cite{cheng2026todiff}.

Our earlier exploratory stage showed that a stated two-field budget can manufacture a representation effect.  Here representation is a secondary, equal-information factor, the budget is unlimited, and edit size is recorded as an outcome.

\section{Problem and Terminology}
\label{sec:problem}

A \emph{task} has five parts:
\begin{itemize}
  \item a consumer requirement $R$;
  \item an old procedure $p_0$ that satisfies $R$ on the old runtime $A_0$;
  \item a finite repair domain $D$ (each procedure is a choice of documented fragments for the template's fields);
  \item a visible input set $X_v$;
  \item a hidden input set $X_h$, disjoint from $X_v$ except where noted.
\end{itemize}
The business obligation stays fixed across the migration.

\begin{definition}[Hidden correctness]
A procedure $p\in D$ is \emph{correct} on the new runtime $A_1$ if, for every $x\in X_h$, two conditions hold:
\begin{itemize}
  \item its output satisfies the acceptance assertions derived from $R$, together with equality to $p_0$ on $A_0$ for the outputs that $R$ constrains;
  \item the files it leaves behind equal those left by $p_0$ (preservation).
\end{itemize}
Hidden contracts are never shown to a repairer.
\end{definition}

\begin{definition}[Release gate and outcomes]
A \emph{release gate} (or \emph{release oracle}) maps a candidate's observations on $X_v$ under $A_1$ to \emph{release} or \emph{block}.  Each attempt ends in exactly one of four outcomes:
\begin{itemize}
  \item \emph{correct completion}: released and correct;
  \item \emph{false acceptance} (FA): released and not correct, that is, a wrong repair passes the gate (FA never means a false alarm in this paper);
  \item \emph{false rejection}: blocked although correct;
  \item \emph{safe non-release}: blocked or abstained, and not correct.
\end{itemize}
A false acceptance is \emph{error-free} when the hidden executions raised no error.
\end{definition}

A \emph{no-safe-repair task} is one in which no $p\in D$ is correct; exhaustive search confirms this.  For such a task the correct disposition is non-release, ideally an explicit abstention.

The ``no safe repair'' label depends on the obligation.  In the earlier study, \VROnlyBoundaryShiftCount{} of its no-safe-repair tasks gain correct programs once equality with the legacy output is dropped and only requirement-level assertions are kept: \VROnlyBoundaryShift{} correct programs (Section~\ref{sec:prior}).

The earlier study compared four gates.  All except ``none'' execute the candidate on $X_v$ and therefore have identical execution cost:
\begin{itemize}
  \item \emph{none};
  \item \emph{runtime/schema}: the program exits normally and prints output of the declared shape and types;
  \item \emph{differential}: runtime/schema plus equality with $p_0$ on $A_0$ for every visible input;
  \item \emph{all consumer assertions}: runtime/schema plus every requirement assertion on every visible input.
\end{itemize}

\emph{Visible-input coverage} is the composition of $X_v$.  The gate on each set is \emph{all consumer assertions} on every input in it:
\begin{itemize}
  \item \emph{T} (typical): the task's typical input or inputs;
  \item \emph{E} (edge input): T plus one author-written edge input derived from the release note;
  \item \emph{W$_k$} (blind writer): T plus the first $k\le 3$ valid inputs of a blind LLM writer that sees only the release note and the visible contract;
  \item \emph{G$_k$} (guarded writer): as W$_k$, but only inputs the writer declares \emph{legacy-binding}, that is, inputs on which the old output is still the required output;
  \item \emph{S} (self-test): T plus the repairing model's own valid legacy-binding inputs, written before it repairs;
  \item \emph{ES}: E plus S.
\end{itemize}
Expected outputs for every added input come from $p_0$ on $A_0$.  Outside the free-form addendum (Section~\ref{sec:freeform}), no model-written code or expected value is ever executed or trusted.  There, model-written programs run only in a container sandbox, and model-stated expected outputs enter only the own-tests verdict, never a gate or the hidden verdict.

Throughout, \emph{silent} refers to the class of the change: calls succeed and the result changes.  \emph{Error-free} refers to an outcome.

\section{Study Design}
\label{sec:design}

The main design was preregistered and frozen twice by SHA-256 hash sets.  The large-menu and hard-distractor stages (Sections~\ref{sec:design-lm} and~\ref{sec:design-hd}) each had their own two freezes, and the free-form addendum (Section~\ref{sec:freeform}) was frozen once, before its first model call; Appendix~\ref{app:freeze} gives the full timeline, including the post-run steps.
\begin{itemize}
  \item \textbf{F0} came before any model call of this study.
  \item \textbf{F-test} came before any repair, self-test or probe call.
\end{itemize}
Each freeze stores a verbatim copy of the design text and hashes its own freeze script.  These are two lessons from the earlier study, whose F0 design text had to be reconstructed afterwards (Appendix~\ref{app:freeze}).

\subsection{Substrate}

\begin{itemize}
  \item \textbf{Runtimes.}  Three installed lineages, executed offline and unchanged since the earlier study: CPython 3.9.6$\to$3.12.15, Node.js 20.20.2$\to$24.21.0 and the SQLite command-line shell 3.51.0$\to$3.54.0.  SQLite 3.54.0 was released on 2026-10-09, after the models' likely training cutoff.
  \item \textbf{Tasks.}  The \FTasks{} tasks are the 55 earlier tasks, imported unchanged, plus \FNewTasks{} new ones.  The new tasks are grounded in verbatim text from already-pinned official pages:
    \begin{itemize}
      \item Python \emph{What's New} 3.10--3.12;
      \item the Node.js deprecation registry and Buffer reference;
      \item the SQLite 3.53/3.54 release logs.
    \end{itemize}
    Examples are the \texttt{random.seed} type restriction, NaN hashing by identity, the PEP~701 tokenizer change, the \texttt{ipaddress} \texttt{is\_private} fix, \texttt{Buffer.poolSize}, and SQLite's right-justified numbers and empty-result headers.
  \item \textbf{Counts.}  \FAffected{} tasks are affected, in \FAffClusters{} clusters (CPython \FCPython{}, Node.js \FNode{}, SQLite \FSQLite{}).  The other \FControls{} are unchanged controls.  By class there are \FAllSilent{} silent, \FAllExplicit{} explicit and \FAllBoundary{} no-safe-repair tasks.
  \item \textbf{Authoring rules.}
    \begin{itemize}
      \item Each new task passes the earlier study's gates unchanged: quote verbatim, determinism, stale failure on hidden inputs, at least one correct fragment choice (none for no-safe-repair tasks), and static safety.
      \item Each has one author edge input that must make the stale procedure fail a visible check: the edge-input QA rule.
      \item Two candidate tasks failed these gates in dry builds and were dropped before F0.
      \item No isomorphic variants were added.
    \end{itemize}
  \item \textbf{Split.}  Earlier clusters keep their earlier partition.  New clusters go 20\% per lineage to development, by hash.  The test partition has \FTestAff{} affected tasks in \FTestClusters{} clusters, \FNewTestClusters{} of them new.
\end{itemize}

\subsection{Visible inputs and gates}

\textbf{Blind writer.}
\begin{itemize}
  \item \textbf{Calls.}  One isolated GPT-5.6-luna call per task (\FWriterTasks{} calls) through the hardened Codex client:
    \begin{itemize}
      \item all MCP servers, plugins, apps and web search are disabled;
      \item the sandbox is read-only, ephemeral, with an empty working directory;
      \item an output schema is enforced;
      \item any use of the model's own tools voids the call.
    \end{itemize}
  \item \textbf{What the writer saw.}  The requirement, the pinned excerpts, the input keys, the typical input with its old output, and the output checks.  It never saw the template, the fragments, the old program, the hidden inputs or any result.
  \item \textbf{Output.}  Up to three inputs, each with a \emph{legacy-binding} flag.
  \item \textbf{Validation} (fail-closed, data only).  An input must have exactly the typical keys and value types.  It must also run cleanly and deterministically under the old procedure on the old runtime, which supplies the expected output.
  \item \textbf{Yield.}  \FWriterKept{} inputs were kept in \FWriterWithKept{} tasks, and \FWriterNonBinding{} of them were flagged non-binding.  \FWriterDropped{} were dropped: \FWriterDropOld{} because the old runtime raised an error, \FWriterDropDup{} as duplicates, and \FWriterDropNondet{} as nondeterministic.  \FWriterOverlap{} kept inputs equal a hidden input; they are kept, as preregistered.
\end{itemize}

\textbf{Coverage-aware gate.}  W$_k$ adds the first $k$ kept inputs.  The preregistered gate G$_k$ adds the first $k$ kept inputs flagged legacy-binding.  The guard was motivated by the earlier study: there, LLM inputs over-rejected on runtime facts that no repair may preserve, such as a module the migration removed.

\textbf{Self-tests.}  Inside each family's run, the repairing model first writes its own inputs with the same prompt and validation (step~0, one call per task).  The S arm uses those flagged legacy-binding.  The unguarded set is analyzed without model calls.

\subsection{Repair protocol, arms and models}

\begin{itemize}
  \item \textbf{Equal information.}  The repair prompt, parser and feedback are the earlier study's, unchanged.  Every arm sees:
    \begin{itemize}
      \item the requirement, the excerpts and the template;
      \item the field domains and the old procedure;
      \item the arm's visible inputs;
      \item ``EDIT BUDGET: unlimited'' and permission to abstain.
    \end{itemize}
  \item \textbf{Arms and cells.}  There are four arms: T, S, E and ES.  Each task and arm has eight cells, crossing patch or full representation, diff view off or on, and listed or reversed fragment order.
  \item \textbf{Steps.}  Step~1 is shared by two branches, \emph{no gate} and \emph{all consumer assertions}.  The gated branch issues one revision if and only if its gate fails, with pass/FAIL feedback and no expected values.
  \item \textbf{Offline gates.}  Every gate W$_1$--W$_3$, G$_1$--G$_3$, S and Sraw is also applied offline to each T-arm step-1 candidate.  Sraw is the unguarded self-test gate: T plus all of the model's valid self-test inputs, including those it flagged non-binding.
  \item \textbf{Models.}
    \begin{itemize}
      \item GPT-5.6-luna via the Codex CLI (reasoning effort high);
      \item Qwen3-8B (4-bit, MLX, greedy);
      \item gpt-oss-20b, qwen3.6-35b-a3b with thinking off, and gemma-4-12b-qat, all in LM Studio at temperature 0.
    \end{itemize}
  \item \textbf{Admission.}  Local families must first pass the unchanged 12-call qualification: 12/12 exact outputs, p5 prefill $\ge$500 tok/s, p5 decode $\ge$32 tok/s, and overhead $\le$0.5\,s.
    \begin{itemize}
      \item Qwen3-8B (\FQwenEightQualExact{}, decode \FQwenEightQualDecode{}), gpt-oss-20b (\FGptOssQualExact{}, \FGptOssQualDecode{}) and qwen3.6-35b-a3b (\FQwenMoeQualExact{}, \FQwenMoeQualDecode{}) were admitted at the first attempt.
      \item gemma-4-12b-qat was \FGemmaAdmitted{}: \FGemmaQualExact{} exact, decode \FGemmaQualDecode{}.  Its answers began with leaked reasoning-channel text (\texttt{<|channel>thought}), a serving-configuration problem rather than evidence about the model's repair ability.  The speed-only retry policy does not cover content failures, so the family was not retried, and we draw no conclusion about gemma.
    \end{itemize}
  \item \textbf{Calls.}  GPT-5.6-luna made \FLunaCalls{} calls, Qwen3-8B \FQwenEightCalls{}, gpt-oss-20b \FGptOssCalls{} and qwen3.6-35b-a3b \FQwenMoeCalls{}.  None was voided.
  \item \textbf{Truncations.}  Answers cut off at the output-token limit are never parsed and count as rejected replies: \FQwenEightTrunc{} for Qwen3-8B, \FGptOssTrunc{} for gpt-oss-20b and \FQwenMoeTrunc{} for qwen3.6-35b-a3b.  The Codex CLI reports no finish reason.
  \item \textbf{Contamination probe.}  Each model answered one step-1 prompt per affected task with the release-note excerpts and diff view removed.
\end{itemize}

\subsection{Hypotheses, decision rules and power}

Statistics are the earlier study's, unchanged:
\begin{itemize}
  \item the unit is the documented-change cluster, with tasks averaged within clusters;
  \item sign-flip permutation $p$-values, exact for $\le$16 clusters and otherwise 200{,}000 Monte Carlo flips;
  \item whole-cluster bootstrap 95\% intervals;
  \item Holm adjustment within each family.
\end{itemize}
Rule thresholds use task means.  \emph{Holm and zero discordance.}  Our frozen analysis code gave a hypothesis without discordant clusters no $p$-value and left it out of the Holm count.  The frozen design text says ``Holm over 3'' and ``over 4''.  We therefore report the strict values, with every hypothesis in $m$ and a zero-discordance hypothesis entering with $p=1$: M4c then has Holm $p$ \FPHMFourCStrict{} (code: \FMFourCHolm{}) and gpt-oss-20b's H2 \FPHGptOssHTwoStrict{} (code: \FGptOssHTwoHolm{}).  No decision changes; tables show the code's values.

\textbf{Family M4 (model-free, complete repair space, Holm over 3).}
\begin{itemize}
  \item \textbf{M4a}, FA(G$_3$)$-$FA(T).  Positive if Holm $p<0.05$, FA(G$_3$) $\le0.10$ and the rise in false rejection is $\le0.05$.
  \item \textbf{M4b}, FR(G$_3$)$-$FR(W$_3$).  Positive if the guard significantly lowers false rejection.
  \item \textbf{M4c}, FA(E)$-$FA(T) on the \FNewTestClusters{} new test clusters only.  This is a fresh replication of HM1, the earlier study's model-free edge-input contrast FA(E)$-$FA(T), with the same thresholds as M4a.
\end{itemize}

\textbf{Family A4 (per model, Holm over 4; per-attempt rates averaged over the eight cells).}
\begin{itemize}
  \item H1: FA(E)$-$FA(T);
  \item H2: FA(S)$-$FA(T);
  \item H3: FA(ES)$-$FA(E);
  \item H4: CC(S)$-$CC(T).
\end{itemize}
H1--H3 are \emph{positive} if Holm $p<0.05$, the difference is negative, and the matching correct-completion difference is $\ge-0.05$.

\textbf{Power and confirmatory status.}
\begin{itemize}
  \item A preregistered simulation resampled the earlier per-cluster rates with eight attempts and the counterfactual effect $r=0.3$.  The frozen design text states the 400-simulation pilot values 0.99, 0.91 and 0.03 for Qwen3-8B, gpt-oss-20b and GPT-5.6-luna; the confirmatory status follows these.  The 1{,}000-simulation record of the same script gives \FPowerQwenEight{}, \FPowerGptOss{} and \FPowerLuna{}.  The first two families are confirmatory; GPT-5.6-luna, near the ceiling of few wrong releases, is descriptive.
  \item The two new families were confirmatory only if their T-arm false-acceptance rate on the \emph{development} partition was at least 0.15.  qwen3.6-35b-a3b had \FQwenMoeDevFA{} and is therefore descriptive.
  \item The 1{,}000-simulation power record was produced after F-test from the frozen script.  This is logged as a deviation.
\end{itemize}

\textbf{Cross-family criteria.}
\begin{itemize}
  \item Edge-input replication: H1 positive in at least three families and in every confirmatory family.
  \item Self-test replication: H2 positive in at least three families.
  \item Model-scale analysis (descriptive): Spearman $\rho$ between capability (T-arm correct completion) and the H1 effect, with an exact permutation $p$.
  \item The thresholds of three of five families were fixed when five families were planned.  With gemma not admitted, they are applied to four.
\end{itemize}

\textbf{AI contract audit.}
\begin{itemize}
  \item \textbf{Calls.}  Two blind GPT-5.6-luna calls per task (\FAuditCalls{} calls).
  \item \textbf{What the auditor saw.}  The requirement, the excerpts, the template, the fragments, the old procedure and the claimed class.  It also saw the contract rules instantiated on the typical input.  It never saw hidden inputs, outputs or results.
  \item \textbf{Rubric.}
    \begin{itemize}
      \item R1: the material documents the change;
      \item R2: the class is right;
      \item R3: legacy equality is a reasonable reading;
      \item R4: no assertion is stricter than the requirement;
      \item R5: the contract is complete.
    \end{itemize}
  \item \textbf{Adjudication.}  An item counts as a ``no'' only if both replicates say no.  Contracts are never changed after the audit.
\end{itemize}

\subsection{Second stage: a large repair menu}
\label{sec:design-lm}

The main external-validity objection is the finite menu: with about five fragments per field, a model can often eliminate wrong options by inspection.  A second, separately frozen stage (F0 \FLFreezeZero{}, F-test \FLFreezeTest{}) therefore enlarges every field domain.
\begin{itemize}
  \item \textbf{Fragments.}  The menu grows from the \FLVFour{} earlier fragments to \FLTotal{}.  Every added fragment comes from an earlier fragment by one or two author-written operators:
    \begin{itemize}
      \item a documented sibling of a module attribute (for example another \texttt{inspect} or \texttt{crypto} function);
      \item a swap among documented builtins, string methods, operators, SQLite functions or CLI modes;
      \item an integer-literal edit.
    \end{itemize}
    Every operator is checked against one of \FLPins{} newly pinned official pages.  Siblings that could act outside the sandbox, start processes or import by name are excluded.
  \item \textbf{Filters.}  A fragment is kept only if the program runs cleanly on the typical input on either runtime (a visible-only filter).  The step-1 prompt must stay within the frozen 2{,}048-token input cap.
  \item \textbf{Coverage.}  \FLSmallTasks{} tasks still have fewer than 15 options, because few documented operators apply to their fragments.  All \FLProcedures{} procedures were evaluated exhaustively.
  \item \textbf{Runs.}  Arms T and E, eight cells each, with the same models, prompts and gates.  The writer's inputs and each family's own self-tests are applied offline.  GPT-5.6-luna made \FLLunaCalls{} calls, Qwen3-8B \FLQwenEightCalls{}, gpt-oss-20b \FLGptOssCalls{} and qwen3.6-35b-a3b \FLQwenMoeCalls{}.  Each local family passed the unchanged qualification again at the first attempt: Qwen3-8B \FQQwenEightQualExact{}, decode \FQQwenEightQualDecode{}; gpt-oss-20b \FQGptOssQualExact{}, \FQGptOssQualDecode{}; qwen3.6-35b-a3b \FQQwenMoeQualExact{}, \FQQwenMoeQualDecode{}.  Truncated answers, never parsed: \FQQwenEightTrunc{}, \FQGptOssTrunc{} and \FQQwenMoeTrunc{}.
  \item \textbf{Hypotheses} (strict Holm over 3 per family, fixed at F0).
    \begin{itemize}
      \item Model-free: LM1 FA(E)$-$FA(T); LM2 FA(W$_3$)$-$FA(T), which makes the earlier descriptive writer result a preregistered test; LM3 FA(T) on the large menu minus FA(T) on the earlier menu.
      \item Per family: LH1 wrong releases per attempt E$-$T; LH2 and LH3, the change in wrong releases and in correct completion with typical inputs relative to the earlier menu.
      \item Thresholds and confirmatory status are as before.  The model-free family was computed between the two freezes, before any model call.
    \end{itemize}
\end{itemize}

\subsection{Third stage: hard distractors and two more writers}
\label{sec:design-hd}

A third frozen stage (F0 \FHFreezeZero{}, F-test \FHFreezeTest{}) addresses the menu-composition problem that LM3 exposed (Section~\ref{sec:results-lm}).
\begin{itemize}
  \item \textbf{Hard distractors.}  The same documented operators generate candidates from every large-menu fragment.  A candidate is kept as a \emph{hard} distractor only if, on the typical inputs, the program's output on the new runtime equals the old program's output on the old runtime.  This is visible information only: the fragment passes the visible differential check.
    \begin{itemize}
      \item \FHHard{} hard fragments were added in \FHHardTasks{} tasks, giving \FHFragments{} fragments and \FHProcedures{} procedures, all evaluated exhaustively.
      \item \FHHardWrong{} of the \FHHardPrograms{} hard-origin programs, over both partitions, turned out hidden-wrong.  The test partition holds \RVHardProgTest{} of them in \RVHardTasksTest{} tasks, \RVHardWrongTest{} hidden-wrong; the development partition holds \RVHardProgDev{} in \RVHardTasksDev{} tasks, \RVHardWrongDev{} hidden-wrong.  Results below use the \RVHardWrongTest{} test-partition hard wrong programs.
    \end{itemize}
  \item \textbf{Writers v2.}  qwen3.6-35b-a3b and Qwen3-8B write up to three inputs each with a new prompt.  It has no legacy-binding flag and asks for inputs that exercise the documented change and its boundary values.  They see the same information as before.  Both passed the qualification (\FHWriterMoeExact{}, decode \FHWriterMoeDecode{}; \FHWriterEightExact{}, \FHWriterEightDecode{}).
  \item \textbf{Runs.}  Arms T and E, eight cells each.  GPT-5.6-luna made \FHLunaCalls{} calls, Qwen3-8B \FHQwenEightCalls{} (decode \FHQwenEightDecode{}), gpt-oss-20b \FHGptOssCalls{} (\FHGptOssDecode{}) and qwen3.6-35b-a3b \FHQwenMoeCalls{} (\FHQwenMoeDecode{}).  Truncated answers: \FHQwenEightTrunc{}, \FHGptOssTrunc{} and \FHQwenMoeTrunc{}.
  \item \textbf{Hypotheses} (strict Holm, fixed at F0).
    \begin{itemize}
      \item HD1 FA(E)$-$FA(T), HD2 FA(W$_3$)$-$FA(T) and HD3 FA(T, hard)$-$FA(T, large), over the complete hard menu.
      \item HA1 E$-$T and HA2 FA$_T$(hard)$-$FA$_T$(large), per repair family.
      \item WR1 and WR2, each new writer's X$_3$ gate against T on the hard menu (X$_3$ is T plus the first three valid inputs of that writer, defined like W$_3$), with the same thresholds as HD1.  WR3 compares the qwen3.6 writer with qwen3.6's own earlier self-tests on the first menu.
    \end{itemize}
    The model-free family HD was computed between the two freezes, before any model call.
\end{itemize}

\textbf{What was known at each freeze.}
\begin{itemize}
  \item M4c depends only on the frozen build.  It was computed during the mock-pipeline test before F0 and is reported as known at F0.
  \item M4a, M4b and the audit were computed between F0 and F-test, as designed.
  \item One amendment followed, before any repair call.  It added offline, descriptive self-test gates and changed no primary rule (Appendix~\ref{app:freeze}).  This concerns the main stage; the later stages' freezes and the free-form addendum's post-run steps are listed in the same appendix.
\end{itemize}

\section{Results}
\label{sec:results}
\sloppy

\subsection{Model-free adequacy over the complete repair space}

\begin{table}[t]
\caption{Model-free release-gate adequacy over the complete finite repair space of the 52 affected test tasks (49 clusters), all-consumer-assertions gate.  Added: mean visible inputs added to the typical ones.  FA: share of hidden-wrong procedures released; FR: share of hidden-correct procedures blocked (task means).  Stale: tasks in which the unrepaired procedure fails visibly.}
\label{tab:curves}
\small
\begin{tabular}{@{}lrrrr@{}}
\toprule
Visible inputs & Added & FA & FR & Stale \\
\midrule
typical only & 0.00 & 0.410 & 0.010 & 34/52 \\
+ blind writer, k=1 & 0.96 & 0.130 & 0.010 & 49/52 \\
+ blind writer, k=2 & 1.85 & 0.075 & 0.010 & 50/52 \\
+ blind writer, k=3 & 2.46 & 0.070 & 0.010 & 50/52 \\
+ guarded writer, k=1 & 0.73 & 0.206 & 0.010 & 42/52 \\
+ guarded writer, k=2 & 1.40 & 0.161 & 0.010 & 42/52 \\
+ guarded writer, k=3 & 1.85 & 0.157 & 0.010 & 42/52 \\
+ author edge input & 1.00 & 0.065 & 0.010 & 52/52 \\
\bottomrule
\end{tabular}
\end{table}

\begin{figure*}[t]
\centering
\includegraphics[width=\textwidth]{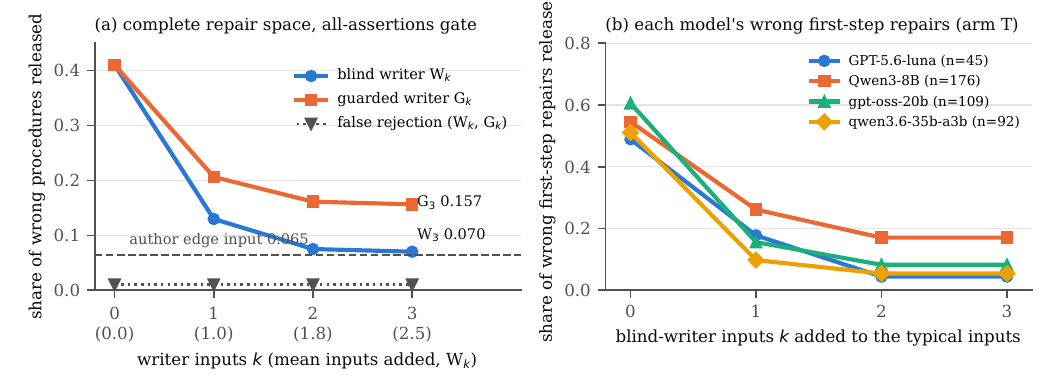}
\Description{Two line charts.  Left: share of wrong procedures released over the complete repair space falls steeply as unguarded blind-writer inputs are added and reaches about the level of the author edge input, shown as a dashed line, while the guarded writer's curve stays higher.  Right: the same gates applied to each model's own wrong first-step repairs show the same ordering.}
\caption{Headline result.  (a) Share of wrong procedures released over the complete finite repair space as blind-writer inputs are added ($k=0$ is typical inputs only; in parentheses, mean inputs actually added per task).  The dashed line marks the author edge input.  Unguarded writer inputs (W$_k$) reach the author level at no cost in false rejection; the preregistered guard (G$_k$) does not.  (b) The same gates applied offline to each model's own hidden-wrong first-step repairs (arm T, test partition; $n$ wrong candidates).  Writer cost: one call per task.}
\label{fig:kcurve}
\end{figure*}

Table~\ref{tab:curves} and Figure~\ref{fig:kcurve}a apply the all-assertions gate to every procedure in the finite domain of each affected test task.  They use the visible inputs of each level.  Tables report task means.  The preregistered tests use cluster means, for example \FMFTypCFA{} and \FMFEdgeCFA{} for T and E; the two never differ by more than 0.005.

\textbf{Author edge input.}
\begin{itemize}
  \item One release-note edge input lowered the share of wrong procedures released from \FMFTypFA{} to \FMFEdgeFA{}.  The cluster-level difference is \FHMAllDiff{}, with \FHMAllDisc{} discordant clusters.  False rejection stayed at \FMFEdgeFR{}.
  \item \textbf{M4c} restricts the contrast to the \FNewTestClusters{} new test clusters, which did not exist when the effect was first observed.  There the rate fell from \FNewFATyp{} to \FNewFAEdge{} (difference \FMFourCDiff{}, 95\% CI [\FMFourCLow, \FMFourCHigh], strict Holm $p$ \FPHMFourCStrict{}).  The decision is \emph{\FMFourCDecision{}}.
  \item M4c is a deterministic function of the build and was known at F0 (Section~\ref{sec:design}).
\end{itemize}

\textbf{Blind writer (descriptive).}  The unguarded writer curve was not a preregistered hypothesis, its values were known before any repair call, and one model wrote all inputs.
\begin{itemize}
  \item Without seeing the program, the fragments or the hidden inputs, the writer's inputs lowered the rate to \FMFWOneFA{} with one input and \FMFWThreeFA{} with up to three.  On average they added \FMFWThreeAdded{} inputs per task.
  \item This is close to the author edge input (\FMFEdgeFA{}), and false rejection did not rise at any $k$.
  \item \textbf{Overlap and similarity (post hoc).}  \FWriterOverlap{} kept writer inputs equal a hidden input, in \FPHNoOverlapTasks{} tasks.  Without them, W$_1$--W$_3$ reach \FPHNoOverlapWOneFA{}, \FPHNoOverlapWTwoFA{} and \FPHNoOverlapWThreeFA{}.  The overlapping inputs carried part of the effect, and the unguarded writer without them remains below the 0.10 ceiling but above the author input.  The mean highest string similarity of the \FPHWriterSimN{} kept writer inputs of the test tasks to a hidden input of the same task is \FPHWriterSimMean{} (median \FPHWriterSimMedian{}; \FPHWriterSimHigh{} at least 0.9), close to that of the author edge inputs of the earlier study (\VPHEdgeRatioMean{}).
  \item The stale procedure fails visibly in \FMFWThreeStale{} of \FTestAff{} tasks at W$_3$.  The writer was not bound by the QA rule that forced \FMFEdgeStale{}/\FTestAff{} for the author input.
\end{itemize}

\textbf{The preregistered guard backfired.}
\begin{itemize}
  \item G$_3$ kept only inputs the writer declared legacy-binding, and reached \FMFGThreeFA{}.  The contrast with T is large (\FMFourADiff{}, Holm $p$ \FMFourAHolm{}).  But the rate exceeds the preregistered ceiling of 0.10, so \textbf{M4a} is \emph{\FMFourADecision{}}.
  \item \textbf{M4b} has \FMFourBDisc{} discordant clusters: the guard removed no false rejection, because the unguarded inputs caused none.
  \item The writer marked \FWriterNonBinding{} of its \FWriterKept{} inputs as non-binding.  These were disproportionately the change-revealing ones: the stale procedure fails visibly in \FMFGThreeStale{} tasks under G$_3$ against \FMFWThreeStale{} under W$_3$.
  \item The judgment ``the old output is no longer required here'' is exactly where the release note is most relevant.  It is also where the writer erred.
\end{itemize}

\begin{table}[t]
\caption{Post-hoc robustness of the model-free effects (affected test tasks; FA task means; cluster-level differences with unadjusted sign-flip $p$).}
\label{tab:robust}
\small
\begin{tabular}{@{}lrrrrr@{}}
\toprule
Subset & FA T & FA E & FA W$_3$ & E$-$T & W$_3-$T \\
\midrule
class S (23 tasks) & 0.571 & 0.080 & 0.091 & $-$0.503 (<0.001) & $-$0.491 (<0.001) \\
class E (25 tasks) & 0.297 & 0.053 & 0.055 & $-$0.235 (<0.001) & $-$0.233 (<0.001) \\
class B (4 tasks) & 0.192 & 0.050 & 0.050 & $-$0.142 (0.500) & $-$0.142 (0.500) \\
without CPython (16 tasks) & -- & -- & -- & $-$0.328 (0.002) & $-$0.368 (0.002) \\
without Node.js (42 tasks) & -- & -- & -- & $-$0.321 (<0.001) & $-$0.313 (<0.001) \\
without SQLite (46 tasks) & -- & -- & -- & $-$0.367 (<0.001) & $-$0.346 (<0.001) \\
\bottomrule
\end{tabular}
\end{table}

\textbf{Robustness by class and lineage (post hoc).}  Table~\ref{tab:robust} splits the model-free contrasts by class and leaves one lineage out.  The edge-input and writer effects hold for silent and explicit changes and in every leave-one-lineage-out subset.  The four no-safe-repair test tasks are too few for a test.

\subsection{Repairs by four model families}

\begin{table*}[t]
\caption{Model repairs on the test partition under the all-consumer-assertions gate: correct completions / false acceptances / false rejections per arm (8 cells per task and arm; 456 records per arm).  T typical inputs; S + the model's own self-tests; E + the author edge input; ES both.  Last column: false acceptances with no gate (arm T).}
\label{tab:arms}
\small
\begin{tabular}{@{}lrrrrr@{}}
\toprule
Model & T & S & E & ES & no gate \\
\midrule
GPT-5.6-luna & 398/22/0 & 403/9/0 & 408/7/0 & 409/0/0 & 45 \\
Qwen3-8B & 268/98/0 & 271/80/0 & 270/24/0 & 268/20/0 & 182 \\
gpt-oss-20b & 320/71/0 & 327/47/1 & 353/8/0 & 351/8/0 & 111 \\
qwen3.6-35b-a3b & 354/49/0 & 363/30/0 & 368/7/0 & 379/8/0 & 97 \\
\bottomrule
\end{tabular}
\end{table*}

\begin{table*}[t]
\caption{Family A4 per model (test clusters; cluster means of per-attempt rates): difference and Holm-adjusted $p$.  H1 FA(E)$-$FA(T); H2 FA(S)$-$FA(T); H3 FA(ES)$-$FA(E); H4 CC(S)$-$CC(T).}
\label{tab:family}
\small
\begin{tabular}{@{}llrrrr@{}}
\toprule
Model & Status & H1 & H2 & H3 & H4 \\
\midrule
GPT-5.6-luna & descriptive & $-$0.038 (0.499) & $-$0.033 (0.749) & $-$0.018 (1.000) & 0.013 (1.000) \\
Qwen3-8B & confirmatory & $-$0.176 (<0.001) & $-$0.046 (0.749) & $-$0.010 (1.000) & 0.010 (0.998) \\
gpt-oss-20b & confirmatory & $-$0.161 (<0.001) & $-$0.064 (0.088) & 0.000 (--) & 0.023 (0.385) \\
qwen3.6-35b-a3b & descriptive & $-$0.102 (0.004) & $-$0.048 (0.256) & 0.003 (1.000) & 0.023 (0.256) \\
\bottomrule
\end{tabular}
\end{table*}

Tables~\ref{tab:arms} and~\ref{tab:family} report the test-partition repairs (\FLunaN{} attempts per arm and model).

\textbf{Edge input (H1).}
\begin{itemize}
  \item Wrong releases per attempt under all assertions fell as follows:
    \begin{itemize}
      \item Qwen3-8B: \FQwenEightHOneB{}$\to$\FQwenEightHOneA{} (CI [\FQwenEightHOneLow, \FQwenEightHOneHigh], Holm $p$ \FQwenEightHOneHolm{});
      \item gpt-oss-20b: \FGptOssHOneB{}$\to$\FGptOssHOneA{} (Holm $p$ \FGptOssHOneHolm{});
      \item qwen3.6-35b-a3b: \FQwenMoeHOneB{}$\to$\FQwenMoeHOneA{} (Holm $p$ \FQwenMoeHOneHolm{}).
    \end{itemize}
  \item All three decisions are positive.  For the two confirmatory families they are confirmatory, and for qwen3.6-35b-a3b descriptive.
  \item Correct completion did not fall: the E$-$T difference was $\ge0$ in every family.
  \item GPT-5.6-luna released \FLunaTFA{} wrong repairs in arm T and \FLunaEFA{} in arm E.  This contrast is descriptive, with Holm $p$ \FLunaHOneHolm{}.
  \item The preregistered edge-input replication criterion is therefore \emph{\FEdgeRepMet{}}.
\end{itemize}

\textbf{Self-tests (H2, H3).}
\begin{itemize}
  \item Self-tests lowered wrong releases in every family: \FLunaTFA{}$\to$\FLunaSFA{}, \FQwenEightTFA{}$\to$\FQwenEightSFA{}, \FGptOssTFA{}$\to$\FGptOssSFA{} and \FQwenMoeTFA{}$\to$\FQwenMoeSFA{} out of \FLunaN{} attempts each.
  \item None of these reductions survived Holm adjustment.  The smallest Holm $p$ was \FPHGptOssHTwoStrict{} (strict; code \FGptOssHTwoHolm{}), for gpt-oss-20b.
  \item The self-test replication criterion is \emph{\FSelfRepMet{}}.
  \item Adding self-tests to the edge input (H3) changed little: ES released \FQwenEightBothFA{}, \FGptOssBothFA{} and \FQwenMoeBothFA{} wrong repairs in the open-weight families, against \FQwenEightEFA{}, \FGptOssEFA{} and \FQwenMoeEFA{} in E.
  \item Self-tests cost one extra call per task.  They did not reduce correct completion: H4 differences were \FQwenEightHFourDiff{} to \FGptOssHFourDiff{}.
\end{itemize}

\begin{table*}[t]
\caption{Self-tests written by each model from the release note and the visible contract: tasks with at least one valid test, valid tests, tests declared legacy-binding, tests identical to a hidden input, mean highest similarity to a hidden input, and the model-free FA of the gate built from the guarded (S) and unguarded (raw) self-tests, with the unguarded gate's FR.}
\label{tab:selftests}
\small
\begin{tabular}{@{}lrrrrrrrr@{}}
\toprule
Model & Tasks & Tests & Binding & =hidden & Sim. & FA (S) & FA (raw) & FR (raw) \\
\midrule
GPT-5.6-luna & 71/73 & 187 & 153 & 5 & 0.69 & 0.153 & 0.054 & 0.031 \\
Qwen3-8B & 30/73 & 49 & 27 & 2 & 0.71 & 0.355 & 0.240 & 0.031 \\
gpt-oss-20b & 57/73 & 128 & 93 & 1 & 0.69 & 0.235 & 0.149 & 0.031 \\
qwen3.6-35b-a3b & 67/73 & 168 & 122 & 3 & 0.68 & 0.215 & 0.100 & 0.031 \\
\bottomrule
\end{tabular}
\end{table*}

\begin{table}[t]
\caption{Offline gates applied to the hidden-wrong first-step candidates of arm T (test partition): share released by the blind-writer gate with 1 and 3 inputs (W1, W3), the guarded writer gate (G3), and the model's own unguarded (Sraw) and guarded (S) self-tests; last column: share of hidden-correct candidates blocked by Sraw.  All other gates blocked no correct candidate.}
\label{tab:step1}
\small
\begin{tabular}{@{}lrrrrrrr@{}}
\toprule
Model & wrong & W1 & W3 & G3 & Sraw & S & FR Sraw \\
\midrule
GPT-5.6-luna & 45 & 0.178 & 0.044 & 0.178 & 0.044 & 0.178 & 0.023 \\
Qwen3-8B & 176 & 0.261 & 0.170 & 0.261 & 0.330 & 0.455 & 0.038 \\
gpt-oss-20b & 109 & 0.156 & 0.083 & 0.266 & 0.339 & 0.376 & 0.030 \\
qwen3.6-35b-a3b & 92 & 0.098 & 0.054 & 0.337 & 0.326 & 0.337 & 0.026 \\
\bottomrule
\end{tabular}
\end{table}

\textbf{Why self-tests helped less.}
\begin{itemize}
  \item \textbf{Fewer valid tests from smaller models.}  The models differed most in how many valid tests they wrote (Table~\ref{tab:selftests}).  GPT-5.6-luna wrote valid tests for \FLunaSelfTasks{} of \FTasks{} tasks; Qwen3-8B managed only \FQwenEightSelfTasks{}.
  \item \textbf{Weaker gates from those tests.}  Over the complete repair space, a gate built from Qwen3-8B's unguarded self-tests released \FQwenEightMFRawFA{} of wrong procedures, against \FMFWThreeFA{} for the blind writer.
  \item \textbf{Weakest exactly where needed.}  The weakness concentrates on each model's own mistakes (Table~\ref{tab:step1}).  Gates were applied offline to the hidden-wrong first-step candidates of arm T:
    \begin{itemize}
      \item the blind writer's gate W$_3$ released \FQwenEightStepWThreeFA{} of Qwen3-8B's wrong candidates and \FGptOssStepWThreeFA{} of gpt-oss-20b's;
      \item the models' own unguarded tests released \FQwenEightStepSelfRawFA{} and \FGptOssStepSelfRawFA{}.
    \end{itemize}
    A model that misreads a release note tends to write tests that agree with its misreading.
  \item \textbf{Matched on test volume (post hoc).}  Restricting to tasks where the family wrote at least one valid self-test removes the volume difference.  On Qwen3-8B's \FPHQwenEightMatchedTasks{} such tasks its own unguarded tests released \FPHQwenEightMatchedStepSraw{} of its \FPHQwenEightMatchedStepN{} wrong first-step repairs, against \FPHQwenEightMatchedStepWThree{} for the blind writer.  For gpt-oss-20b the figures are \FPHGptOssMatchedStepSraw{} against \FPHGptOssMatchedStepWThree{}, and for qwen3.6-35b-a3b \FPHQwenMoeMatchedStepSraw{} against \FPHQwenMoeMatchedStepWThree{}.  For GPT-5.6-luna, which is also the blind writer, they are equal (\FPHLunaMatchedStepSraw{}).
  \item \textbf{Writer or independence? (post hoc, descriptive).}  Each family's self-test step used exactly the blind-writer prompt and validation.  Its unguarded self-tests are therefore that family's blind-writer inputs, and every family's tests can gate every family's repairs without new calls.
    \begin{itemize}
      \item \textbf{Coverage differs.}  The number of tasks with at least one valid test differs by writer, from 23 for Qwen3-8B to 52 for GPT-5.6-luna, and tasks without tests fall back to typical inputs.  The full matrix (Table~\ref{tab:matrix}) therefore mixes coverage with test quality.
      \item \textbf{Common tasks.}  Table~\ref{tab:matrixcommon} restricts the matrix to the \FQCommonTasks{} tasks where every writer has tests.  There the GPT-5.6-luna writer run gates every family best or tied for best.  The mean same-model cell is \FQDiagCommon{} against \FQOffCommon{} for cross-model cells.
      \item \textbf{Reading.}  This is consistent with writer capability mattering more than independence, and gives no sign that a model's own tests are specifically blind to its own errors.  It rests on four families, one prompt and few wrong repairs per cell, so we treat it as a hypothesis.
    \end{itemize}
  \item \textbf{The guard hurt again.}  The guard (S, G$_3$) again let more wrong candidates through than its unguarded counterpart in every open-weight family.
\end{itemize}

\begin{figure}[t]
\centering
\includegraphics[width=\columnwidth]{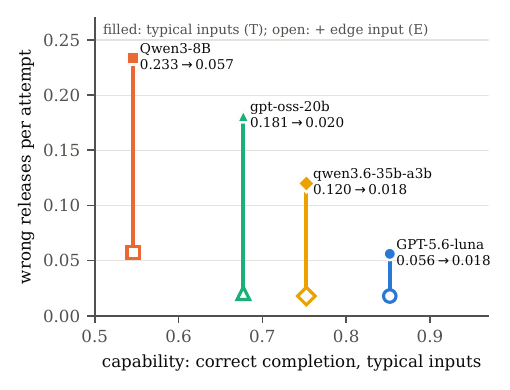}
\Description{Scatter plot of wrong releases per attempt against model capability for four model families.  Filled points (typical inputs) fall as capability rises; open points (with the edge input) lie near zero for every family.}
\caption{Wrong releases per attempt under all assertions with typical inputs (filled) and with the author edge input (open), against each family's capability (test partition, cluster means).}
\label{fig:scale}
\end{figure}

\begin{table*}[t]
\caption{Post-hoc writer $\times$ repairer matrix (zero calls).  Each family's self-test step used exactly the blind-writer prompt and validation, so its unguarded self-tests are that family's blind-writer inputs.  Cells: share of the repairer's hidden-wrong first-step candidates (arm T, test partition) released by the all-assertions gate on typical inputs plus the writer's inputs.  Tasks: affected test tasks with $\ge$1 valid input.  Last column: model-free FA over the complete repair space on the 21 tasks where every writer has inputs (typical only: 0.510).}
\label{tab:matrix}
\footnotesize
\begin{tabular}{@{}lrrrrrr@{}}
\toprule
Writer $\downarrow$ / repairer $\rightarrow$ & Tasks & GPT-5.6-luna & Qwen3-8B & gpt-oss-20b & qwen3.6-35b-a3b & model-free \\
\midrule
GPT-5.6-luna (writer-001) & 50 & 0.044 & 0.170 & 0.083 & 0.054 & 0.069 \\
GPT-5.6-luna & 52 & 0.044 & 0.176 & 0.128 & 0.065 & 0.075 \\
Qwen3-8B & 23 & 0.356 & 0.330 & 0.385 & 0.337 & 0.109 \\
gpt-oss-20b & 42 & 0.178 & 0.330 & 0.339 & 0.272 & 0.088 \\
qwen3.6-35b-a3b & 48 & 0.311 & 0.210 & 0.202 & 0.326 & 0.078 \\
\bottomrule
\end{tabular}
\end{table*}

\begin{table*}[t]
\caption{Post-hoc writer $\times$ repairer matrix restricted to the 21 affected test tasks where every writer has at least one valid input (removes the coverage difference; 30--70 wrong first-step repairs per repairer).  Cells: share of the repairer's hidden-wrong first-step candidates (arm T) released by typical inputs plus the writer's inputs.  Tasks: tasks with $\ge$1 valid input overall.  Descriptive.}
\label{tab:matrixcommon}
\footnotesize
\begin{tabular}{@{}lrrrrr@{}}
\toprule
Writer \textbackslash{} repairer & Tasks & GPT-5.6-luna & Qwen3-8B & gpt-oss-20b & qwen3.6-35b-a3b \\
\midrule
GPT-5.6-luna (writer-001) & 50 & 0.000 & 0.029 & 0.000 & 0.000 \\
GPT-5.6-luna & 52 & 0.000 & 0.157 & 0.125 & 0.021 \\
Qwen3-8B & 23 & 0.400 & 0.143 & 0.179 & 0.106 \\
gpt-oss-20b & 42 & 0.200 & 0.257 & 0.268 & 0.021 \\
qwen3.6-35b-a3b & 48 & 0.400 & 0.157 & 0.214 & 0.234 \\
\bottomrule
\end{tabular}
\end{table*}

\textbf{Model scale} (Figure~\ref{fig:scale}).
\begin{itemize}
  \item Across the four admitted families, the absolute edge-input effect shrank as capability grew: Spearman $\rho=\FScaleRhoAbs$, exact $p=\FScalePAbs$ with $n=4$.
  \item The relative effect did not ($\rho=\FScaleRhoRel$).  The edge input removed most of whatever wrong releases a family produced.  Stronger models simply produced fewer.
  \item With four families this is descriptive.
\end{itemize}

\subsection{Large-menu replication}
\label{sec:results-lm}

\begin{table}[t]
\caption{Large-menu stage, model-free: share of hidden-wrong (FA) and hidden-correct blocked (FR) procedures over the complete large-menu space of the affected test tasks (1986 procedures in all tasks), all-consumer-assertions gate, task means.}
\label{tab:lm}
\small
\begin{tabular}{@{}lrr@{}}
\toprule
Visible inputs & FA & FR \\
\midrule
typical only & 0.278 & 0.014 \\
+ author edge input & 0.045 & 0.014 \\
+ blind writer, k=1 & 0.089 & 0.036 \\
+ blind writer, k=2 & 0.054 & 0.043 \\
+ blind writer, k=3 & 0.053 & 0.043 \\
+ guarded writer, k=1 & 0.135 & 0.036 \\
+ guarded writer, k=2 & 0.107 & 0.043 \\
+ guarded writer, k=3 & 0.106 & 0.043 \\
\bottomrule
\end{tabular}
\end{table}

\begin{table*}[t]
\caption{Large-menu stage, model repairs (test partition, 456 attempts per arm): correct completions / wrong releases under all consumer assertions for arms T and E; LH1 wrong releases per attempt T$\to$E (strict Holm $p$); LH2 change in wrong releases per attempt with typical inputs relative to the v4 menu (strict Holm $p$); LH3 change in correct completion relative to the v4 menu.}
\label{tab:la}
\small
\begin{tabular}{@{}llrrrrr@{}}
\toprule
Model & Status & T & E & LH1 & LH2 & LH3 \\
\midrule
GPT-5.6-luna & descriptive & 391/29 & 401/12 & 0.074$\to$0.031 (0.378) & 0.018 (0.378) & $-$0.018 \\
Qwen3-8B & confirmatory & 241/108 & 244/27 & 0.253$\to$0.062 (<0.001) & 0.019 (0.634) & $-$0.047 \\
gpt-oss-20b & confirmatory & 310/85 & 334/15 & 0.212$\to$0.038 (<0.001) & 0.031 (0.335) & $-$0.026 \\
qwen3.6-35b-a3b & descriptive & 327/72 & 344/9 & 0.177$\to$0.023 (<0.001) & 0.057 (0.047) & $-$0.069 \\
\bottomrule
\end{tabular}
\end{table*}

\textbf{Model-free (Table~\ref{tab:lm}).}
\begin{itemize}
  \item On the large menu, the edge input lowered the share of wrong procedures released from \FLMFTypFA{} to \FLMFEdgeFA{} (\textbf{LM1} \emph{\FLLMOneDecision{}}, strict Holm $p$ \FLLMOneHolm{}).
  \item The unguarded blind writer lowered it to \FLMFWThreeFA{}, with false rejection \FLMFWThreeFR{} against \FLMFTypFR{} (\textbf{LM2} \emph{\FLLMTwoDecision{}}, Holm $p$ \FLLMTwoHolm{}).  The writer result thus passes a preregistered test on a larger space, for a single writer model (GPT-5.6-luna).
  \item The guarded writer again lagged (\FLMFGThreeFA{}).
  \item \textbf{LM3} is \emph{\FLLMThreeDecision{}} and goes the other way.  The typical-input gate released a smaller share of wrong procedures on the large menu than on the earlier one (\FLLMThreeB{}$\to$\FLLMThreeA{}, cluster means).  Most added wrong fragments fail visibly on typical inputs.  The \emph{share} of wrong programs released therefore depends on how the menu is built.  The headline shares (\FMFTypFA{}, \FLMFTypFA{}) hold only for their own menus, and are not properties of a gate.
  \item \textbf{Menu-independent view (post hoc, Table~\ref{tab:menufree}).}  We count, per task, the wrong programs that the visible inputs cannot tell apart from a correct one.  With typical inputs this \emph{number} grows with the menu, from \FQSmallTypTotal{} to \FQLargeTypTotal{} over the test tasks, and \FQSmallTypAny{} or \FQLargeTypAny{} tasks have at least one such program.  With the edge input it is \FQSmallEdgeTotal{} and \FQLargeEdgeTotal{}, in \FQSmallEdgeAny{} and \FQLargeEdgeAny{} tasks; with the blind writer \FQSmallWThreeTotal{} and \FQLargeWThreeTotal{}.  The task-level view of model repairs agrees.  For example, Qwen3-8B released a wrong repair in \FQQwenEightSmallTAny{} tasks with typical inputs and \FQQwenEightSmallEAny{} with the edge input on the earlier menu, and in \FQQwenEightLargeTAny{} and \FQQwenEightLargeEAny{} on the large menu.
\begin{table}[t]
\caption{Post-hoc, menu-independent view (52 affected test tasks).  Top: wrong programs that pass the all-assertions gate on the visible inputs, i.e. that the visible inputs cannot tell apart from a correct one (total, mean per task in parentheses), and tasks with at least one such program.  Bottom: tasks in which a model released at least one wrong repair in any of its 8 cells under all assertions, on the earlier and the large menu.}
\label{tab:menufree}
\small
\begin{tabular}{@{}lrrrrrrr@{}}
\toprule
& Programs & \multicolumn{2}{c}{T} & \multicolumn{2}{c}{E} & \multicolumn{2}{c}{W$_3$} \\
Menu & median & wrong & tasks & wrong & tasks & wrong & tasks \\
\midrule
earlier menu & 5 & 113 (2.17) & 41 & 14 (0.27) & 11 & 15 (0.29) & 12 \\
large menu & 30 & 445 (8.56) & 42 & 46 (0.88) & 17 & 62 (1.19) & 15 \\
\bottomrule
\end{tabular}

\medskip
\begin{tabular}{@{}lrrrr@{}}
\toprule
& \multicolumn{2}{c}{earlier menu} & \multicolumn{2}{c}{large menu} \\
Model & T & E & T & E \\
\midrule
GPT-5.6-luna & 5 & 1 & 7 & 3 \\
Qwen3-8B & 19 & 5 & 20 & 5 \\
gpt-oss-20b & 16 & 1 & 16 & 2 \\
qwen3.6-35b-a3b & 11 & 2 & 17 & 2 \\
\bottomrule
\end{tabular}
\end{table}

\end{itemize}

\textbf{Model repairs (Table~\ref{tab:la}).}
\begin{itemize}
  \item \textbf{LH1.}  The edge input cut wrong releases per attempt for Qwen3-8B (\FLQwenEightHOneB{}$\to$\FLQwenEightHOneA{}), gpt-oss-20b (\FLGptOssHOneB{}$\to$\FLGptOssHOneA{}) and qwen3.6-35b-a3b (\FLQwenMoeHOneB{}$\to$\FLQwenMoeHOneA{}).  All three decisions are positive, both confirmatory families included, so the preregistered cross-family criterion is \emph{\FLRepMet{}}.  GPT-5.6-luna again stayed near the ceiling (\FLLunaHOneB{}$\to$\FLLunaHOneA{}, Holm $p$ \FLLunaHOneHolm{}).
  \item \textbf{LH2 and LH3.}  Unlike the model-free space, the models' own typical-input wrong releases rose slightly on the large menu in every family, and their correct completion fell slightly.  Only qwen3.6-35b-a3b, a descriptive family, reached Holm $p<0.05$ (LH2 \FLQwenMoeHTwoDiff{}, $p$ \FLQwenMoeHTwoHolm{}).  These are descriptive changes, not significant in the confirmatory families.  The edge input still closed the gate.
  \item \textbf{Offline gates.}  The blind writer's gate W$_3$ released \FLQwenEightStepWThree{}, \FLGptOssStepWThree{} and \FLQwenMoeStepWThree{} of the open-weight families' wrong first-step repairs.  Their own unguarded self-tests released \FLQwenEightStepSraw{}, \FLGptOssStepSraw{} and \FLQwenMoeStepSraw{}.
\end{itemize}

\subsection{Hard distractors and two more writers}
\label{sec:results-hd}

\begin{table}[t]
\caption{Hard-distractor stage, model-free over the complete hard-menu space of the affected test tasks (2310 procedures in all tasks): FA and FR task means under all consumer assertions, and FA restricted to hard-origin programs (wrong options that match the legacy output on the typical inputs).}
\label{tab:hd}
\small
\begin{tabular}{@{}lrrr@{}}
\toprule
Visible inputs & FA & FR & FA hard \\
\midrule
typical only & 0.337 & 0.010 & 0.990 \\
+ blind writer (GPT-5.6-luna), k=3 & 0.058 & 0.042 & 0.153 \\
+ guarded writer, k=3 & 0.135 & 0.042 & 0.492 \\
+ author edge input & 0.058 & 0.010 & 0.168 \\
+ blind writer v2 (qwen3.6), k=3 & 0.094 & 0.083 & -- \\
+ blind writer v2 (Qwen3-8B), k=3 & 0.182 & 0.049 & -- \\
\bottomrule
\end{tabular}
\end{table}

\begin{table*}[t]
\caption{Hard-distractor stage, model repairs (test partition, 456 attempts per arm): correct completions / wrong releases under all consumer assertions in arms T and E; HA1 wrong releases per attempt T$\to$E (strict Holm $p$); HA2 change of typical-input wrong releases relative to the large menu.  Last four columns (descriptive, offline): share of each model's hidden-wrong first-step repairs released by the typical inputs plus three inputs of the GPT-5.6-luna writer, the qwen3.6 and Qwen3-8B writers (prompt v2), and the model's own self-tests.  $^{c}$ confirmatory family.}
\label{tab:ha}
\footnotesize
\begin{tabular}{@{}lrrrrrrrr@{}}
\toprule
Model & T & E & HA1 & HA2 & luna & qwen3.6 & Qwen3-8B & own \\
\midrule
GPT-5.6-luna & 385/42 & 394/12 & 0.107$\to$0.031 (0.063) & 0.033 (0.094) & 0.117 & 0.433 & 0.300 & 0.117 \\
Qwen3-8B$^{c}$ & 245/105 & 244/22 & 0.250$\to$0.050 (<0.001) & $-$0.003 (1.000) & 0.173 & 0.303 & 0.378 & 0.395 \\
gpt-oss-20b$^{c}$ & 306/88 & 339/17 & 0.218$\to$0.043 (<0.001) & 0.006 (0.757) & 0.104 & 0.392 & 0.384 & 0.416 \\
qwen3.6-35b-a3b & 317/80 & 347/11 & 0.198$\to$0.028 (<0.001) & 0.020 (0.225) & 0.048 & 0.373 & 0.349 & 0.317 \\
\bottomrule
\end{tabular}
\end{table*}

\textbf{Model-free (Table~\ref{tab:hd}).}
\begin{itemize}
  \item \textbf{The menu is harder.}  The hard distractors raised the typical-input gate's FA above the large menu's (HD3 \emph{\FHHDThreeDecision{}}, \FHHDThreeB{}$\to$\FHHDThreeA{} cluster means).  Typical inputs released \FHMFTypHard{} of the hard wrong programs, as constructed.
  \item \textbf{The edge input reduces but does not close the gap.}  It released \FHMFEdgeFA{} of all wrong programs (HD1 \emph{\FHHDOneDecision{}}, Holm $p$ \FHHDOneHolm{}), but \FHMFEdgeHard{} of the hard wrong programs still passed, above the 0.10 ceiling that the overall rate meets.  This is a mean over the \PVHardTasks{} test tasks that have hard wrong programs.  A post-hoc, descriptive analysis (Section~\ref{sec:posthoc-v2}) adds two points.  Pooled over all \PVHardWrong{} test-partition hard wrong programs the share is \PVHardPassEPooled{} (\PVHardPassE{} programs, all in \PVHardPassETasks{} tasks).  And it is a worst case over the menu, not a failure rate of model repairs: across all four families, the edge-input arm released a hard wrong fragment only \PVHardRelE{} times.  HD1 passes because hard programs are a minority of the wrong ones.  In absolute terms, the wrong programs that the visible inputs cannot tell apart from a correct one number \FFHardIndTyp{} with typical inputs, \FFHardIndEdge{} with the edge input and \FFHardIndWThree{} with the writer's inputs.  Under the edge input this count grows with the menu: \FQSmallEdgeTotal{}, \FQLargeEdgeTotal{} and \FFHardIndEdge{} on the three menus.
  \item \textbf{The GPT-5.6-luna writer behaves alike}: \FHMFWThreeFA{} overall and \FHMFWThreeHard{} of the hard wrong programs (HD2 \emph{\FHHDTwoDecision{}}).
  \item \textbf{The two v2 writers are weaker.}  The qwen3.6 writer reached \FHWOneFA{}, below the 0.10 ceiling.  But its inputs raised false rejection to \FHWOneFR{} against \FHTypFRTask{}, beyond the 0.05 tolerance, so WR1 is \emph{\FHWROneDecision{}}.  The Qwen3-8B writer reached only \FHWTwoFA{} (WR2 \emph{\FHWRTwoDecision{}}).
  \item \textbf{WR3.}  We found no evidence that the v2 prompt helps: on the first menu, the qwen3.6 writer with the new prompt released \FHWRThreeA{} against \FHWRThreeB{} for qwen3.6's own earlier self-tests (two-sided, Holm $p$ \FHWRThreeHolm{}).  With the earlier prompt, the GPT-5.6-luna writer outperformed qwen3.6's self-tests.
  \item Writer yield: the qwen3.6 writer kept \FHWriterMoeKept{} inputs in \FHWriterMoeTasks{} tasks (\FHWriterMoeParse{} unparseable answers); the Qwen3-8B writer kept \FHWriterEightKept{} in \FHWriterEightTasks{} (\FHWriterEightParse{} unparseable).
\end{itemize}

\textbf{Model repairs (Table~\ref{tab:ha}).}
\begin{itemize}
  \item \textbf{HA1.}  The edge input cut wrong releases per attempt for Qwen3-8B (\FHQwenEightHAOneB{}$\to$\FHQwenEightHAOneA{}), gpt-oss-20b (\FHGptOssHAOneB{}$\to$\FHGptOssHAOneA{}) and qwen3.6-35b-a3b (\FHQwenMoeHAOneB{}$\to$\FHQwenMoeHAOneA{}), all with strict Holm $p<0.001$.  The cross-family criterion is \emph{\FHRepMet{}}.  GPT-5.6-luna: \FHLunaHAOneB{}$\to$\FHLunaHAOneA{} (Holm $p$ \FHLunaHAOneHolm{}, descriptive).
  \item \textbf{HA2.}  Typical-input wrong releases did not change detectably relative to the large menu in any family.
  \item \textbf{Offline.}  Applied offline to the models' own wrong first-step repairs, the GPT-5.6-luna writer's inputs again let through the fewest.  The two v2 writers and the models' own self-tests let through several times more.  Across all writers we tested, the strongest writer gated best.  This is descriptive, and it means that the writer result depends on the commercial writer (GPT-5.6-luna); we tested no higher-priced GPT-5.6 tier as a writer or repairer.
\end{itemize}

\subsection{AI contract audit}

\begin{itemize}
  \item The two blind replicates agreed on \FAuditAgree{} of item verdicts ($\kappa=\FAuditKappa$).
  \item The yes-rates were R1 \FAuditYesOne{}, R2 \FAuditYesTwo{}, R3 \FAuditYesThree{}, R4 \FAuditYesFour{} and R5 \FAuditYesFive{}.
  \item Under the frozen rule, \FAuditFlagged{} of \FTasks{} tasks have at least one item on which both replicates said no.  The items were R4 for \FAuditNoFour{} tasks, R3 for \FAuditNoThree{}, R5 for \FAuditNoFive{} and R2 for \FAuditNoTwo{}.
  \item \textbf{R4 objections.}  Most R4 objections concern the file-preservation check p0, which the requirements do not state.  In a post-hoc blind review of R4 on a sample of contracts by three human reviewers (the first author and two non-author software practitioners), no reviewer upheld any of the \HRAuditRFourNo{} sampled R4 objections; it is not a human contract sign-off (Section~\ref{sec:threats}, Construct validity).
  \item \textbf{R2 objections.}  The R2 objections are substantive.  For two Node.js tasks the quoted entry documents a deprecation history rather than the removal itself.
  \item \textbf{Contracts unchanged.}  Contracts were not changed, as preregistered.
  \item \textbf{Post-hoc check of the objections.}  Redefining hidden correctness without p0 changes \FPostNoPChanged{} of the exhaustive hidden verdicts of the test tasks.  So does dropping p0 and legacy equality d0 together: \FPostNoPDChanged{} changes.  The checks p0 and d0 never decide a verdict that the consumer assertions do not already decide.
  \item \textbf{Sensitivity analysis.}  Excluding the \FAuditFlaggedTest{} flagged test tasks entirely leaves the edge-input contrast at \FAuditHMDiff{} ($p$ \FAuditHMP{}, \FAuditHMClusters{} clusters).
  \item \textbf{Strong-evidence criterion.}  Our predeclared strong-evidence criterion fails on two counts: M4a is not positive, and more than 10\% of tasks are AI-flagged.  The human sign-off of the earlier study is also not claimed.
\end{itemize}

\subsection{Contamination probe}

\begin{table}[t]
\caption{Contamination probe: hidden-correct first-step repairs without any release-note text (in parentheses: with the excerpts, typical inputs, full representation, no diff view, listed order).  Uniform random choice from each finite domain would be expected to be hidden-correct in 9.9/45, 3.8/13, 0.8/4 and 1.0/4 tasks.}
\label{tab:probe}
\small
\begin{tabular}{@{}lrrrr@{}}
\toprule
Model & CPython & Node.js & SQLite 3.53 & SQLite 3.54 \\
\midrule
GPT-5.6-luna & 37/45 (37) & 10/13 (10) & 3/4 (2) & 3/4 (3) \\
Qwen3-8B & 19/45 (24) & 6/13 (9) & 1/4 (0) & 1/4 (0) \\
gpt-oss-20b & 29/45 (28) & 7/13 (8) & 2/4 (1) & 3/4 (3) \\
qwen3.6-35b-a3b & 32/45 (34) & 10/13 (6) & 2/4 (3) & 2/4 (2) \\
\bottomrule
\end{tabular}
\end{table}

Without any release-note text, GPT-5.6-luna produced hidden-correct first-step repairs in \FLunaProbeCPy{} CPython tasks.  It solved \FLunaProbeSQFiveFour{} of the post-cutoff SQLite 3.54 tasks.  The open-weight families solved fewer, but every family was well above uniform random choice (\FRandCPy{} for CPython).  The template, requirement and visible inputs often suffice to infer the repair.  The probe therefore bounds what release notes add to \emph{repair}.  It does not bear on the gate results, which concern what reaches release.

\subsection{Post-hoc analyses added in this version}
\label{sec:posthoc-v2}

The analyses in this subsection were run after all results above were known.  They are descriptive, read only the frozen records, make no model call, and are not preregistered.

\textbf{What still passes the edge input.}  All \PVHardPassE{} hard wrong programs that pass the edge input sit in \PVHardPassETasks{} tasks, and they follow four mechanisms, all forms of point-versus-region coverage: one release-note input checks one point of the changed input region, while hidden inputs probe others (Table~\ref{tab:hard-mech}).  The edge input and the blind writer's inputs are complementary: only \PVHardPassBoth{} of \PVHardWrong{} hard wrong programs (\PVHardPassBothPooled{}) pass both.  What remains are near-miss constants and over-general guards, which call for boundary values around the documented limit rather than more inputs of the same kind.  Models rarely chose these programs: across the four families, hard wrong fragments were released \PVHardRelT{} times on typical inputs and \PVHardRelE{} times with the edge input.

\begin{table}[t]
\caption{Mechanisms of hard wrong programs that pass the release-note edge input (post hoc).}
\label{tab:hard-mech}
\small
\begin{tabularx}{\linewidth}{@{}>{\raggedright\arraybackslash}p{0.30\linewidth}>{\raggedright\arraybackslash}p{0.36\linewidth}Y@{}}
\toprule
Mechanism & Tasks (passing / hard wrong) & Caught by writer inputs \\
\midrule
Near-miss threshold or constant & py-int-max-str-digits (5/7), node-writable-highwatermark (1/9) & no \\
Over-general guard & py-math-pow-inf (2/10) & no \\
Untested input sub-region & py-fromisoformat-legacy (5/10), py-gzip-filename (8/8), py-randrange-float (5/6) & yes \\
Vacuous or inverted check & py-unittest-assertdictcontainssubset (5/10) & 3 of 5 \\
\bottomrule
\end{tabularx}
\end{table}

\textbf{The edge input on top of a model's own tests.}  Coding agents usually write their own tests.  On the test partition we took each family's wrong first-step candidates that pass its own self-test inputs (the S arm's inputs before the legacy-binding guard) and applied the edge input offline (Table~\ref{tab:own-tests}).  The edge input rejected most of them in every open-weight family.  For gpt-oss-20b and qwen3.6, the model's own inputs added nothing once the edge input was present.  In this study self-tests are inputs judged by the legacy-differential check, not agent-written assertions; the free-form addendum (Section~\ref{sec:freeform}) checks model-stated expected outputs, as data rather than test code.

\begin{table}[t]
\caption{Wrong first-step candidates that pass the model's own test inputs, and how many the edge input still rejects (test partition, post hoc).}
\label{tab:own-tests}
\small
\begin{tabular}{@{}lrrrr@{}}
\toprule
Family & wrong & pass own tests & also pass edge input & edge input rejects \\
\midrule
GPT-5.6-luna & \PVOwnLunaWrong{} & \PVOwnLunaPassOwn{} & \PVOwnLunaPassOwnE{} & \PVOwnLunaCaught{} (\PVOwnLunaCaughtRate{}) \\
Qwen3-8B & \PVOwnQwenEightWrong{} & \PVOwnQwenEightPassOwn{} & \PVOwnQwenEightPassOwnE{} & \PVOwnQwenEightCaught{} (\PVOwnQwenEightCaughtRate{}) \\
gpt-oss-20b & \PVOwnGptOssWrong{} & \PVOwnGptOssPassOwn{} & \PVOwnGptOssPassOwnE{} & \PVOwnGptOssCaught{} (\PVOwnGptOssCaughtRate{}) \\
qwen3.6-35b-a3b & \PVOwnQwenThreeSixWrong{} & \PVOwnQwenThreeSixPassOwn{} & \PVOwnQwenThreeSixPassOwnE{} & \PVOwnQwenThreeSixCaught{} (\PVOwnQwenThreeSixCaughtRate{}) \\
\bottomrule
\end{tabular}
\end{table}

\textbf{Why the local writers failed their tests.}  The qwen3.6 writer met the false-acceptance ceiling and failed only on false rejection, which two tasks drive.  In \texttt{py-imp-removed}, its inputs have a legacy output that depends on the removed \texttt{imp} module, so no correct repair can reproduce it.  In \texttt{py-asyncio-timeouterror}, its inputs of 0.049 and 0.051\,s straddle a 50\,ms deadline and do not reproduce reliably on the new runtime.  The Qwen3-8B writer failed mainly on output format (parse failures in 21 of 73 tasks), and about half of its false acceptance comes from tasks where it produced no valid input.  Both failure kinds have mechanical remedies (schema-constrained output, a determinism filter and a legacy-achievability filter); whether they suffice must be tested on held-out tasks.

\section{Free-form repairs}
\label{sec:freeform}

All stages above restrict repairs to documented fragments.  This section reports a preregistered addendum in which models write the whole new program themselves.  It was added in this version, after all results above were known.

\textbf{Preregistration and freeze.}  The addendum, its runner, evaluator and freeze script, the container executor and the container qualification records were hashed (SHA-256) and made read-only before the first free-form model call.  The addendum fixes the question, substrate, prompt contents, sampling, estimands, predictions and decision rules, and requires the earlier stages' statistical procedures.  One departure is disclosed: the analysis script was written after the runs had finished, and after the descriptive GPT-5.6-luna counts had been inspected, but before it was run on any of the data (Appendix~\ref{app:freeze}).  It implements only the frozen estimands, with the earlier stages' procedures: the cluster bootstrap, and the cluster sign-flip test, exact for at most 16 clusters and otherwise 200{,}000 Monte Carlo draws with a fixed seed, as in the fragment stages (Section~\ref{sec:design}).  The script's own record lists the Monte Carlo test as a deviation, but it is the procedure the addendum refers to.

\textbf{Design.}
\begin{itemize}
  \item \textbf{Substrate.}  The \VSTasks{} affected test tasks of the earlier stages (\VSClusters{} clusters), with their typical inputs, edge input E, three blind-writer inputs W$_3$ and hidden contracts unchanged.
  \item \textbf{Prompt.}  Each model sees the same information as in the fragment arm: the requirement, the release-note excerpts, the old program, and the typical input with its old output.  It never sees E, W$_3$ or a hidden input.  It returns one JSON object with the full new program and 1--5 test cases, each an input and the output it believes the new program must produce.  These \emph{own tests} are data; no model-written test code exists.
  \item \textbf{Models and sampling.}  gpt-oss-20b (confirmatory; reasoning low), qwen3.6-35b-a3b (secondary; thinking off) and GPT-5.6-luna via the Codex CLI (descriptive; reasoning effort high).  Each model drew \VSK{} independent samples per task, with at most \VSMaxTok{} output tokens and no revision round.  The local models ran at temperature \VSTemp{} with seeds 1--\VSK{}.  The Codex CLI exposes no temperature or seed, so GPT-5.6-luna's samples are independent stateless calls at the service's default sampling.
  \item \textbf{Execution.}  Model-written programs run only in a container sandbox, never on the host: a new container per execution, no network, read-only root file system, all capabilities dropped, no new privileges, 64 processes, 512\,MB, one CPU, unprivileged user.  The container images pin the same runtime versions as the host.
  \item \textbf{Gates.}  Every clean candidate is executed once per input, and the gates T, E and W$_3$ and the hidden verdict use exactly the rules of the earlier stages.  The own-tests verdict O passes if the program's output equals the model's stated output on every own test and the run is clean.  A candidate that contains a literal copy of the typical input's old output is flagged as hard-coded.
\end{itemize}

\textbf{Container qualification.}  Before the freeze, every gate verdict and the hidden verdict of \VSQualCand{} author-written candidates were computed both on the host and in the container.  Host and container verdicts differed in \VSQualDiff{} cases, and two host runs differed in \VSQualHostDiff{}.  At the level of raw outputs, \VSQualRawDiff{} of \VSQualExec{} executions differed, all in \VSQualRawTasks{} tasks whose output varies from run to run (temporary paths and names, the hash of NaN, an unseeded shuffle).  The frozen rule excludes tasks whose container verdicts differ; it excluded \VSExcluded{} tasks.

\textbf{Estimands.}  As before, a \emph{false acceptance} (FA) means that a gate releases a hidden-wrong repair.  Here FA$(g)$ is the share of candidates that run cleanly on the typical input, are hidden-wrong and still pass gate $g$, averaged over clusters.
\begin{itemize}
  \item \textbf{FF-H1} (primary, gpt-oss-20b): $\Delta=$ FA(E)$-$FA(T).  It is supported if $\Delta<0$ with Holm $p<0.05$.  The prediction was FA(T) $\ge0.25$ and FA(E) $\le0.10$, and the hypothesis is falsified if FA(E) $>0.15$ or the 95\% CI of $\Delta$ includes 0.
  \item \textbf{FF-H2} (gpt-oss-20b): the incremental catch rate ICR, the share of hidden-wrong candidates that pass T and their own tests and that E still rejects.  It is supported if the lower 95\% bound exceeds 0.25.  The prediction was ICR $\ge0.5$.
  \item The Holm family is \{FF-H1, FF-H2\} per model.  qwen3.6 forms a separate secondary family.  GPT-5.6-luna is descriptive.  A model with fewer than 20 clusters that have a hidden-wrong clean candidate is reported descriptively.
  \item The directional expectation, stated in advance, was that free-form FA(T) would exceed the \FMFTypFA{} of the exhaustive fragment space.
\end{itemize}

\begin{table}[t]
\caption{Free-form repairs (preregistered addendum).  FA is a cluster mean over clean hidden-wrong candidates; the CI is a 95\% cluster bootstrap; ICR is pooled over candidates that pass T and their own tests.  FF-H2 $p$ is a one-sided cluster-bootstrap $p$ against 0.25.  Wrong releases are per attempt (all responses).  Parenthesized values are descriptive.}
\label{tab:freeform}
\small
\begin{tabular}{@{}lrrr@{}}
\toprule
 & gpt-oss-20b & qwen3.6-35b-a3b & GPT-5.6-luna \\
 & (confirmatory) & (secondary) & (descriptive) \\
\midrule
Responses / parsed & \VSGptOssResp{} / \VSGptOssParsed{} & \VSQwenThreeSixResp{} / \VSQwenThreeSixParsed{} & \VSLunaResp{} / \VSLunaParsed{} \\
Clean on typical input & \VSGptOssClean{} & \VSQwenThreeSixClean{} & \VSLunaClean{} \\
Hidden-wrong (clusters) & \VSGptOssWrong{} (\VSGptOssWrongClusters{}) & \VSQwenThreeSixWrong{} (\VSQwenThreeSixWrongClusters{}) & \VSLunaWrong{} (\VSLunaWrongClusters{}) \\
FA(T) $\to$ FA(E) & \VSGptOssFAT{} $\to$ \VSGptOssFAE{} & \VSQwenThreeSixFAT{} $\to$ \VSQwenThreeSixFAE{} & \VSLunaFAT{} $\to$ \VSLunaFAE{} \\
FF-H1 $\Delta$ & \VSGptOssDelta{} & \VSQwenThreeSixDelta{} & \VSLunaDelta{} \\
\quad 95\% CI & [\VSGptOssDeltaLow, \VSGptOssDeltaHigh] & [\VSQwenThreeSixDeltaLow, \VSQwenThreeSixDeltaHigh] & [\VSLunaDeltaLow, \VSLunaDeltaHigh] \\
\quad Holm $p$ & \VSGptOssHOneHolm{} & \VSQwenThreeSixHOneHolm{} & (\VSLunaHOneHolm{}) \\
FF-H2 ICR ($n$) & \VSGptOssICR{} (\VSGptOssICRn{}) & \VSQwenThreeSixICR{} (\VSQwenThreeSixICRn{}) & \VSLunaICR{} (\VSLunaICRn{}) \\
\quad lower 95\% bound & \VSGptOssICRLow{} & \VSQwenThreeSixICRLow{} & \VSLunaICRLow{} \\
\quad Holm $p$ & \VSGptOssHTwoHolm{} & \VSQwenThreeSixHTwoHolm{} & (\VSLunaHTwoHolm{}) \\
Decision FF-H1 / FF-H2 & \VSGptOssHOneDec{} / \VSGptOssHTwoDec{} & \VSQwenThreeSixHOneDec{} / \VSQwenThreeSixHTwoDec{} & \VSLunaHOneDec{} \\
Wrong releases, T $\to$ E & \VSGptOssWrongRelT{} $\to$ \VSGptOssWrongRelE{} & \VSQwenThreeSixWrongRelT{} $\to$ \VSQwenThreeSixWrongRelE{} & \VSLunaWrongRelT{} $\to$ \VSLunaWrongRelE{} \\
Correct released, T / E & \VSGptOssCorrRelT{} / \VSGptOssCorrRelE{} & \VSQwenThreeSixCorrRelT{} / \VSQwenThreeSixCorrRelE{} & \VSLunaCorrRelT{} / \VSLunaCorrRelE{} \\
Vacuous own tests / hard-coded & \VSGptOssVacuous{} / \VSGptOssHardcoded{} & \VSQwenThreeSixVacuous{} / \VSQwenThreeSixHardcoded{} & \VSLunaVacuous{} / \VSLunaHardcoded{} \\
\bottomrule
\end{tabular}
\end{table}

\textbf{Post-run re-evaluation of timeouts.}  After the runs we found that the frozen executor, on a timeout, stopped only the container client: containers running looping programs kept running and could slow, and time out, executions that followed them.  We therefore decided, after the runs, to re-evaluate every response offline, without model calls, with an executor that is identical except that it names each container, kills it on a timeout and records the timeout and wall time of every execution.  The re-evaluation ran serially, one container at a time; every execution that timed out was run once more.  A verdict was to be replaced only where it differed from the original and the difference was explained by timeouts; a program that timed out again would keep its fail.  Over \VSReExec{} executions, \VSReTOFirst{} timed out, all of one qwen3.6 candidate (\VSReTOConf{} again on the repeat, a true infinite loop), and none in the other two models.  No gate verdict, hidden verdict or own-tests verdict differed from the original evaluation in any model (\VSReDiff{} differing candidates), so the analysis of the corrected evaluations equals the original one, which we report.

\textbf{Results against the predictions} (Table~\ref{tab:freeform}).
\begin{itemize}
  \item \textbf{gpt-oss-20b (confirmatory).}  \VSGptOssWrong{} clean candidates in \VSGptOssWrongClusters{} clusters were hidden-wrong.  The edge input lowered FA from \VSGptOssFAT{} to \VSGptOssFAE{} ($\Delta$ \VSGptOssDelta{}, 95\% CI [\VSGptOssDeltaLow, \VSGptOssDeltaHigh], Holm $p$ \VSGptOssHOneHolm{}), so FF-H1 is \emph{\VSGptOssHOneDec{}}.  The predicted FA(T) $\ge0.25$ was \VSGptOssPredFAT{}.  The predicted FA(E) $\le0.10$ was \emph{\VSGptOssPredFAE{}}: \VSGptOssFAE{} lies between the prediction and the falsification bound of 0.15, so the hypothesis is \VSGptOssFalsified{}, but the point prediction failed.  E rejected \VSGptOssICRCaught{} of the \VSGptOssICRn{} wrong candidates that passed T and their own tests (ICR \VSGptOssICR{}, lower bound \VSGptOssICRLow{}, Holm $p$ \VSGptOssHTwoHolm{}), so FF-H2 is \emph{\VSGptOssHTwoDec{}}, on a small sample.
  \item \textbf{qwen3.6-35b-a3b (secondary).}  FA fell from \VSQwenThreeSixFAT{} to \VSQwenThreeSixFAE{} ($\Delta$ \VSQwenThreeSixDelta{}, CI [\VSQwenThreeSixDeltaLow, \VSQwenThreeSixDeltaHigh], Holm $p$ \VSQwenThreeSixHOneHolm{}): FF-H1 is \VSQwenThreeSixHOneDec{}.  FA(E) $\le0.10$ was again \VSQwenThreeSixPredFAE{} (\VSQwenThreeSixFAE{}; \VSQwenThreeSixFalsified{}).  E rejected \VSQwenThreeSixICRCaught{} of \VSQwenThreeSixICRn{} candidates that passed T and their own tests (ICR \VSQwenThreeSixICR{}, lower bound \VSQwenThreeSixICRLow{}), so FF-H2 is \emph{\VSQwenThreeSixHTwoDec{}}.
  \item \textbf{GPT-5.6-luna (descriptive).}  Only \VSLunaWrong{} candidates in \VSLunaWrongClusters{} clusters were hidden-wrong, below the 20-cluster rule even apart from the descriptive role.  These wrong programs are hard for the edge input: FA went from \VSLunaFAT{} to \VSLunaFAE{}, and E rejected only \VSLunaICRCaught{} of the \VSLunaICRn{} that passed T and their own tests (ICR \VSLunaICR{}).  Applied to this model, the FF-H1 prediction would be \emph{\VSLunaFalsified{}} (FA(E) $>0.15$).  The bootstrap CI of $\Delta$ excludes 0, but the sign-flip test over \VSLunaWrongClusters{} clusters gives $p$ \VSLunaHOneP{}.  Its wrong releases per attempt stay low (\VSLunaWrongRelT{}$\to$\VSLunaWrongRelE{}) because it rarely writes a wrong program.
  \item \textbf{Directional expectation and collateral cost.}  Free-form FA(T) exceeded the exhaustive fragment-space value of \FMFTypFA{} in all three models.  The two figures have different denominators (model-written programs versus every documented program), so this is a coarse comparison.  Every hidden-correct clean candidate passed both T and E in every model, so the edge input cost no correct completion, and wrong releases per attempt fell in all three.
  \item \textbf{Own tests.}  Own tests that the stale old program also passes (vacuous) were rare (\VSGptOssVacuous{}, \VSQwenThreeSixVacuous{} and \VSLunaVacuous{} candidates), and \VSQwenThreeSixHardcoded{} qwen3.6 candidates hard-coded the typical output.
\end{itemize}

\textbf{Limitations of the addendum.}
\begin{itemize}
  \item \textbf{Sampling differs from the earlier stages.}  The fragment stages decode greedily (temperature 0).  The addendum samples at temperature \VSTemp{} with \VSK{} samples per task, as an independent review recommended (temperature 0 would give one candidate per task); the vendors' model cards also recommend non-greedy sampling.  GPT-5.6-luna uses the service default.  Free-form and fragment results therefore differ in decoding as well as in repair format.
  \item \textbf{One call for program and tests.}  The program and its own tests come from a single call.  Asking for tests may change the program, and own tests are input--output data rather than agent-written assertion code.
  \item \textbf{No revision round}, unlike the gated branch of the fragment stages.
  \item \textbf{Parse failures.}  Only \VSGptOssParsed{} of \VSGptOssResp{} gpt-oss-20b responses (\VSGptOssParseRate{}) and \VSQwenThreeSixParsed{} of \VSQwenThreeSixResp{} qwen3.6 responses (\VSQwenThreeSixParseRate{}) were valid JSON programs.  FA is conditional on parsed, clean candidates, so it describes the programs a model did deliver.  Unparsed responses count as non-releases in the per-attempt rates.
  \item \textbf{Small ICR samples.}  The FF-H2 samples are \VSGptOssICRn{} and \VSQwenThreeSixICRn{} candidates in a few clusters, so their bootstrap bounds are coarse.
  \item \textbf{Three secondary outcomes computed late.}  Three preregistered descriptive outcomes were missing from the analysis output and were computed afterwards by a separate script (descriptive, no tests).  \emph{Unconditional FA}, without conditioning on a clean typical run, was \VSSecGptOssUncFAT{}$\to$\VSSecGptOssUncFAE{} for gpt-oss-20b (\VSSecGptOssUncN{} hidden-wrong candidates), \VSSecQwenThreeSixUncFAT{}$\to$\VSSecQwenThreeSixUncFAE{} for qwen3.6 (\VSSecQwenThreeSixUncN{}) and \VSSecLunaUncFAT{}$\to$\VSSecLunaUncFAE{} for GPT-5.6-luna (\VSSecLunaUncN{}), the same pattern as the conditional FA.  \emph{Own tests rejected hidden-correct candidates} often: \VSSecGptOssORej{} of \VSSecGptOssORejN{} (\VSSecGptOssORejShare{}) for gpt-oss-20b, \VSSecQwenThreeSixORej{} of \VSSecQwenThreeSixORejN{} (\VSSecQwenThreeSixORejShare{}) for qwen3.6 and \VSSecLunaORej{} of \VSSecLunaORejN{} (\VSSecLunaORejShare{}) for GPT-5.6-luna, because the models' stated expected outputs were wrong; own tests are therefore a noisy release gate for the local models.  In the \emph{SQLite 3.54 subset} (\VSSecSqlTasks{} tasks, after the models' likely training data), the edge input helped less: of the hidden-wrong clean candidates, \VSSecGptOssSqlPassT{} of \VSSecGptOssSqlWrong{} passed T and \VSSecGptOssSqlPassE{} passed E for gpt-oss-20b, and \VSSecQwenThreeSixSqlPassT{} of \VSSecQwenThreeSixSqlWrong{} and \VSSecQwenThreeSixSqlPassE{} for qwen3.6; GPT-5.6-luna's \VSSecLunaSqlWrong{} wrong SQLite candidates passed neither.  These counts are small.
  \item \textbf{Executor timeouts.}  The re-evaluation above was decided after the runs; its executor is a wrapper of the frozen one, not a frozen file.  It changed no verdict, so the preregistered analysis stands as run.
\end{itemize}

\section{Prior Evidence}
\label{sec:prior}

Four earlier layers of this project motivated the present design.  They keep their own denominators and are never pooled with it.

\subsection{Preregistered coverage study (the direct predecessor)}

The direct predecessor used the first 55 of the present tasks, three model families and three coverage levels: typical, author edge input, and LLM-written inputs.
\begin{itemize}
  \item \textbf{Model-free.}  Over the complete repair space, the author edge input lowered the all-assertions false-acceptance rate from \VHMOneB{} to \VHMOneA{} (cluster means; Holm $p$ \VHMOneHolm{}).
  \item \textbf{LLM inputs.}  LLM inputs went further, to \VHMTwoA{}, but raised false rejection by \VHMTwoFRInc{}.
  \item \textbf{Model families.}  Wrong releases per attempt fell for Qwen3-8B (\VQwenHOneB{}$\to$\VQwenHOneA{}) and gpt-oss-20b (\VGptOssHOneB{}$\to$\VGptOssHOneA{}).
\end{itemize}
Its freeze history, edge-input QA rule and contract review were open points that the present study addresses (Section~\ref{sec:design} and Appendix~\ref{app:freeze}).

\subsection{Typical-input study on two lineages}

The first, exploratory study used 33 tasks on CPython and Node.js, with one typical visible input per task and six release gates.  Its main findings were:
\begin{itemize}
  \item \textbf{Exhaustive gate adequacy.}  Over the complete repair space, runtime/schema checks released \SAAdqRuntimeFA{} of wrong repairs (task mean), and all consumer assertions and differential testing released \SAAdqConsumerCompleteFA{}.  The two coincided because each task had a single visible input.
  \item \textbf{GPT-5.6-luna.}  It was near the ceiling, and no preregistered contrast was significant.
  \item \textbf{Qwen3-8B.}  It released \SAQwenRuntimeFA{} wrong repairs under runtime checks and \SAQwenConsumerCompleteFA{} under all assertions, out of \SAQwenN{} attempts.  This difference was descriptive: Holm $p=\SAQwenHOneHolm$.
\end{itemize}

Two post-hoc analyses followed, neither of which required any model call:
\begin{itemize}
  \item \textbf{Requirement-only sensitivity.}  When only requirement-derived assertions are kept, the gate ordering is unchanged.  Differential testing then over-rejects (\ZRDifferentialFR{}), and the no-safe-repair labels shift: \VROnlyBoundaryShift{} correct programs appear.
  \item \textbf{Counterfactual.}  Moving one hidden edge input into the visible set and re-judging the existing candidates lowered the all-assertions false-acceptance rate to \VCFConsumerFA{}, with no false rejection.
\end{itemize}
The counterfactual reuses inputs that also define correctness, so it is optimistic.  The coverage study above and the present study test the same mechanism prospectively, with new inputs that are never hidden.

\subsection{Sealed deterministic study and visible-completion failures}

Table~\ref{tab:prior-main} reproduces the scope-neutral comparison from the sealed deterministic study (24 lineages, 120 task units):
\begin{itemize}
  \item the stale procedure completed \PriorBZeroJoint{} held-out units, and a further \PriorBZeroSilent{} were accepted but wrong;
  \item scope-neutral reconstruction (B1R) and local repair (B4) each completed \PriorBFourJoint{};
  \item the injected B1/B1C evaluator-qualification rows are listed in Appendix Table~\ref{tab:evaluator-qualification}.
\end{itemize}

\begin{table*}[t]
\caption{Prior evidence: sealed deterministic study, held-out test split (20 lineages, 100 task units).  Scope-neutral reconstruction B1R and local repair B4 share the finite candidate base and the 24-probe ceiling; B1R also reads the complete synthetic new-version document.  Resource columns are the only resources this layer measured (means per unit); tokens, latency and graph-construction effort were not measured.  The injected B1/B1C rows are in Appendix~\ref{app:evaluator-qualification}.}
\label{tab:prior-main}
\footnotesize\setlength{\tabcolsep}{4pt}
\begin{tabular}{@{}lrrrrrrrr@{}}
\toprule
Condition & Joint & affected & controls & silent & abstain & probes & doc.\ bytes & edits \\
\midrule
stale procedure (B0) & 24/100 & -- & -- & 48 & 0 & 0.00 & 0.00 & 0.00 \\
scope-neutral reconstruction (B1R) & 96/100 & 72 & 24 & 0 & 4 & 1.04 & 119.83 & 1.00 \\
contract-guided local repair (B4) & 96/100 & 72 & 24 & 0 & 4 & 2.60 & 11.16 & 1.00 \\
\bottomrule
\end{tabular}
\end{table*}

In a blind sample of six public APIFlow-Bench tasks, three repairs met the visible-completion rule but failed the upstream validator; Table~\ref{tab:external-cases} reads them mechanically from the retained records.  In each case the requirement asked for re-authentication and the candidate reused a stale token that the mock accepted.  The visible check verified form and availability, not the obligation.

\begin{table*}[t]
\caption{The three public tasks that met the frozen visible-completion rule (``the main request completes without an error status and every declared output field can be derived'') but failed the upstream validator.  Columns are read mechanically from retained records; the validator's per-predicate reasons were stored only as hashes, so which predicates failed is unknown and left blank.  Exploratory, post hoc.}
\label{tab:external-cases}
\small
\begin{tabularx}{\textwidth}{@{}lrrXlX@{}}
\toprule
Task & predicates & partial score & candidate operations & verdicts & decidable missing obligation \\
\midrule
\texttt{w02} & 4 & 0.75 & 1 call(s); auth header reused; edits 0 & visible: yes; upstream: fail & re-authentication via \texttt{/v3/oauth/token} requested but not performed (stale bearer accepted by the mock) \\
\texttt{w08} & 3 & 0.67 & 1 call(s); auth header reused; edits 0 & visible: yes; upstream: fail & re-authentication via \texttt{/v3/oauth/token} requested but not performed (stale bearer accepted by the mock) \\
\texttt{w14} & 2 & 0.50 & 1 call(s); auth header reused; edits 0 & visible: yes; upstream: fail & re-authentication via \texttt{/v3/oauth/token} requested but not performed (stale bearer accepted by the mock); requested second record; answer is the first record \\
\bottomrule
\end{tabularx}
\end{table*}

\subsection{Edit budget}

On ten official migration sources, a stated two-field edit budget produced a local-repair advantage that disappeared under a three-field budget: \PHSBudgetFullDiffTest{} each in a post-hoc replay.  A prospective replication on a 12-source bank (Appendix Table~\ref{tab:studyB}) is consistent with this reading.  With the budget unlimited, full and patch output were indistinguishable for GPT-5.6-luna (\SBBInfFullBeh{} versus \SBBInfPatchBeh{}).  Under the two-field budget the model abstained instead of exceeding the budget.  The present study therefore states an unlimited budget.

\section{Discussion}
\label{sec:discussion}

\textbf{What makes a release gate adequate here is its inputs, not its assertion family.}  On typical inputs, the all-assertions gate released \FMFTypFA{} of wrong repairs.  It released \FMFEdgeFA{} once one input exercised the documented change.  The effect replicated on \FNewTestClusters{} clusters built after it was first observed.  It held in every open-weight family without materially lowering correct completion (at most \RVMaxCCDrop{} of \RVMaxCCDropN{} attempts in any family and menu, within the preregistered $-0.05$ tolerance).

\textbf{The inputs can come from a blind commercial model.}  The objection to the earlier study was that the authors wrote both the edge inputs and the hidden inputs from the same note.  A writer that saw neither the program nor the hidden inputs, GPT-5.6-luna (a low-cost commercial tier, not a flagship model), reached \FMFWThreeFA{} with three inputs, at one call per task.  For practitioners this suggests a cheap pre-release step:
\begin{enumerate}
  \item ask an isolated model for a few release-note edge inputs;
  \item compute their expected outputs on the old runtime;
  \item gate the repair on them.
\end{enumerate}

\textbf{Do not let the writer decide which old outputs still bind.}  Our preregistered guard asked the writer to flag inputs on which the legacy output is no longer required.  It discarded exactly the inputs that revealed the change.  On the first menu it also prevented no false rejection, because the unguarded inputs caused none.  On the large and hard menus the writer's inputs did raise false rejection (W$_3$ \FLMFWThreeFR{} against \FLMFTypFR{} with typical inputs, Table~\ref{tab:lm}; \FHMFWThreeFR{} against \FHMFTypFR{}, Table~\ref{tab:hd}), and the guard removed none of it (G$_3$ \FLMFGThreeFR{} and \FHMFGThreeFR{}).  In the earlier study, LLM inputs did over-reject.  There the cause was runtime facts that no repair may preserve, which a stricter validity check on the old runtime may handle better than a model's judgment.

\textbf{Who writes the tests may matter more than whether they are independent.}  A natural hypothesis is that a model's own tests share its misreading of the release note.  Our post-hoc data do not support it.  On the \FQCommonTasks{} tasks where every writer has tests, the strongest writer gates best, and same-model cells are not worse than cross-model cells (\FQDiagCommon{} versus \FQOffCommon{}).  CodeT ranks programs by agreement with tests from the \emph{same} model~\cite{chen2023codet}.  Independence is also weaker than it looks: across twelve models, implementations from different models still fail on the same tests far more often than independence predicts~\cite{nogueira2026failureindependence}, so a second model is not by itself an independent check.  Our results are consistent with that design working when the model is strong.  For release gating, a weak repairer gains most from a strong test writer, whichever model repairs.  This matrix is post hoc and uses one prompt; it is a hypothesis for a preregistered test, not a confirmed result.

\textbf{Stronger models need the gate less, but not never.}  The absolute benefit of the edge input shrank with capability.  GPT-5.6-luna produced few wrong releases in any arm.  The relative benefit stayed high, and GPT-5.6-luna still shipped \FLunaTFA{} wrong repairs on typical inputs versus \FLunaBothFA{} with both input sources.

\textbf{A checklist for release gates after a runtime or API migration.}  Table~\ref{tab:checklist} collects the practical consequences of the results above.  Each row names the evidence it rests on; none of it requires running model-written code.

\begin{table}[t]
\caption{Practitioner checklist for gating LLM repairs after a documented migration, with the supporting result in this study.}
\label{tab:checklist}
\small
\begin{tabularx}{\linewidth}{@{}>{\raggedright\arraybackslash}p{0.47\linewidth}Y@{}}
\toprule
Do & Evidence \\
\midrule
Add at least one input that exercises each documented change, written from the release note & typical inputs released \FMFTypFA{} of wrong repairs; one edge input \FMFEdgeFA{} (\FNewFAEdge{} on unseen clusters) \\
Take expected outputs from the legacy program on the legacy runtime, not from a model & every added input in this study; no model-written oracle was needed \\
If no one can write the edge input, ask an isolated capable model that sees only the note & three blind GPT-5.6-luna inputs reached \FMFWThreeFA{} (descriptive on the first menu; preregistered tests LM2 and HD2 positive on the larger menus); two local-model writers failed \\
Do not let the model decide which legacy outputs still bind & the guarded writer discarded exactly the change-revealing inputs \\
Prefer the strongest available test writer over a second, weaker ``independent'' one & same-model cells were not worse than cross-model cells (post hoc, descriptive) \\
Keep the gate even for strong repairers & GPT-5.6-luna as repairer still shipped \FLunaTFA{} wrong repairs on typical inputs \\
Pair the release-note input with two or three writer inputs from the changed region & of \PVHardWrong{} hard wrong programs, \PVHardPassE{} pass the edge input, \PVHardPassW{} the writer's inputs, and only \PVHardPassBoth{} both (post hoc) \\
Do not treat a passing gate as proof & \FHMFEdgeHard{} of hard, legacy-matching wrong programs (task mean; \PVHardPassEPooled{} pooled, post hoc) still passed the edge input \\
\bottomrule
\end{tabularx}
\end{table}

\section{Threats to Validity}
\label{sec:threats}

Table~\ref{tab:threats} summarizes the main threats and what this study does about each.

\begin{table*}[t]
\caption{Threats to validity and how they are addressed.}
\label{tab:threats}
\small
\begin{tabularx}{\textwidth}{@{}lYY@{}}
\toprule
Threat & What we did & What remains \\
\midrule
Author-written edge and hidden inputs & blind writer that never sees program or hidden inputs; \FWriterOverlap{} of \FWriterKept{} inputs equal a hidden input & hidden inputs are still author-written \\
Effect seen before it was tested & fresh replication on \FNewTestClusters{} new clusters; freeze copies of the design text & M4c computable from the build before F0 (disclosed) \\
Contracts stricter than requirements & two blind AI audits per contract; post-hoc removal of p0 and d0 changes \FPostNoPDChanged{} verdicts & post-hoc blind review of R4 (assertion strictness) only, by three human reviewers (the first author and two non-author software practitioners), not preregistered: none upheld any of the audit's \HRAuditRFourNo{} R4 rejections in the sample, and a majority rejected \HRMajorityNo{} contract; R1--R3 and R5 reviewed only by a blind LLM; R2 class-label objections unresolved \\
One model in three roles & writer $\times$ repairer matrix (post hoc); two further preregistered writers (qwen3.6, Qwen3-8B) & only GPT-5.6-luna's writer passed; auditor and writer are the same model \\
Menu composition & hard-distractor stage (\FHHard{} legacy-matching distractors); menu-independent counts & distractors remain author-generated by documented operators \\
Strong-evidence criterion & reported against its frozen definition & not met: M4a not positive; audit flags $>$10\% \\
Finite repair menu & second frozen stage with \FLTotal{} documented fragments; exhaustive model-free measurement & still author-written fragments; \FLSmallTasks{} tasks below 15 options; the free-form addendum uses other decoding, one call for program and tests and no revision round \\
Model coverage & four admitted families from three providers; power analysis & GPT-5.6-luna, a low-cost commercial tier, near the ceiling; no flagship-tier model tested as writer or repairer (GPT-5.6-sol only re-ran the audit, post hoc); gemma not admitted \\
Contamination & post-cutoff SQLite 3.54 subset; no-documentation probe & public Python and Node.js notes \\
Ecosystem breadth & three lineages, \FAffClusters{} documented changes & no web APIs; no further lineages \\
\bottomrule
\end{tabularx}
\end{table*}

\textbf{Construct validity.}
\begin{itemize}
  \item Hidden correctness treats legacy output as part of the obligation.  The AI audit judged several contracts stricter than their requirements, mainly through the file-preservation check.  Removing it, and legacy equality as well, changed no verdict, and none of the three human reviewers below upheld any of the sampled R4 objections.
  \item The R2 objections, that a quote documents a deprecation history rather than the removal, are not resolved by that analysis.  We report the flagged tasks in the artifacts.
  \item Human review covers one criterion.  Three human reviewers judged R4 (no assertion stricter than the requirement) blind for \HRItems{} contracts, \HRSampleFlagged{} flagged and \HRSampleUnflagged{} not flagged by the audit, shown in random order without the audit's conclusions.  They were the first author and two non-author software practitioners, who worked independently on blank sheets.  The review was decided after all results were known and was not preregistered.  The first author co-wrote the tasks and their contracts and knew the study's results when reviewing, so the review is not a human contract sign-off.
  \begin{itemize}
    \item On R4 the first author rejected \HRAuthorRFourNo{} contract and the two other reviewers \HRFOneRFourNo{} and \HRFTwoRFourNo{} (the second also marked \HRFTwoUnclear{} as unclear).  The frozen AI audit rejected \HRAuditRFourNo{} on R4, and no reviewer upheld any of them, so each reviewer agreed with the audit on \HRAgreeHumanAuditMin{}--\HRAgreeHumanAuditMax{} of \HRItems{} contracts (rejected versus not rejected).  The reviewers agreed pairwise on \HRAgreeAuthorFOne{}, \HRAgreeAuthorFTwo{} and \HRAgreeFOneFTwo{} of \HRItems{} contracts.  Fleiss' $\kappa$ among them is \HRFleissKappa{} (\HRFleissKappaUnclear{} if unclear counts as a rejection); with so few rejections it is dominated by chance agreement and is unstable, so we do not interpret it.
    \item An independent blind LLM review of the same contracts (a different model from the auditor, not a human) rejected none on R4 and agreed with each reviewer on \HRAgreeHumanLLMMin{}--\HRAgreeHumanLLMMax{} of \HRItems{}.  The auditor was GPT-5.6-luna (reasoning effort high), and an audit rejection on R4 required both of its independent replicates to reject; the LLM reviewer was Claude Opus 5.5.  On this criterion the human judgments were much closer to the latter.  With three reviewers, one criterion, \HRItems{} contracts and a single run of each reviewer, we report this only as a descriptive observation, not as evidence about either model.  A post-hoc re-run of the identical frozen audit prompt, rubric and adjudication rule with GPT-5.6-sol (OpenAI's flagship tier of GPT-5.6)\footnote{OpenAI model page, \url{https://developers.openai.com/api/docs/models/gpt-5.6-sol} (accessed 2026-11-12): ``GPT-5.6 flagship model for complex professional work''.}, two replicates on the same \HRItems{} contracts, rejected \HRSolRFourNo{} on R4, upholding \HRSolUpheld{} of the \HRAuditRFourNo{} rejections of the frozen GPT-5.6-luna audit, and agreed with the human majority on \HRSolAgreeMajority{} of \HRItems{} contracts.  So the strictness gap is not specific to the low-cost tier: on R4 the human reviewers and the Claude reviewer judged the contracts far more leniently than either GPT-5.6 tier.  This re-run is descriptive, with a single run per replicate, and was not preregistered.
    \item Two of the three reviewers rejected one contract, \texttt{\HRMajorityTask}, which the audit had rejected under R3 and marked soft under R4 and R5.  It is in the test partition and, as an AI-flagged task, is already excluded by the sensitivity analysis that drops flagged tasks (Section~\ref{sec:results}).  Re-running the frozen analyses of all four stages without it changes \HRExclChanged{} of the \HRExclChecked{} preregistered decision outputs we compared (hypothesis decisions, Holm-adjusted significance, family status and replication criteria), so no preregistered conclusion changes; effect estimates move by at most \HRExclMaxDiff{}.  Two descriptive GPT-5.6-luna results weaken: its hard-menu HA1 Holm $p$ rises from \HRExclLunaHAOneHolmAll{} to \HRExclLunaHAOneHolm{}, and the bootstrap 95\% CI of its free-form $\Delta$ becomes [\HRExclLunaCILow, \HRExclLunaCIHigh] (from [\VSLunaDeltaLow, \VSLunaDeltaHigh]), so its upper bound reaches 0; its sign-flip test was already not significant.
    \item The first author's rejection, \texttt{sqlite-indexes-pattern}, was also flagged by the audit and the LLM reviewer (both under R3) and marked unclear by one other reviewer: its legacy-equality check requires whole-output equality, although the requirement accepts any listing layout.  That task is in the development partition, so no test-partition result uses it.  Excluding it alone changes qwen3.6's development false-acceptance rate from \HRDevFAAll{} to \HRDevFAExcl{}; both are below the 0.15 threshold for confirmatory status, so that family's preregistered status is unchanged.  Removing all \HRAnyHumanNo{} contracts that any reviewer rejected changes \HRExclAnyChanged{} of the preregistered decision outputs.
    \item R1--R3 and R5 were reviewed only by the LLM reviewer, which flagged \HRLLMFlagged{} contracts on any criterion against \HRAuditFlagged{} for the audit (task-level agreement \HRLLMTaskAgree{}).  No human has checked those criteria, including the R2 class-label objections, and three reviewers on one criterion are not a full human sign-off.
  \end{itemize}
  \item Only the commercial writer (GPT-5.6-luna, the low-cost tier of GPT-5.6) passed its tests; both local-model writers failed theirs, so the writer result is specific to that model.  Results may differ for flagship models, such as higher-priced GPT-5.6 tiers, which we did not test as writers or repairers.
  \item GPT-5.6-luna is the blind writer, the auditor and one of the repairers.  The audit may share the writer's reading of the release notes.  The writer's advantage on its own repairs is not an independent test, and the matrix shows the writer's lead holds on the other three families.
\end{itemize}

\textbf{Internal validity.}
\begin{itemize}
  \item The authors wrote the tasks, typical inputs, author edge inputs and hidden inputs.  Under the QA rule, the author edge input must make the stale procedure fail visibly.  The blind writer was not bound by that rule and reached a similar level, which limits but does not remove this concern.
  \item Model-free M4c was known at F0.  M4a, M4b and the audit were known before any repair call.  In the main stage one descriptive amendment followed, before any repair call; the large-menu and hard-distractor stages had their own freezes.  In the free-form addendum, the analysis script was written after the runs, and the timeout re-evaluation, the blind LLM review and the human review of the audit were all decided after the results were known.  The full timeline is in Appendix~\ref{app:freeze}.
\end{itemize}

\textbf{Statistical conclusion validity.}
\begin{itemize}
  \item The test partition has \FTestClusters{} clusters.  In the fragment stages, GPT-5.6-luna and qwen3.6-35b-a3b are descriptive by preregistered rules.  In the free-form addendum, gpt-oss-20b is confirmatory, qwen3.6 forms a separate secondary family with its own preregistered tests, and GPT-5.6-luna is descriptive.
  \item The scale analysis has four families.
  \item The cross-family thresholds were fixed for five families and applied to the four admitted.
  \item In the free-form addendum, the own-tests samples behind FF-H2 are small (\VSGptOssICRn{} and \VSQwenThreeSixICRn{} candidates), and gpt-oss-20b returned a parsable program in only \VSGptOssParseRate{} of responses.
\end{itemize}

\textbf{External validity.}
\begin{itemize}
  \item In the main stages, repairs are choices among documented fragments.  The free-form addendum (Section~\ref{sec:freeform}) shows that the edge-input effect holds for free-form programs of two local models but is much weaker for GPT-5.6-luna's wrong programs.  It samples at temperature \VSTemp{} rather than greedily, its programs run in a qualified container rather than on the host, and its analysis script was written after the runs.
  \item Three runtime lineages share release ancestries within a lineage.  We make no claim about web APIs or other ecosystems.
\end{itemize}

\textbf{Contamination.}  Python and Node.js release notes are public.  The SQLite 3.54.0 subset post-dates the models' likely training.  The no-documentation probe is descriptive.

\textbf{Run conditions.}
\begin{itemize}
  \item Local runs passed their admission checks.
  \item The launcher records only point readings of machine load.
  \item gemma-4-12b-qat was not admitted.
  \item In the free-form addendum, the frozen container executor stopped only the client of a timed-out execution, so looping programs kept running and could slow later executions.  A post-run re-evaluation with a corrected executor changed no verdict (Section~\ref{sec:freeform}).
\end{itemize}

\section{Conclusion}

On \FAffClusters{} documented changes in three real runtimes, the release check's inputs decided whether wrong LLM repairs shipped:
\begin{itemize}
  \item one release-note edge input cut the share of wrong repairs released from \FMFTypFA{} to \FMFEdgeFA{} over the complete repair space;
  \item the effect replicated on unseen changes, in three open-weight families, on a repair menu enlarged to \FLTotal{} documented fragments, and with \FHHard{} hard distractors added;
  \item a blind low-cost commercial writer (GPT-5.6-luna) supplied inputs of nearly the same value without seeing the program, while two local-model writers did not pass;
  \item letting the writer decide which legacy outputs still bind helped less, and so did weaker models' self-tests; a post-hoc comparison points to writer capability rather than lack of independence.
\end{itemize}
The contracts passed an AI audit only in part.  The disputed preservation and legacy-equality checks change no verdict, the class-label disputes remain open, and our predeclared strong-evidence criterion is not met.  In a preregistered free-form addendum, the edge input also cut false acceptance of model-written wrong programs for gpt-oss-20b and qwen3.6, missing the predicted level narrowly, but it caught few of GPT-5.6-luna's rarer wrong programs; closing that gap is future work.

\section*{Data Availability}

Task specifications, build records (tasks, visible oracles, hidden contracts and exhaustive verdicts), pinned release-note pages, prompts, every model call with its response (including the free-form programs and the container executor that ran them), run receipts, freeze hash sets with verbatim design copies, analysis outputs and the scripts that render every number and figure are available from the authors on request.  They are not posted on a public code host.  The hidden contracts and hidden inputs are the evaluation key of the benchmark, and publishing them would let future models be trained or tuned on them.  The records also contain verbatim responses of a commercial model service whose redistribution terms we have not cleared.  We will share the complete package with reviewers and with researchers who request it for reproduction.

\begin{acks}
We thank two independent reviewers for the blind contract review.
\end{acks}

\section*{Use of AI Tools}
The study design, preregistration decisions and interpretation of results were made by the authors. LLM-based coding and writing assistants were used to implement and test experiment and analysis scripts and to draft and edit parts of the manuscript; the authors reviewed all of their outputs and take full responsibility for the content. AI tools are not authors. The language models evaluated in the experiments are objects of study, not authors or assistants.

\bibliographystyle{ACM-Reference-Format}
\bibliography{references-v4}

\appendix
\section{Tasks}
\label{app:tasks}

{\footnotesize
\begin{longtable}{@{}llllllll@{}}
\caption{The 73 tasks.  Class: S silent change, E explicit error, B no-safe-repair task, C unchanged control.  Correct: hidden-correct procedures in the finite domain.  Last two columns: whether the stale procedure fails a visible all-assertions check with typical inputs and with the author edge input added (by the edge-input QA rule always yes for affected tasks).}\label{tab:tasks}\\
\toprule
Task & Runtime & Origin & Class & Split & Correct & typical & edge \\
\midrule
\endfirsthead
\toprule
Task & Runtime & Origin & Class & Split & Correct & typical & edge \\
\midrule
\endhead
\texttt{py-binascii-hqx} & cpython & v2 & B & test & 0/6 & yes & yes \\
\texttt{py-distutils-looseversion} & cpython & v2 & B & test & 0/18 & yes & yes \\
\texttt{py-shuffle-random-param} & cpython & v2 & B & test & 0/6 & yes & yes \\
\texttt{py-control-csv} & cpython & v2 & C & development & 1/4 & no & no \\
\texttt{py-control-json-canonical} & cpython & v2 & C & development & 1/4 & no & no \\
\texttt{py-control-sha256} & cpython & v2 & C & test & 2/4 & no & no \\
\texttt{py-collections-abc-alias} & cpython & v2 & E & test & 3/6 & yes & yes \\
\texttt{py-imp-removed} & cpython & v2 & E & test & 2/10 & yes & yes \\
\texttt{py-inspect-getargspec} & cpython & v2 & E & development & 1/5 & yes & yes \\
\texttt{py-int-max-str-digits} & cpython & v2 & E & test & 1/5 & yes & yes \\
\texttt{py-locale-format} & cpython & v2 & E & test & 2/6 & yes & yes \\
\texttt{py-random-sample-set} & cpython & v2 & E & test & 2/6 & yes & yes \\
\texttt{py-randrange-float} & cpython & v2 & E & test & 1/5 & yes & yes \\
\texttt{py-safeconfigparser} & cpython & v2 & E & development & 1/9 & yes & yes \\
\texttt{py-unittest-assertequals} & cpython & v2 & E & test & 2/6 & yes & yes \\
\texttt{py-asyncio-timeouterror} & cpython & v2 & S & test & 11/32 & no & yes \\
\texttt{py-enum-contains-value} & cpython & v2 & S & test & 2/6 & no & yes \\
\texttt{py-intenum-str} & cpython & v2 & S & test & 2/7 & yes & yes \\
\texttt{py-intflag-invert} & cpython & v2 & S & test & 1/6 & no & yes \\
\texttt{py-mkdtemp-absolute} & cpython & v2 & S & test & 1/6 & yes & yes \\
\texttt{py-strenum-format} & cpython & v2 & S & test & 2/6 & yes & yes \\
\texttt{py-sum-neumaier} & cpython & v2 & S & test & 1/6 & no & yes \\
\texttt{py-ssl-match-hostname} & cpython & v3 & B & development & 0/5 & yes & yes \\
\texttt{py-unicodedata-15} & cpython & v3 & B & test & 0/5 & no & yes \\
\texttt{py-asyncio-coroutine} & cpython & v3 & E & test & 1/5 & yes & yes \\
\texttt{py-asyncio-loop-param} & cpython & v3 & E & development & 3/15 & yes & yes \\
\texttt{py-configparser-readfp} & cpython & v3 & E & development & 1/5 & yes & yes \\
\texttt{py-gettext-lgettext} & cpython & v3 & E & test & 1/5 & yes & yes \\
\texttt{py-gzip-filename} & cpython & v3 & E & test & 1/5 & yes & yes \\
\texttt{py-inspect-formatargspec} & cpython & v3 & E & development & 1/5 & yes & yes \\
\texttt{py-parser-module} & cpython & v3 & E & test & 3/15 & yes & yes \\
\texttt{py-re-inline-flags} & cpython & v3 & E & test & 6/20 & yes & yes \\
\texttt{py-unittest-assertdictcontainssubset} & cpython & v3 & E & test & 2/5 & yes & yes \\
\texttt{py-flag-iteration} & cpython & v3 & S & test & 1/5 & no & yes \\
\texttt{py-asynchat-producer} & cpython & v4 & E & test & 2/15 & yes & yes \\
\texttt{py-chr-decimal-index} & cpython & v4 & E & development & 1/5 & yes & yes \\
\texttt{py-formatter-dumbwriter} & cpython & v4 & E & development & 1/10 & yes & yes \\
\texttt{py-os-bytearray-path} & cpython & v4 & E & development & 2/5 & yes & yes \\
\texttt{py-pkgutil-impimporter} & cpython & v4 & E & test & 1/5 & yes & yes \\
\texttt{py-random-seed-tuple} & cpython & v4 & E & test & 1/5 & yes & yes \\
\texttt{py-sqlite3-optimizedunicode} & cpython & v4 & E & test & 2/5 & yes & yes \\
\texttt{py-traceback-etype} & cpython & v4 & E & test & 2/5 & yes & yes \\
\texttt{py-zipimport-find-module} & cpython & v4 & E & test & 1/5 & yes & yes \\
\texttt{py-fromisoformat-legacy} & cpython & v4 & S & test & 1/5 & no & yes \\
\texttt{py-ipaddress-private} & cpython & v4 & S & test & 1/5 & no & yes \\
\texttt{py-math-pow-inf} & cpython & v4 & S & test & 1/5 & no & yes \\
\texttt{py-nan-hash-buckets} & cpython & v4 & S & test & 1/5 & no & yes \\
\texttt{py-tokenize-fstring} & cpython & v4 & S & test & 2/5 & no & yes \\
\texttt{node-createcipher} & node & v2 & B & development & 0/5 & yes & yes \\
\texttt{node-control-path-posix} & node & v2 & C & test & 2/4 & no & no \\
\texttt{node-control-sha256} & node & v2 & C & test & 1/4 & no & no \\
\texttt{node-fs-truncate-fd} & node & v2 & E & test & 2/5 & yes & yes \\
\texttt{node-process-assert} & node & v2 & E & test & 4/15 & yes & yes \\
\texttt{node-util-isdate} & node & v2 & E & test & 3/6 & yes & yes \\
\texttt{node-buffer-max-length} & node & v2 & S & test & 1/6 & no & yes \\
\texttt{node-dirent-path} & node & v2 & S & test & 2/6 & yes & yes \\
\texttt{node-global-navigator} & node & v2 & S & test & 3/6 & yes & yes \\
\texttt{node-stream-highwatermark} & node & v2 & S & test & 1/5 & no & yes \\
\texttt{node-zlib-bytesread} & node & v2 & S & development & 2/5 & yes & yes \\
\texttt{node-createdecipher} & node & v3 & B & development & 0/4 & yes & yes \\
\texttt{node-writable-highwatermark} & node & v3 & S & test & 1/5 & no & yes \\
\texttt{node-timers-active} & node & v4 & E & test & 1/5 & yes & yes \\
\texttt{node-buffer-poolsize} & node & v4 & S & test & 3/5 & no & yes \\
\texttt{sqlite-control-group-sum} & sqlite & v3 & C & test & 1/4 & no & no \\
\texttt{sqlite-control-json-extract} & sqlite & v3 & C & test & 2/4 & no & no \\
\texttt{sqlite-dotcmd-semicolon} & sqlite & v3 & E & development & 1/5 & yes & yes \\
\texttt{sqlite-go-terminator} & sqlite & v3 & E & test & 1/5 & yes & yes \\
\texttt{sqlite-rowid-alias-column} & sqlite & v3 & E & test & 1/5 & yes & yes \\
\texttt{sqlite-blob-zero-bytes} & sqlite & v3 & S & test & 2/5 & no & yes \\
\texttt{sqlite-fp-17-digits} & sqlite & v3 & S & test & 1/6 & no & yes \\
\texttt{sqlite-indexes-pattern} & sqlite & v3 & S & development & 1/5 & no & yes \\
\texttt{sqlite-column-numeric-align} & sqlite & v4 & S & test & 1/5 & yes & yes \\
\texttt{sqlite-empty-result-header} & sqlite & v4 & S & test & 1/5 & no & yes \\
\bottomrule
\end{longtable}

}

\FloatBarrier
\section{Freeze timeline and deviations}
\label{app:freeze}

\textbf{This study.}  All times are UTC on 2026-10-17.
\begin{center}
\small
\begin{tabular}{@{}ll@{}}
\toprule
Time & Event \\
\midrule
03:10 & F0 hash set written (351 files; design copy stored) \\
03:15 & blind writer finished (\FWriterTasks{} calls) \\
03:24 & AI audit finished (\FAuditCalls{} calls) \\
03:43 & writer levels attached; model-free family M4 computed \\
03:48 & F-test hash set written (794 files; design copy stored) \\
04:38 & GPT-5.6-luna run finished \\
later & local windows: Qwen3-8B, gpt-oss-20b, qwen3.6-35b-a3b admitted; gemma-4-12b-qat not admitted \\
\bottomrule
\end{tabular}
\end{center}

Between F0 and F-test, two hashed files changed:
\begin{itemize}
  \item the design document gained a disclosed amendment reporting the writer and audit results;
  \item the evaluator gained an offline, descriptive self-test gate.
\end{itemize}
No hypothesis, arm or decision rule changed.  Deviations:
\begin{itemize}
  \item the 1{,}000-simulation power record was produced after F-test from the frozen script;
  \item the cross-family thresholds are applied to four instead of five families;
  \item the contract-strictness analysis without p0 and d0 is post hoc.
\end{itemize}

\textbf{The large-menu stage.}  F0 (\FLFreezeZero{}; \FQFreezeZeroFiles{} files) was written on \FQFreezeZeroDate{} at \FQFreezeZeroTime{} UTC, before the build and before any call.  F-test (\FLFreezeTest{}; \FQFreezeTestFiles{} files) was written at \FQFreezeTestTime{} UTC, after the build and the model-free family LM and before any model call.  No hashed file changed between the two; only the build files were added.  The model-free result file itself lies outside the F-test hash set; its inputs (the build) are hashed.  Both freezes store a copy of the addendum.

\textbf{The hard-distractor stage.}  F0 (\FHFreezeZero{}; \FHFreezeZeroFiles{} files) was written at \FHFreezeZeroTime{} UTC, before the build and before any call.  F-test (\FHFreezeTest{}; \FHFreezeTestFiles{} files) was written at \FHFreezeTestTime{} UTC, after the build and family HD and before any call.  No hashed file changed between them.

\textbf{The free-form addendum and post-run steps.}  All times are UTC.
\begin{center}
\small
\begin{tabularx}{\linewidth}{@{}lY@{}}
\toprule
Time & Event \\
\midrule
\RVReviewSampleTime{} & sample for the blind contract reviews drawn (\HRSampleFlagged{} audit-flagged and \HRSampleUnflagged{} unflagged contracts, shuffled); key with the audit's verdicts sealed \\
\RVFFFreezeTime{} & free-form addendum frozen: hash set of \RVFFFreezeFiles{} files (addendum, runner, evaluator, freeze script, container executor, SQLite image recipe, container qualification records, task plan), SHA-256 of the hash file \RVFFFreezeHash{}; no free-form model call before \\
\RVFFLunaDone{} & GPT-5.6-luna free-form run and evaluation finished \\
\RVFFLocalDone{} & gpt-oss-20b and qwen3.6 runs and evaluations finished \\
\RVFFAnalysisTime{} & analysis script (SHA-256 \RVFFAnalysisHash{}) written after the runs and run once; implements the frozen estimands \\
\RVFFReevalTime{} & post-run timeout re-evaluation finished (decided after the runs; executor wrapper \RVFFReevalExecHash{}); no verdict changed; the three missing secondary outcomes computed post hoc on the same day \\
\RVLLMReviewTime{} & blind LLM review (Claude Opus 5.5) of the sampled contracts finished; post hoc \\
\RVHumanReviewTime{} & first author's review of R4 on the sampled contracts finished; post hoc, not preregistered \\
\HRFriendsDoneTime{} & both non-author reviews of R4 (independent, blank sheets, without the key) finished; post hoc, not preregistered \\
\bottomrule
\end{tabularx}
\end{center}
The frozen addendum is unchanged; notes after the freeze went only to the run log.

\textbf{The predecessor study.}  Between its two freezes, the design file and run driver changed.  The third family and its strong-evidence criterion were added after its model-free family had been computed.  Its F0 design text, not archived at the time, was reconstructed byte-exactly from the hashes.  The present study archives the design text at every freeze and hashes the freeze script itself.

\FloatBarrier
\section{Evaluator qualification (deterministic study)}
\label{app:evaluator-qualification}

\begin{table}[h]
\caption{Evaluator qualification (not a method comparison): the sealed B1 row and the post-audit B1C row contain an injected, scope-marker-triggered preservation perturbation.  Their rejections show that the hidden preservation evaluator detects it.}
\label{tab:evaluator-qualification}
\small
\begin{tabular}{@{}lrrr@{}}
\toprule
Row & Joint & silent & discordant with B4 \\
\midrule
B1 (sealed) & 88/100 & 12 & 8 \\
B1C (post-audit) & 88/100 & -- & 8 \\
\bottomrule
\end{tabular}
\end{table}

The sealed B1 row and the post-audit B1C row contain an injected, scope-marker-triggered preservation perturbation.  Their rejections show that the hidden preservation evaluator detects it.  They do not measure a repair-method effect.

\FloatBarrier
\section{Edit-budget replication (negative result)}
\label{app:studyB}

\begin{table}[t]
\caption{Study B (appendix, descriptive): stated edit budget $\times$ representation on the 12-source minimal-reproduction bank, Codex.  Joint: hidden acceptance, preservation and forbidden-effect contracts all hold.}
\label{tab:studyB}
\small
\begin{tabular}{@{}lllrrrr@{}}
\toprule
Budget & Repr. & Feedback & Joint & silent & abstain & mean edits \\
\midrule
2 & full & behavioral & 45/48 & 0 & 3 & 1.24 \\
2 & full & none & 45/48 & 0 & 3 & 1.24 \\
2 & patch & behavioral & 46/48 & 0 & 2 & 1.28 \\
2 & patch & none & 46/48 & 0 & 2 & 1.28 \\
$\infty$ & full & behavioral & 48/48 & 0 & 0 & 1.40 \\
$\infty$ & full & none & 48/48 & 0 & 0 & 1.40 \\
$\infty$ & patch & behavioral & 47/48 & 0 & 1 & 1.36 \\
$\infty$ & patch & none & 47/48 & 0 & 1 & 1.36 \\
\bottomrule
\end{tabular}
\end{table}

Behavioral feedback never triggered a revision.  With an unlimited budget, full and patch output were indistinguishable; with a two-field budget, shortfalls appear as abstentions.  This is consistent with the earlier ``local wins'' difference being a budget artifact.

\FloatBarrier
\section{Reproducibility}
\label{app:repro}

Everything below is available from the authors on request.  Nothing is published on a public code host.
\begin{itemize}
  \item \textbf{Runtimes.}  CPython 3.9.6 (\texttt{/usr/bin/python3}) and 3.12.15 (Homebrew); Node.js 20.20.2 and 24.21.0; SQLite CLI 3.51.0 (system) and 3.54.0 (Homebrew).  Binary hashes are recorded in each freeze.
  \item \textbf{Frozen materials.}  Task specifications, build records (tasks, visible oracles, hidden contracts, exhaustive verdicts), pinned release-note pages with retrieval time and SHA-256, prompts, parser, gates, analysis code and every freeze hash set (two per fragment stage, one for the free-form addendum) with verbatim design copies.
  \item \textbf{Model records.}  Every call with prompt hash, response text, usage and timing.  Also the qualification records of all five families, including the non-admitted one, the writer and audit outputs, and the self-tests of each family.
  \item \textbf{Derived records.}  The hidden evaluation, the independent checker's receipts, per-run analyses, the cross-family analysis, the post-hoc contract analysis, and the render manifests that hash every input of every number and figure in this paper.
  \item \textbf{Container images.}  \texttt{python:3.9.6-slim}, \texttt{python:3.12.15-slim}, \texttt{node:20.20.2-slim}, \texttt{node:24.21.0-slim}, and a locally built SQLite image with both pinned shells compiled from the official source tarballs; its recipe is in the free-form freeze.
  \item \textbf{Executors.}  The frozen container executor (v1) used for all free-form runs, and the post-run wrapper (v2, SHA-256 \RVFFReevalExecHash{}) used only for the timeout re-evaluation, with its equivalence check and per-execution timeout and wall-time records.
  \item \textbf{Free-form records.}  Every response, the parsed programs and own tests, the original, re-evaluated and corrected evaluation records, the primary and re-evaluated analysis outputs, and the post-hoc secondary outcomes.
  \item \textbf{Review records.}  The blind-review sample and sealed key, the LLM reviewer's and the three human reviewers' sheets, the post-hoc GPT-5.6-sol re-audit of the sampled contracts (\texttt{sol-audit-001}: plan, driver, call records, adjudication and comparison), and the agreement scripts and summary, which hash their inputs.
  \item \textbf{Replay without model access.}  All model-free results, the evaluation, the check and the figures can be recomputed from the records on the stated runtimes without any model call.
\end{itemize}

\FloatBarrier
\section{Artifacts}

Run directories are create-only.  An independent checker recomputes every release decision and hidden verdict with separately implemented assertion semantics; all four admitted runs pass.  All numbers in this paper are rendered by scripts from the frozen records.  Task records, contracts, writer and audit outputs, run records and receipts are available from the authors on request.

\end{document}